\documentclass[letterpaper,english,notitlepage,twocolumn, superscriptaddress]{revtex4-1}
\usepackage[T1]{fontenc}
\usepackage[latin9]{inputenc}
\usepackage{mathrsfs}
\usepackage{mathtools}
\usepackage{amsmath}
\usepackage{amssymb}
\usepackage{graphicx}
\usepackage{esint}

\makeatletter

\pdfpageheight\paperheight
\pdfpagewidth\paperwidth

\usepackage{braket}

\makeatother

\usepackage{babel}
\begin{document}
\title{Hierarchical structure of fluctuation theorems for a driven system
in contact with multiple heat reservoirs}
\author{Jin-Fu Chen}
\affiliation{School of Physics, Peking University, Beijing, 100871, China}
\author{H. T. Quan}
\thanks{Corresponding author: htquan@pku.edu.cn}
\affiliation{School of Physics, Peking University, Beijing, 100871, China}
\affiliation{Collaborative Innovation Center of Quantum Matter, Beijing 100871,
China}
\affiliation{Frontiers Science Center for Nano-optoelectronics, Peking University,
Beijing, 100871, China}
\date{\today}
\begin{abstract}
For driven open systems in contact with multiple heat reservoirs,
we find the marginal distributions of work or heat do not satisfy
any fluctuation theorem, but only the joint distribution of work and
heat satisfies a family of fluctuation theorems. A hierarchical structure
of these fluctuation theorems is discovered from microreversibility
of the dynamics by adopting a step-by-step coarse-graining procedure
in both classical and quantum regimes. Thus, we put all fluctuation
theorems concerning work and heat into a unified framework. We also
propose a general method to calculate the joint statistics of work
and heat in the situation of multiple heat reservoirs via the Feynman-Kac
equation. For a classical Brownian particle in contact with multiple
heat reservoirs, we verify the validity of the fluctuation theorems
for the joint distribution of work and heat.
\end{abstract}
\maketitle

\section{Introduction}

Work and heat are two fundamental quantities in thermodynamics, and
usually coexist in generic thermodynamic processes. In stochastic
thermodynamics, the definitions of work and heat are extended from
ensemble average quantities to trajectory functionals \citep{Jarzynski1997,Jarzynski1997a,Sekimoto1998}.
A remarkable achievement of the stochastic thermodynamics is the discovery
of the fluctuation theorems, which reformulate various versions of
the second law of thermodynamics from inequalities into equalities
\citep{Seifert2008,Esposito2009,Campisi2011,Sekimoto2011,Jarzynski2011,Seifert2012,LucaPeliti2021},
for example, the Jarzynski equality for work \citep{Jarzynski1997,Jarzynski1997a},
the exchange fluctuation theorem for heat \citep{Jarzynski_2004},
and the fluctuation theorem for entropy production \citep{Crooks1999,Seifert2005}.

Despite the coexistence of work and heat in generic thermodynamic
processes, on most occasions the fluctuation theorems of work and
those of heat were studied separately in the past. Either the system
is in contact with a single reservoir and meanwhile driven by an external
agent, or the system is in contact with multiple heat reservoirs but
without driving. For the first case, by defining the work as a trajectory
functional in stochastic thermodynamics, the work statistics and the
nonequilibrium work fluctuation theorems have been extensively studied
in various classical systems \citep{Jarzynski1997a,Sekimoto1998,Mazonka1999,Jarzynski2000,Hummer2001,Zon2003,Narayan2003,Speck2004,Kawai2007,Imparato2007,Maragakis2008,Sagawa2010,Speck2011,Kwon2013,Saha2014,Holubec2015,Gong2015,Gong2016,Hoang2018,Pagare2019,Salazar2020,Taniguchi2007,Taniguchi2008a,PhysRevE.74.026106,Then2008,Engel2009,PhysRevE.79.021122,Baiesi2006,PhysRevE.83.011104,PhysRevE.90.042146,PhysRevE.77.022105,PhysRevE.100.012127,Talkner2009,Jarzynski1997,Nicolis2017}.
The nonequilibrium work fluctuation theorems were later extended to
the quantum realm based on the two-point measurement definition of
the quantum fluctuating work \citep{Kurchan2000,Tasaki2000,Talkner2007,Deffner2008,Andrieux2008,Esposito2009,Campisi2009,Campisi2011,Hekking2013,Liu2014,Funo2018a}.
The consistency of the two seemingly unrelated definitions is justified
by the quantum-classical correspondence principle for the work statistics
\citep{Jarzynski2015,GarciaMata2017,Brodier2020,Funo2018a,PhysRevE.93.062108,PhysRevE.98.012132}.
A hierarchical structure of fluctuation theorems concerning work in
the case of a single heat reservoir has been clarified (see the supplemental
material of Ref. \citep{Hoang2018}). For the second case, the heat
statistics has been widely explored in various thermal transport models,
where the system is usually in contact with multiple heat reservoirs
\citep{Saito2007,Dubi2011,Ren2010,Ren2012,Thingna2012,Wang2013,Fogedby2014,Li2015,Kilgour2019,Aurell2020,Santos2020,Levy2020,Gupta2021}.
In the absence of external driving, the system finally reaches a nonequilibrium
steady state, where the heat exchange satisfies the exchange fluctuation
theorem \citep{Jarzynski_2004} and/or the Gallavotti-Cohen fluctuation
theorem \citep{Gallavotti1995}. The heat statistics has also been
studied in relaxation processes in the case of a single heat reservoir
but without driving \citep{Zon2004,Fogedby2009,Chatterjee2010,GomezSolano2011,Salazar2016,Funo2018,Imparato2007,Denzler2018,Salazar2019,Fogedby2020,Popovic2021,Pagare2019,Paraguassu2021a,Chen2021b,PhysRevE.95.052138}.
Besides the above two cases, there is the third case. That is, when
the system is in contact with multiple heat reservoirs and meanwhile
driven by an external agent. In this case, the fluctuation theorems
concerning work and/or heat has been largely unexplored so far (but
see Refs. \citep{Murashita2016,Pal2017,Lee2018}).

In this article, we study fluctuation theorems of the third case.
The system is weakly coupled to multiple heat reservoirs, and meanwhile
is driven by an external agent. We find that in this case the marginal
distributions of work or heat do not satisfy any fluctuation theorems.
But only the joint distribution of work and heat satisfies a family
of fluctuation theorems, which are derived in both classical and quantum
regimes. We discover a hierarchical structure of fluctuation theorems
for the joint distribution of work and heat from microreversibility
\citep{Campisi2011} of the dynamics by adopting a step-by-step coarse-graining
procedure. Thus, we put all fluctuation theorems into a unified framework.
This is an exhaustive list of all fluctuation theorems concerning
work and heat for a driven system in contact with multiple heat reservoirs.\textbf{
}Especially, the Jarzynski equality, the Crooks relation, the exchange
fluctuation theorem, and the Clausius inequality can all be recovered
under specific conditions. In addition, we also discover some new
fluctuation theorems that have not been reported previously.

We also propose a general method to calculate the joint statistics
of work and heat via the Feynman-Kac equation, and illustrate this
method with a classical Brownian particle in contact with multiple
heat reservoirs. For the breathing harmonic oscillator, analytical
results of the joint statistics of work and heat are obtained in both
the highly underdamped and the overdamped regimes, which recover the
known results of work distribution in the highly underdamped and the
overdamped regimes \citep{Salazar2020,Kwon2013}. Moreover, we can
also calculate the joint statistics of work and heat in the generic
underdamped regime, which has not been explored so far. The fluctuation
theorems for the joint distribution of work and heat are verified
through the characteristic function of work and heat.

This article is organized as follows. In Sec. \ref{sec:Joint-fluctuation-theorems},
we derive a family of fluctuation theorems organized in a hierarchy
for the joint distribution of work and heat in the situation of multiple
heat reservoirs. These fluctuation theorems can be grouped into three
categories, the detailed ones (at the trajectory level), the differential
ones (at the distribution level), and the integral ones. In Sec. \ref{sec:Joint-statistics-of},
we propose a general method to calculate the joint statistics of work
and heat, and illustrate this method with a classical Brownian particle
in contact with multiple heat reservoirs. The conclusion is given
in Sec. \ref{sec:Conclusion}.

\section{Fluctuation theorems of work and heat \label{sec:Joint-fluctuation-theorems}}

\begin{figure}
\includegraphics[width=5.5cm]{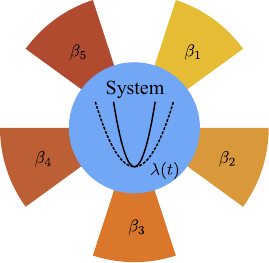}

\caption{Schematic of an open system under external driving. The system (illustrated
as a harmonic oscillator) is in contact with five heat reservoirs
simultaneously or sequentially. The inverse temperatures of the heat
reservoirs are $\beta_{\nu}$, $\nu=1,2,...,5$. The external driving
is depicted by the change of the frequency $\lambda(t)$ of the harmonic
potential.\label{fig:Schematic-of-driven}}
\end{figure}

As shown in Fig. \ref{fig:Schematic-of-driven}, a system of interest
described by the Hamiltonian $H_{S}(\gamma_{S}(t),\lambda(t))$ is
in contact with $N$ different heat reservoirs described by the Hamiltonians
$H_{\nu}(\gamma_{\nu}(t))$, $1\leq\nu\leq N$, where $\gamma_{S}(t)$
and $\gamma_{\nu}(t)$ denote phase-space points of the system and
the $\nu$-th heat reservoir. The external driving $\lambda(t),\;0\leq t\leq\tau$
is only applied to the system. We suppose that we can establish or
break the interaction between the system and any of the reservoirs
as we choose \citep{Jarzynski2000}. The initial state of the total
system is a product state $\rho_{\mathrm{tot}}^{\mathrm{i}}(\Gamma(0))=\rho_{S}^{\mathrm{i}}(\gamma_{S}(0))\otimes\pi_{1}(\gamma_{1}(0))\otimes...\otimes\pi_{N}(\gamma_{N}(0))$,
where $\rho_{S}^{\mathrm{i}}(\gamma_{S}(0))$ is the initial distribution
of the system, and $\pi_{\nu}(\gamma_{\nu}(0))=\exp[-\beta_{\nu}H_{\nu}(\gamma_{\nu}(0))]/Z_{\nu}(\beta_{\nu})$
is the canonical distributions of the $\nu$-th heat reservoir with
the inverse temperatures $\beta_{\nu}$ and the partition functions
$Z_{\nu}(\beta_{\nu})$.

For the classical dynamics, the state of the total system at time
$t$ is represented by a point $\Gamma(t)$ in the phase space. An
arbitrary trajectory is denoted as $\Gamma=(\gamma_{S},\gamma_{1},...,\gamma_{N})$,
where $\gamma_{S}=(x_{S},p_{S})$ and $\gamma_{\nu}=(x_{\nu},p_{\nu})$
with the position $x$ and the momentum $p$ are the trajectories
of the system and the $\nu$-th heat reservoir. For a specific external
driving $\lambda(t)$, the deterministic trajectory $\Gamma_{\mathrm{d}}$
is fully determined by the initial condition $\Gamma(0)$ of the total
system.

Through joint measurements of the internal energies of the system
and the heat reservoirs at the beginning and the end \citep{Talkner2009},
the heat exchange with the $\nu$-th heat reservoir along the trajectory
$\Gamma$ is given by
\begin{equation}
q_{\nu}(\Gamma)\coloneqq H_{\nu}(\gamma_{\nu}(0))-H_{\nu}(\gamma_{\nu}(\tau)),
\end{equation}
 and the trajectory work performed by the external driving according
to the first law is 
\begin{equation}
w(\Gamma)\coloneqq\int_{0}^{\tau}\dot{\lambda}\frac{\partial H_{S}}{\partial\lambda}dt.
\end{equation}
We assume the interactions between the system and the heat reservoirs
are weak and can be neglected in defining the work and the heat, and
the first law holds at the trajectory level by definition \citep{Sekimoto1998}

\begin{equation}
H_{S}(\gamma_{S}(\tau),\lambda(\tau))-H_{S}(\gamma_{S}(0),\lambda(0))=w(\Gamma)+\sum_{\nu}q_{\nu}(\Gamma).\label{eq:first_law_trajectory}
\end{equation}
In the following, we formulate the fluctuation theorems based on the
classical deterministic dynamics. We remark that the following results
can be parallel formulated in quantum systems weakly coupled to multiple
heat reservoirs (see Appendix \ref{sec:Joint-fluctuation-theorems_quantum-setup}).

\subsection{Detailed fluctuation theorems}

To formulate the most detailed fluctuation theorem, we define the
conditional probability density $\mathcal{P}(\Gamma|\Gamma(0))$ in
the trajectory space, where $\Gamma$ can be an arbitrary trajectory
(not necessarily the deterministic trajectory $\Gamma_{\mathrm{d}}$
determined by the equation of motion). The deterministic evolution
implies $\mathcal{P}(\Gamma|\Gamma(0))$ is nonzero only when $\Gamma=\Gamma_{\mathrm{d}}$.
In the reverse process, the trajectory is represented by $\tilde{\Gamma}(t)=\Theta[\Gamma(\tau-t)]$
with the time-reversal operation $\Theta$; the position and the momentum
are $\tilde{x}_{S(\nu)}(t)=\Theta[x_{S(\nu)}(\tau-t)]=x_{S(\nu)}(\tau-t)$
and $\tilde{p}_{S(\nu)}(t)=\Theta[p_{S(\nu)}(\tau-t)]=-p_{S(\nu)}(\tau-t)$;\textbf{
}the Hamiltonians are $\tilde{H}_{S}(\tilde{\gamma}_{S}(t),\tilde{\lambda}(t))\coloneqq\Theta[H(\gamma_{S}(\tau-t),\lambda(\tau-t))]$
and $\tilde{H}_{\nu}(\tilde{\gamma}_{\nu}(t))\coloneqq\Theta[H(\gamma_{\nu}(\tau-t))]$.
The control parameter is assumed to be even parity under the time-reversal
operation, and in the reverse process it changes as $\tilde{\lambda}(t)=\lambda(\tau-t)$.
As a consequence of microreversibility, the conditional probability
density of the trajectory in the reverse process satisfies

\begin{equation}
\tilde{\mathcal{P}}(\tilde{\Gamma}|\tilde{\Gamma}(0))=\mathcal{P}(\Gamma|\Gamma(0)),\label{eq:microreversibility}
\end{equation}
which can be regarded as the most detailed fluctuation theorem. Various
fluctuation theorems at different levels can be derived from Eq. (\ref{eq:microreversibility})
by adopting a step-by-step coarse-graining procedure.

We also prepare the heat reservoirs in their equilibrium states at
the initial time in the reverse processes $\tilde{\pi}_{\nu}(\tilde{\gamma}_{\nu}(0))=\Theta[\pi_{\nu}(\gamma_{\nu}(\tau))]$.
For given initial value $\gamma_{\nu}(0)$ and final value $\gamma_{\nu}(\tau)$
of the reservoir trajectories, the conditional probability density
of the system trajectory in the forward process is

\begin{align}
 & \mathcal{P}_{S}(\gamma_{S};\{\gamma_{\nu}(\tau)\},\{\gamma_{\nu}(0)\}|\gamma_{S}(0))\nonumber \\
 & \quad\coloneqq\sum_{\{\gamma_{\nu}\}}\mathcal{P}(\gamma_{S},\{\gamma_{\nu}\}|\gamma_{S}(0),\{\gamma_{\nu}(0)\})\prod_{\nu}\pi_{\nu}(\gamma_{\nu}(0)),\label{eq:2}
\end{align}
where $\mathcal{P}(\gamma_{S},\{\gamma_{\nu}\}|\gamma_{S}(0),\{\gamma_{\nu}(0)\})=\mathcal{P}(\Gamma|\Gamma(0))$,
and the summation over the trajectory $\gamma_{\nu}$ of the heat
reservoirs, as the path integral, is conditioned on the initial values
$\gamma_{\nu}(0)$ and the final values $\gamma_{\nu}(\tau)$. It
is similar to define $\mathcal{\tilde{P}}_{S}(\tilde{\gamma}_{S};\{\tilde{\gamma}_{\nu}(\tau)\},\{\tilde{\gamma}_{\nu}(0)\}|\tilde{\gamma}_{S}(0))$
for the reverse process. According to Eq. (\ref{eq:microreversibility}),
$\mathcal{\tilde{P}}_{S}$ and $\mathcal{P}_{S}$ are both nonzero
simultaneously. Then, we integrate over $\gamma_{\nu}(0)$ and $\gamma_{\nu}(\tau)$
and obtain the coarse-grained conditional probability density $\mathcal{P}_{S}(\gamma_{S};\{q_{\nu}\}|\gamma_{S}(0))\coloneqq\dotsint\prod_{\nu}\{d\gamma_{\nu}(0)d\gamma_{\nu}(\tau)\delta[q_{\nu}-H_{\nu}(\gamma_{\nu}(0))+H_{\nu}(\gamma_{\nu}(\tau))]\}\mathcal{P}_{S}(\gamma_{S};\{\gamma_{\nu}(0)\},\{\gamma_{\nu}(\tau)\}|\gamma_{S}(0))$.
The ratio of the probability densities in the reverse and the forward
processes is

\begin{equation}
\frac{\mathcal{\tilde{P}}_{S}(\tilde{\gamma}_{S};\{-q_{\nu}\}|\tilde{\gamma}_{S}(0))}{\mathcal{P}_{S}(\gamma_{S};\{q_{\nu}\}|\gamma_{S}(0))}=e^{\sum_{\nu}\beta_{\nu}q_{\nu}}.\label{eq:trajectoryprobabilityconditional}
\end{equation}
Here, $q_{\nu}$ ($-q_{\nu}$) denotes the heat exchange with the
$\nu$-th heat reservoir in the forward (reverse) process. Equation
(\ref{eq:trajectoryprobabilityconditional}) can be regarded as the
generalization of its single-reservoir version $\mathcal{\tilde{P}}_{S}(\tilde{\gamma}_{S};-q|\tilde{\gamma}_{S}(0))=\mathcal{P}_{S}(\gamma_{S};q|\gamma_{S}(0))\exp(\beta q)$
\citep{Kurchan1998,Crooks2000,Jarzynski2000,Maes2003,Maes2004,Seifert2005,Andrieux2007,GomezMarin2008a,Parrondo2009,Crooks2011}
to the situation of multiple heat reservoirs, and a coarse-grained
version of Eq. (\ref{eq:trajectoryprobabilityconditional}) has been
previously obtained in Ref. \citep{Jarzynski2000} (see Eq. (23) therein).

Together with the initial distribution $\rho_{S}^{\mathrm{i}}(\gamma_{S}(0))$
of the system, we obtain the complete trajectory probability density
$\mathcal{P}_{S}(\gamma_{S};\{q_{\nu}\})\coloneqq\mathcal{P}_{S}(\gamma_{S};\{q_{\nu}\}|\gamma_{S}(0))\rho_{S}^{\mathrm{i}}(\gamma_{S}(0))$
of observing the system trajectory $\gamma_{S}$ associated with the
heat exchange $q_{\nu}$ with the $\nu$-th heat reservoir. The ratio
of the complete trajectory probability densities is

\begin{equation}
\frac{\mathcal{\tilde{P}}_{S}(\tilde{\gamma}_{S};\{-q_{\nu}\})}{\mathcal{P}_{S}(\gamma_{S};\{q_{\nu}\})}=e^{\sum_{\nu}\beta_{\nu}q_{\nu}}\frac{\tilde{\rho}_{S}^{\mathrm{i}}(\tilde{\gamma}_{S}(0))}{\rho_{S}^{\mathrm{i}}(\gamma_{S}(0))},\label{eq:fluctuationtheorem_complete}
\end{equation}
The initial distributions $\rho_{S}^{\mathrm{i}}(\gamma_{S}(0))$
and $\tilde{\rho}_{S}^{\mathrm{i}}(\tilde{\gamma}_{S}(0))$ of the
system can be arbitrarily chosen. By choosing the equilibrium states
at the inverse temperature $\beta_{S}$ as the initial distributions
$\rho_{S}^{\mathrm{i}}(\gamma_{S}(0))=\pi_{S}^{\mathrm{i}}(\gamma_{S}(0))$
and $\tilde{\rho}_{S}^{\mathrm{i}}(\tilde{\gamma}_{S}(0))=\tilde{\pi}_{S}^{\mathrm{i}}(\tilde{\gamma}_{S}(0))$,
we can express the detailed fluctuation theorems concerning work and
heat at the trajectory level

\begin{align}
\frac{\mathcal{\tilde{P}}_{S}(\tilde{\gamma}_{S};\{-q_{\nu}\})}{\mathcal{P}_{S}(\gamma_{S};\{q_{\nu}\})} & =e^{-\beta_{S}[w(\gamma_{S})-\Delta F_{S}]+\sum_{\nu}(\beta_{\nu}-\beta_{S})q_{\nu}},\label{eq:trajectoryprobabilityobserved}
\end{align}
where $\Delta F_{S}=-\ln[Z_{S}^{\mathrm{f}}(\beta_{S})/Z_{S}^{\mathrm{i}}(\beta_{S})]/\beta_{S}$
is the free energy difference of the system. In the classical regime,
the work $w(\gamma_{S})$ is solely determined by the system trajectory.
The detailed fluctuation theorems (\ref{eq:trajectoryprobabilityconditional})
and (\ref{eq:trajectoryprobabilityobserved}) are the fine-grained
versions of the differential fluctuation theorems \citep{Jarzynski2000,Maragakis2008}.

\subsection{Differential fluctuation theorems}

By grouping the system trajectories $\gamma_{S}$ according to the
work $w$, the initial and final values $\gamma_{S}(0)$ and $\gamma_{S}(\tau)$
of the phase-space points, we obtain the conditional joint distribution
$P(w,\{q_{\nu}\},\gamma_{S}(\tau)|\gamma_{S}(0))\coloneqq\sum_{\gamma_{S}}\mathcal{P}_{S}(\gamma_{S};\{q_{\nu}\}|\gamma_{S}(0))\delta(w-\int_{0}^{\tau}\dot{\lambda}\partial_{\lambda}H_{S}dt)$.
The same coarse-graining procedure is applied to the reverse process.
The ratio of the conditional joint distributions of the reverse and
the forward processes follows from Eq. (\ref{eq:trajectoryprobabilityconditional})
as

\begin{equation}
\frac{\tilde{P}(-w,\{-q_{\nu}\},\tilde{\gamma}_{S}(\tau)|\tilde{\gamma}_{S}(0))}{P(w,\{q_{\nu}\},\gamma_{S}(\tau)|\gamma_{S}(0))}=e^{\sum_{\nu}\beta_{\nu}q_{\nu}}.\label{eq:differential_FT_classical_conditional}
\end{equation}
Similarly, the complete joint distribution of $w$, $q_{\nu}$, $\gamma_{S}(\tau)$
and $\gamma_{S}(0)$ follows as $P(w,\{q_{\nu}\},\gamma_{S}(\tau),\gamma_{S}(0))\coloneqq\sum_{\gamma_{S}}\mathcal{P}_{S}(\gamma_{S};\{q_{\nu}\})\delta(w-\int_{0}^{\tau}\dot{\lambda}\partial_{\lambda}H_{S}dt)$.
Please note that a coarse-grained version of Eq. (\ref{eq:differential_FT_classical_conditional})
has been previously obtained in Ref. \citep{Jarzynski2000} (see Eq.
(4) therein). For initial equilibrium states of the system, the ratio
of the complete joint distributions becomes

\begin{equation}
\frac{\tilde{P}(-w,\{-q_{\nu}\},\tilde{\gamma}_{S}(\tau),\tilde{\gamma}_{S}(0))}{P(w,\{q_{\nu}\},\gamma_{S}(\tau),\gamma_{S}(0))}=e^{-\beta_{S}[w-\Delta F_{S}]+\sum_{\nu}(\beta_{\nu}-\beta_{S})q_{\nu}}.\label{eq:differential_FT_classical_complete}
\end{equation}
Equations (\ref{eq:differential_FT_classical_conditional}) and (\ref{eq:differential_FT_classical_complete})
generalize the results in Refs. \citep{Jarzynski2000} and \citep{Maragakis2008}
to the situation of multiple heat reservoirs, respectively. These
two differential fluctuation theorems are the most detailed ones that
can be verified in experiments \citep{Hoang2018}.

From the complete joint distribution $P(w,\{q_{\nu}\},\gamma_{S}(\tau),\gamma_{S}(0))$,
we integrate over the initial and the final phase-space points and
obtain the joint distribution of work and heat $P(w,\{q_{\nu}\})\coloneqq\iint d\gamma_{S}(\tau)d\gamma_{S}(0)P(w,\{q_{\nu}\},\gamma_{S}(\tau),\gamma_{S}(0))$.
Since the right-hand side of Eq. (\ref{eq:differential_FT_classical_complete})
is independent of the phase-space points, the joint distributions
of work and heat in the reverse and the forward processes satisfy
a generalized Crooks relation

\begin{equation}
\frac{\tilde{P}(-w,\{-q_{\nu}\})}{P(w,\{q_{\nu}\})}=e^{-\beta_{S}[w-\Delta F_{S}]+\sum_{\nu}(\beta_{\nu}-\beta_{S})q_{\nu}},\label{eq:crook_workandheat_classical}
\end{equation}
We remark that Eq. (\ref{eq:crook_workandheat_classical}) has been
obtained previously in Refs. \citep{Murashita2016,Pal2017}, and a
simplified version for the case of a single heat reservoir has been
obtained in Ref. \citep{Talkner2009}. But the fine-grained versions
of Eq. (\ref{eq:crook_workandheat_classical}), Eqs. (\ref{eq:trajectoryprobabilityconditional})-(\ref{eq:differential_FT_classical_complete})
have not been reported so far.

By integrating over $\gamma_{S}(0)$, $w$ and $q$ in Eq. (\ref{eq:differential_FT_classical_complete}),
we obtain a generalized Hummer-Szabo relation

\begin{equation}
\left.\left\langle e^{-\beta_{S}[w-\Delta F_{S}]+\sum_{\nu}(\beta_{\nu}-\beta_{S})q_{\nu}}\right\rangle \right\vert _{_{\gamma_{S}(\tau)}}=\frac{\tilde{\pi}_{S}^{\mathrm{i}}(\tilde{\gamma}_{S}(0))}{\rho_{S}^{\mathrm{f}}(\gamma_{S}(\tau))},\label{eq:Hummer-Szabo_relation_classical}
\end{equation}
which generalizes the results in Ref. \citep{Hummer2001} to the situation
of multiple heat reservoirs. Here, $\left.\left\langle \cdot\right\rangle \right\vert _{\gamma_{S}(\tau)}$
denotes the average being conditioned on the given final value $\gamma_{S}(\tau)$
of the system phase-space point, and the marginal distribution $\rho_{S}^{\mathrm{f}}(\gamma_{S}(\tau))$
is obtained by integrating over other variables in $P(w,\{q_{\nu}\},\gamma_{S}(\tau),\gamma_{S}(0))$.

By integrating over $\gamma_{S}(\tau)$, $w$ and $q$ in Eq. (\ref{eq:differential_FT_classical_complete}),
we obtain a generalized Jarzynski equality for an initial $\delta$
distribution

\begin{equation}
\left.\left\langle e^{-\beta_{S}[w-\Delta F_{S}]+\sum_{\nu}(\beta_{\nu}-\beta_{S})q_{\nu}}\right\rangle \right\vert _{\gamma_{S}(0)}=\frac{\tilde{\rho}_{S}^{\mathrm{f}}(\tilde{\gamma}_{S}(\tau))}{\pi_{S}^{\mathrm{i}}(\gamma_{S}(0))}.\label{eq:generalized_Jarzynski_equality_classical}
\end{equation}
In Refs. \citep{Kawai2007,Gong2015,Hoang2018}, similar fluctuation
theorems were obtained for a single heat reservoir. Here we generalize
the results to the situation of multiple heat reservoirs. In Eqs.
(\ref{eq:Hummer-Szabo_relation_classical}) and (\ref{eq:generalized_Jarzynski_equality_classical}),
$\rho_{S}^{\mathrm{f}}(\gamma_{S}(\tau))$ and $\tilde{\rho}_{S}^{\mathrm{f}}(\tilde{\gamma}_{S}(\tau))$
are the final nonequilibrium distributions in the forward and the
reverse processes, respectively.

\subsection{Integral fluctuation theorems}

\begin{figure*}
\includegraphics[width=16cm]{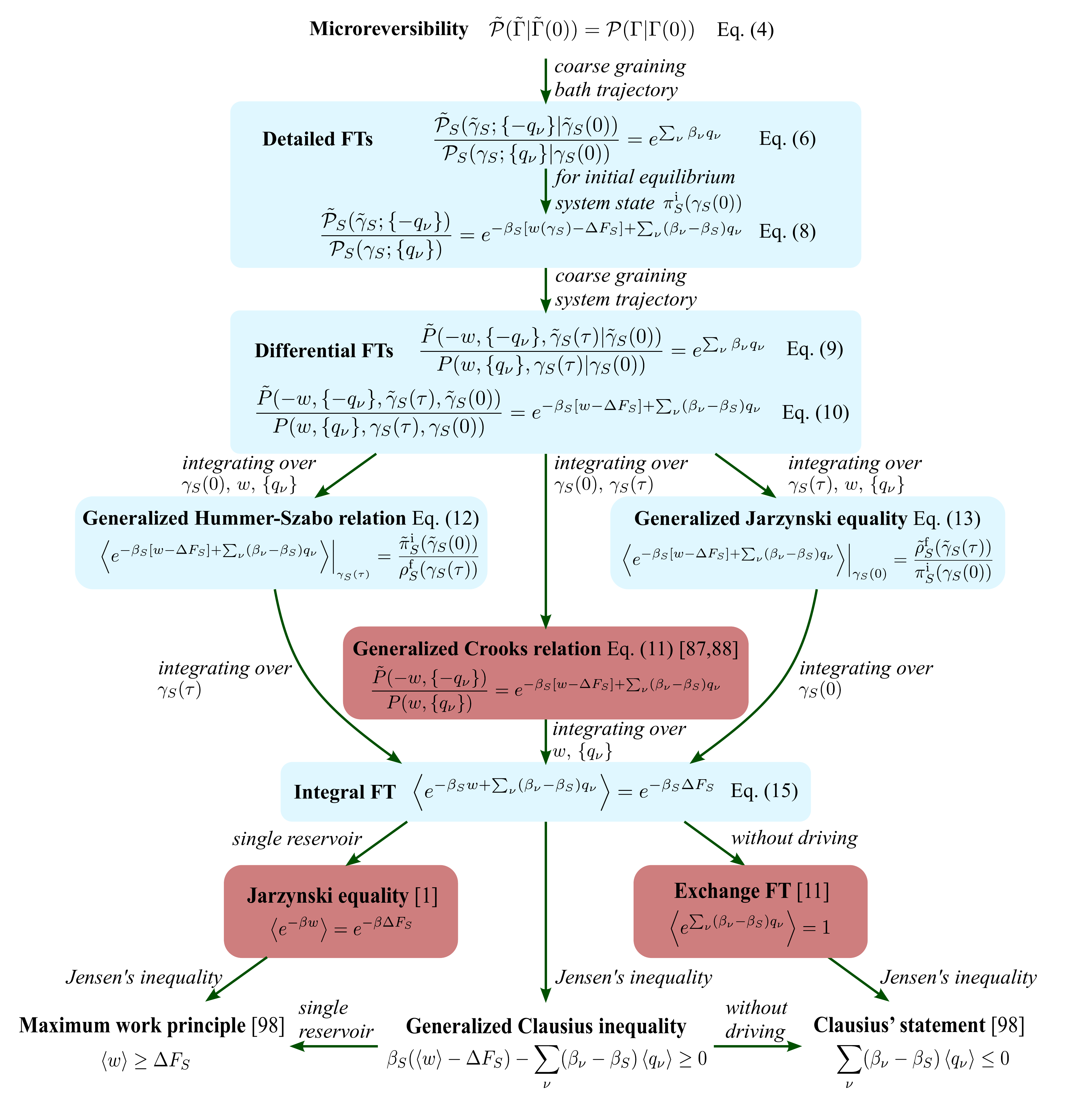}

\caption{Hierarchical structure of fluctuation theorems (FTs) for a driven
system in contact with multiple heat reservoirs. Fluctuation theorems
at different levels can be derived from microreversibility \citep{Campisi2011}
of the dynamics by adopting a step-by-step coarse-graining procedure.
The fluctuation theorems with red (dark) background have been obtained
previously, but these with blue (light) background have not been reported
so far. Please note that Eq. (11) has been previously obtained in
Refs. \citep{Murashita2016,Pal2017}, and coarse-grained versions
of Eqs. (\ref{eq:differential_FT_classical_conditional}) and (\ref{eq:trajectoryprobabilityconditional})
have been obtained in Ref. \citep{Jarzynski2000} (see Eqs. (4) and
(23) therein). A similar hierarchical structure of FTs for the work
distribution in the situation of a single heat reservoir can be found
in the supplemental material of Ref. \citep{Hoang2018}. \label{fig:Fluctuation-theorems-(FTs)}}
\end{figure*}

According to Eq. (\ref{eq:fluctuationtheorem_complete}), we can obtain
the integral fluctuation theorem

\begin{equation}
\left\langle e^{\sum_{\nu}\beta_{\nu}q_{\nu}}\frac{\tilde{\rho}_{S}^{\mathrm{i}}(\tilde{\gamma}_{S}(0))}{\rho_{S}^{\mathrm{i}}(\gamma_{S}(0))}\right\rangle =1.\label{eq:integral_fluctuation_theorem_general}
\end{equation}
This is a generalization of the unified integral fluctuation theorem
\citep{Seifert2008} to the situation of multiple heat reservoirs.
In case the system is initially prepared in an equilibrium state at
the inverse temperature $\beta_{S}$, we can express the integral
fluctuation theorem of work and heat as

\begin{equation}
\left\langle e^{-\beta_{S}w+\sum_{\nu}(\beta_{\nu}-\beta_{S})q_{\nu}}\right\rangle =e^{-\beta_{S}\Delta F_{S}},\label{eq:integral_fluctuation_theorem_work=000026heat}
\end{equation}
which can also be obtained from the above differential fluctuation
theorems (\ref{eq:differential_FT_classical_complete})-(\ref{eq:generalized_Jarzynski_equality_classical})
by integrating over the rest variables (see Fig. \ref{fig:Fluctuation-theorems-(FTs)}).
We would like to emphasize that previously it was believed that in
order to construct a fluctuation theorem for a driven open system,
e.g., the Jarzynski equality, the system is required to be initially
prepared in an equilibrium state whose temperature is the same as
that of the heat reservoir. But in Eq. (\ref{eq:integral_fluctuation_theorem_work=000026heat}),
we loosen this constraint, i.e., the initial temperature of the system
can be different from that (those) of the heat reservoir(s). Thus,
we extend the Jarzynski equality to a broader domain.

If initially the system has the same inverse temperature as those
of the heat reservoirs $\beta_{\nu}=\beta_{S}=\beta$ or the system
is isolated from the heat reservoir after the initial preparation,
the equality (\ref{eq:integral_fluctuation_theorem_work=000026heat})
is reduced to the Jarzynski equality $\left\langle \exp(-\beta w)\right\rangle =\exp(-\beta\Delta F_{S})$
\citep{Jarzynski1997}. On the other hand, if there is no external
driving, the equality (\ref{eq:integral_fluctuation_theorem_work=000026heat})
is reduced to the exchange fluctuation theorem of heat $\left\langle \exp[\sum_{\nu}(\beta_{\nu}-\beta_{S})q_{\nu}]\right\rangle =1$
\citep{Jarzynski_2004}. From the above analysis we can see that Eq.
(\ref{eq:integral_fluctuation_theorem_work=000026heat}) unifies the
Jarzynski equality \citep{Jarzynski1997} and the exchange fluctuation
theorem of heat \citep{Jarzynski_2004}. From Jensen's inequality,
the Jarzynski equality and the exchange fluctuation theorem lead to
the maximum work principle and Clausius' statement of the second law
\citep{BlundellBook2009}, respectively (see Fig. \ref{fig:Fluctuation-theorems-(FTs)}).

As a self-consistent check, one can derive the Clausius inequality
\citep{BlundellBook2009} from Eq. (\ref{eq:integral_fluctuation_theorem_work=000026heat}).
From Jensen's inequality, the integral fluctuation theorem (\ref{eq:integral_fluctuation_theorem_work=000026heat})
leads to a generalized Clausius inequality $-\beta_{S}\left(\left\langle w\right\rangle +\sum_{\nu}\left\langle q_{\nu}\right\rangle \right)+\sum_{\nu}\beta_{\nu}\left\langle q_{\nu}\right\rangle \leq-\beta_{S}\Delta F_{S}$.
In an irreversible cycle considered by Clausius, the free energy difference
is zero $\Delta F_{S}=0$ due to $\lambda(\tau)=\lambda(0)$, and
the state of the system returns to its initial state at the end of
the cycle, i.e., $\left\langle H_{S}(\gamma_{S}(\tau),\lambda(\tau))\right\rangle -\left\langle H_{S}(\gamma_{S}(0),\lambda(0))\right\rangle =\left\langle w\right\rangle +\sum_{\nu}\left\langle q_{\nu}\right\rangle =0$.
Under these two constraints, the inequality obtained from the integral
fluctuation theorem (\ref{eq:integral_fluctuation_theorem_work=000026heat})
is reduced to the Clausius inequality $\sum_{\nu}\beta_{\nu}\left\langle q_{\nu}\right\rangle \leq0$.

We remark that a fluctuation theorem relevant to Eq. (\ref{eq:integral_fluctuation_theorem_work=000026heat})
expressed by the internal energy change has been experimentally verified
quite recently \citep{Gomez2021}. Also, an integral fluctuation theorem
for the joint distribution of work and heat was reported for the cyclic
operation of heat engines \citep{Sinitsyn2011,Campisi2014,Chen2021a}.

We also notice in Ref. \citep{Jarzynski1999} an integral fluctuation
theorem is obtained as

\begin{equation}
\left\langle e^{\beta_{S}H_{S}(\gamma_{S}(0),\lambda(0))-\beta_{S}^{\prime}H_{S}(\gamma_{S}(\tau),\lambda(\tau))+\sum_{\nu}\beta_{\nu}q_{\nu}}\right\rangle =\frac{Z_{S}^{\mathrm{f}}(\beta_{S}^{\prime})}{Z_{S}^{\mathrm{i}}(\beta_{S})},\label{eq:Jarzynski_ft}
\end{equation}
which can be derived from Eq. (\ref{eq:integral_fluctuation_theorem_general})
by choosing $\rho_{S}^{\mathrm{i}}(\gamma_{S}(0))$ and $\tilde{\rho}_{S}^{\mathrm{i}}(\tilde{\gamma}_{S}(0))$
as two equilibrium states with different inverse temperatures $\beta_{S}$
and $\beta_{S}^{\prime}$. Only by setting $\beta_{S}^{\prime}=\beta_{S}$,
can the internal energy change in Eq. (\ref{eq:Jarzynski_ft}) be
rewritten into the combination of work and heat.

The hierarchical structure of fluctuation theorems are summarized
in Fig. \ref{fig:Fluctuation-theorems-(FTs)}. Figure \ref{fig:Fluctuation-theorems-(FTs)}
is an exhaustive list of fluctuation theorems concerning work and
heat for a driven system in contact with multiple heat reservoirs.
All the fluctuation theorems at different levels can be derived from
microreversibility \citep{Campisi2011} of the dynamics {[}Eq. (\ref{eq:microreversibility}){]}
by adopting a step-by-step coarse-graining procedure. The arrows indicate
that the fluctuation theorems in lower panels can be derived from
those in upper panels after the coarse-graining procedure, but the
reverse is not true. From Fig. \ref{fig:Fluctuation-theorems-(FTs)},
one can see how the previous known fluctuation theorems (with red
(dark) background) and the new fluctuation theorems (with blue (light)
background) discovered by us can be fitted into the hierarchical structure
of fluctuation theorems. In Appendix \ref{sec:Joint-fluctuation-theorems_quantum-setup},
we parallel formulate a similar hierarchical structure of fluctuation
theorems of work and heat in the quantum regime. We remark that the
detailed {[}see Eqs. (\ref{eq:trajectoryprobabilityconditional})
and (\ref{eq:trajectoryprobabilityobserved}){]} and the differential
fluctuation theorems {[}see Eqs. (\ref{eq:differential_FT_classical_conditional})
and (\ref{eq:differential_FT_classical_complete}){]} coincide in
the quantum regime {[}see Eqs. (\ref{eq:differential_FT_quantum_conditional})
and (\ref{eq:differential_FT_quantum_complete}){]} since a quantum
trajectory is defined in the two-point measurement scheme. Generally,
several different fluctuation theorems can be ascribed to the fluctuation
theorems for entropy production \citep{Crooks1999,Seifert2005,Yang2020,Manzano2018,Rao2018}.
In Appendix \ref{sec:fluctuation_theorems_entropy}, we formulate
the hierarchical structure of fluctuation theorems for entropy production.

\section{Joint statistics of work and heat in the situation of multiple heat
reservoirs\label{sec:Joint-statistics-of}}

Recently, the joint distribution of thermodynamic quantities and their
associated fluctuation theorems attract more and more attention \citep{Jarzynski1999,Talkner2009,Sinitsyn2011,GarciaGarcia2012,Campisi2014,Murashita2016,Pal2017,Miller2021,Gomez2021,Denzler2021,Chen2021a,Lee2018}.
The joint distribution of work and heat is a quantity of significant
importance to verify the above fluctuation theorems, but has not been
calculated for a nonequilibrium driving process previously. In the
following we study the calculation of the joint distribution function
of work and heat.

We propose a general method to calculate the joint statistics of work
and heat when the system is in contact with multiple heat reservoirs.
We would like to emphasize that our method can recover the known results
of the work distribution in the highly underdamped and the overdamped
regimes \citep{Salazar2020,Kwon2013}. Moreover, we can calculate
the joint statistics of work and heat in the generic underdamped regime.
Our method is illustrated via a classical Brownian particle with mass
$m$ moving in a time-dependent potential $\mathcal{U}(x,\lambda(t))$
with the control parameter $\lambda(t)$. We consider the Brownian
particle is in contact with multiple heat reservoirs with the inverse
temperatures $\beta_{\nu}$ and the friction coefficients $\kappa_{\nu}$.
From the point of view of the probability distribution $\rho(x,p,t)$,
the stochastic dynamics is described by the Kramers equation \citep{Kramers1940}
\begin{equation}
\frac{\partial\rho}{\partial t}=\mathscr{L}\left[\rho\right]+\sum_{\nu}\mathscr{D}_{\nu}\left[\rho\right],\label{eq:Kramers_equation}
\end{equation}
where $\mathscr{L}$ characterizes the deterministic evolution 
\begin{equation}
\mathscr{L}\left[\rho\right]=-\frac{\partial}{\partial x}\left(\frac{p}{m}\rho\right)+\frac{\partial}{\partial p}\left(\frac{\partial\mathcal{U}}{\partial x}\rho\right),
\end{equation}
and $\mathscr{D}_{\nu}$ characterizes the dissipation induced by
the $\nu$-th heat reservoir

\begin{equation}
\mathscr{D}_{\nu}\left[\rho\right]=\frac{\partial}{\partial p}\left(\kappa_{\nu}p\rho+\frac{\kappa_{\nu}m}{\beta_{\nu}}\frac{\partial\rho}{\partial p}\right).
\end{equation}
Even through we consider weak coupling between the system and the
heat bath, the system-bath interaction can be strong in the classical
Brownian motion model, especially in the overdamped situation. Our
results are still valid for the classical Brownian motion model. The
validity of the Kramers equation (\ref{eq:Kramers_equation}) is ensured
by short bath correlation time, and $\mathcal{U}(x,\lambda(t))$ represents
a renormalized potential that the system particle feels \citep{RevModPhys.92.041002}.
A consistent thermodynamic structure is restored by defining work
and heat based on the renormalized system Hamiltonian $H_{S}(x,p,\lambda)=p^{2}/(2m)+\mathcal{U}(x,\lambda)$.

In the following, we study the joint statistics of work and heat for
this specific model, and verify the generalized Crooks relation (\ref{eq:crook_workandheat_classical})
and the integral fluctuation theorem (\ref{eq:integral_fluctuation_theorem_work=000026heat}).

\subsection{Feynman-Kac equation for work and heat}

In the classical Brownian motion model, both work and heat are random
variables. With the joint distribution of work and heat $P(w,\{q_{\nu}\})$,
the characteristic function of work and heat is defined as $\chi^{w,\{q_{\nu}\}}(s,\{u_{\nu}\})\coloneqq\left\langle \exp[i(sw+\sum_{\nu}u_{\nu}q_{\nu})]\right\rangle $.
The characteristic function at time $\tau$ can be calculated through

\begin{equation}
\chi^{w,\{q_{\nu}\}}(s,\{u_{\nu}\})=\iint_{-\infty}^{\infty}\eta(x,p,\tau)dxdp,\label{eq:chiWQ}
\end{equation}
where $\eta(x,p,t)$ is a distribution function in the phase space
depending on the values of $s$ and $u_{\nu}$. The evolution of $\eta(x,p,t)$
is governed by the Feynman-Kac equation (also called the twisted Fokker-Planck
equation) \citep{Ren2012,Liu2014}
\begin{equation}
\frac{\partial\eta}{\partial t}=\mathscr{L}\left[\eta\right]+\sum_{\nu}e^{iu_{\nu}H_{S}}\mathscr{D}_{\nu}\left[e^{-iu_{\nu}H_{S}}\eta\right]+is\dot{\lambda}\frac{\partial\mathcal{U}}{\partial\lambda}\eta.\label{eq:Feynman-Kac_equation_joint}
\end{equation}
The initial condition is 
\begin{equation}
\eta(x,p,0)=\rho(x,p,0)=\frac{e^{-\beta_{S}H_{S}(x,p,\lambda(0))}}{Z_{S}^{\mathrm{i}}(\beta_{S})},\label{eq:initial_condition_underdamped}
\end{equation}
with the partition function $Z_{S}^{\mathrm{i}}(\beta_{S})=\iint_{-\infty}^{\infty}\exp[-\beta_{S}H_{S}(x,p,\lambda(0))]dxdp$
at the initial time. Previously, the Feynman-Kac equation was used
to calculate the work statistics and to prove the Jarzynski equality
\citep{Hummer2001}.

With the characteristic function, the integral fluctuation theorem
(\ref{eq:integral_fluctuation_theorem_work=000026heat}) can be rewritten
as 
\begin{equation}
\chi^{w,\{q_{\nu}\}}(i\beta_{S},\{i(\beta_{S}-\beta_{\nu})\})=e^{-\beta_{S}\Delta F_{S}}.\label{eq:chi_tau_WQ}
\end{equation}
Such an equality can be easily verified by noting that the solution
to the Feynman-Kac equation (\ref{eq:Feynman-Kac_equation_joint})
with $s=i\beta_{S}$ and $u_{\nu}=i(\beta_{S}-\beta_{\nu})$ is 
\begin{equation}
\eta(x,p,t)=\frac{e^{-\beta_{S}H_{S}(x,p,\lambda(t))}}{Z_{S}^{\mathrm{i}}(\beta_{S})}.\label{eq:eta_solution-1}
\end{equation}
We rewrite the generalized Crooks relation (\ref{eq:crook_workandheat_classical})
in terms of the characteristic function as

\begin{equation}
\frac{\tilde{\chi}^{w,\{q_{\nu}\}}(-s,\{-u_{\nu}\})}{\chi^{w,\{q_{\nu}\}}(i\beta_{S}+s,\{i(\beta_{S}-\beta_{\nu})+u_{\nu}\})}=e^{\beta_{S}\Delta F_{S}},\label{eq:crook_relation_by_characteristic_function}
\end{equation}
where $\tilde{\chi}^{w,\{q_{\nu}\}}(s,\{u_{\nu}\})$ is the characteristic
function in the reverse process. The equality (\ref{eq:crook_relation_by_characteristic_function})
can also be proven from the Feynman-Kac equation (\ref{eq:Feynman-Kac_equation_joint}),
and the proof is given in Appendix \ref{sec:Proof-of-Eq.}.

\subsection{Example: breathing harmonic oscillator}

As an example, we study the joint statistics of work and heat for
a Brownian particle in a breathing harmonic potential $\mathcal{U}(x,\lambda(t))=m\lambda^{2}(t)x^{2}/2$,
where the control parameter $\lambda(t)$ is the frequency. We consider
the situation of a single heat reservoir with the inverse temperature
$\beta$ and the friction coefficient $\kappa$. The system is initially
prepared in an equilibrium state with the inverse temperature $\beta_{S}$.
We would like to emphasize that the extension of the following calculation
to multiple heat reservoirs is straightforward.

In this situation, we assume $\eta(x,p,t)$ in a quadratic form
\begin{equation}
\eta(x,p,t)=\frac{\beta_{S}\lambda(0)}{2\pi}e^{-\frac{a}{2}\frac{p^{2}}{m}-\frac{b}{2}m\lambda^{2}(t)x^{2}-c\lambda(t)xp-\Lambda}.\label{eq:assumption_breathing}
\end{equation}
Substituting Eq. (\ref{eq:assumption_breathing}) into the Feynman-Kac
equation (\ref{eq:Feynman-Kac_equation_joint}), we obtain the following
set of time-dependent ordinary differential equations

\begin{align}
\dot{\Lambda} & =-\kappa\left(1-\frac{a+iu}{\beta}\right),\label{eq:Lambdadot}\\
\dot{a} & =2\kappa(a+iu)\left(1-\frac{a+iu}{\beta}\right)-2\lambda(t)c,\label{eq:adotequation}\\
\dot{b} & =2c\left(\lambda(t)-\frac{\kappa}{\beta}c\right)-2\left(b+is\right)\frac{\dot{\lambda}(t)}{\lambda(t)},\label{eq:bdotequation}\\
\dot{c} & =\lambda(t)(a-b)-2\frac{\kappa}{\beta}(a+iu)c+\left(\kappa-\frac{\dot{\lambda}(t)}{\lambda(t)}\right)c.\label{eq:cdotequation}
\end{align}
The initial conditions are $\Lambda(0)=0$, $a(0)=\beta_{S}$, $b(0)=\beta_{S}$
and $c(0)=0$ according to Eq. (\ref{eq:initial_condition_underdamped}).
The characteristic function of work and heat follows from Eq. (\ref{eq:chiWQ})
as

\begin{align}
\chi^{w,q}(s,u) & =\frac{\lambda(0)}{\lambda(\tau)}\frac{\beta_{S}e^{-\Lambda(\tau)}}{\sqrt{a(\tau)b(\tau)-c(\tau)^{2}}}.\label{eq:chi_tauWQexact_numericaljointcharacteristic}
\end{align}

In the highly underdamped regime $\kappa\ll\lambda(t)$, the dynamics
and the work statistics can be calculated with the method of stochastic
differential equation of energy \citep{Salazar2016,Salazar2020,Chen2021a}.
For the breathing harmonic oscillator in the highly underdamped regime,
the kinetic energy and the potential energy are approximately equal
$a\approx b$ (Virial theorem), and the correlation can be neglected
$c\approx0$. The differential Eqs. (\ref{eq:Lambdadot})-(\ref{eq:cdotequation})
can be reduced to

\begin{align}
\dot{\Lambda} & =-\kappa\left(1-\frac{a+iu}{\beta}\right),\label{eq:42}\\
\dot{a} & =\kappa\left(a+iu\right)\left(1-\frac{a+iu}{\beta}\right)-\frac{\dot{\lambda}(t)}{\lambda(t)}\left(a+is\right),\label{eq:43}
\end{align}
and the characteristic function can be simplified into

\begin{align}
\chi_{\mathrm{under}}^{w,q}(s,u) & =\frac{\lambda(0)}{\lambda(\tau)}\frac{\beta_{S}e^{-\Lambda(\tau)}}{a(\tau)}.\label{eq:10-1-1}
\end{align}

As has been shown previously \citep{Salazar2020,Chen2021a}, we can
even obtain analytical results of the work statistics if we adopt
the exponential protocol of the control parameter

\begin{equation}
\lambda(t)=\lambda(0)\exp(\alpha t),
\end{equation}
where $\alpha$ is a constant determining the tuning rate of the control
parameter. For this protocol, the analytical result of the characteristic
function can be obtained as\begin{widetext}

\begin{equation}
\chi_{\mathrm{under}}^{w,q}(s,u)=\frac{\exp[(\kappa-\alpha)\tau/2]}{\cosh(\Omega\tau)+[\beta\Omega/(\beta_{S}\kappa)-(\alpha\beta-\kappa\beta+2i\kappa u)(\alpha\beta-\kappa\beta+2i\kappa u+2\kappa\beta_{S})/(4\beta\beta_{S}\kappa\Omega)]\sinh(\Omega\tau)},\label{eq:jointcharacteristicfunction_underdamped}
\end{equation}
\end{widetext}where $\Omega=\sqrt{(\kappa-\alpha)^{2}/4-i\alpha\kappa(s-u)/\beta}$.
The free energy difference is $\Delta F_{S}=\alpha\tau/\beta_{S}$.
One can check that the analytical expression (\ref{eq:jointcharacteristicfunction_underdamped})
satisfies the differential fluctuation theorem (\ref{eq:crook_relation_by_characteristic_function}).

We can similarly consider the overdamped regime. In the overdamped
regime $\kappa\gg\lambda(t)$, the relaxation timescales of momentum
and position are separated. The relaxation timescale of the momentum
is much less than that of the position, and their joint distribution
is in a product form $\rho(x,p,t)=\rho_{M}(p)\cdot\hat{\rho}(x,t)$.
The momentum distribution $\rho_{M}(p)=\sqrt{\beta/(2\pi m)}\exp[-\beta p^{2}/(2m)]$
is assumed to be the Maxwellian distribution, while the position distribution
$\hat{\rho}(x,t)=\int_{-\infty}^{\infty}\rho(x,p,t)dp$ is effectively
governed by the Smoluchowski equation $\partial_{t}\hat{\rho}=\hat{\mathscr{D}}[\hat{\rho}]$
due to the separation of the relaxation timescales \citep{Smoluchowski1916,Bocquet1997,Pan2018}.
The dissipative operator is

\begin{equation}
\hat{\mathscr{D}}\left[\hat{\rho}\right]=\frac{1}{m\kappa}\frac{\partial}{\partial x}\left(\frac{\partial\mathcal{U}}{\partial x}\hat{\rho}+\frac{1}{\beta}\frac{\partial\hat{\rho}}{\partial x}\right).\label{eq:smoluchowski_eq}
\end{equation}

Similar to Eq. (\ref{eq:chiWQ}), the characteristic function can
be calculated through

\begin{equation}
\hat{\chi}_{\mathrm{over}}^{w,q}(s,u)=\int_{-\infty}^{\infty}\hat{\eta}(x,\tau)dx,\label{eq:chitauWQ_overdamped}
\end{equation}
where the distribution function $\hat{\eta}(x,t)$ satisfies the Feynman-Kac
equation

\begin{equation}
\frac{\partial\hat{\eta}}{\partial t}=e^{iu\mathcal{U}}\hat{\mathscr{D}}\left[e^{-iu\mathcal{U}}\hat{\eta}\right]+is\dot{\lambda}\frac{\partial\mathcal{U}}{\partial\lambda}\hat{\eta},\label{eq:Feynman-Kac_overdamped}
\end{equation}
with the initial condition 
\begin{equation}
\hat{\eta}(x,0)=\hat{\rho}(x,0)=\frac{e^{-\beta_{S}\mathcal{U}(x,\lambda(0))}}{\hat{Z}_{S}^{\mathrm{i}}(\beta_{S})}.
\end{equation}
The normalized constant of the initial position distribution is $\hat{Z}_{S}^{\mathrm{i}}(\beta_{S})=\int_{-\infty}^{\infty}\exp[-\beta_{S}\mathcal{U}(x,\lambda(0))]dx$.

We assume the distribution function $\hat{\eta}$ in the quadratic
form 
\begin{equation}
\hat{\eta}(x,t)=\lambda(0)\sqrt{\frac{\beta_{S}m}{2\pi}}e^{-\frac{\hat{b}}{2}m\lambda^{2}(t)x^{2}-\hat{\Lambda}}.
\end{equation}
The Feynman-Kac equation (\ref{eq:Feynman-Kac_overdamped}) leads
to the differential equations
\begin{align}
\dot{\hat{\Lambda}} & =-\frac{\lambda^{2}(t)}{\kappa}\left(1-\frac{\hat{b}+iu}{\beta}\right),\label{eq:Lambdadot_overdamped}\\
\dot{\hat{b}} & =2\frac{\lambda^{2}(t)}{\kappa}\left(\hat{b}+iu\right)\left(1-\frac{\hat{b}+iu}{\beta}\right)-2\frac{\dot{\lambda}(t)}{\lambda(t)}(\hat{b}+is).\label{eq:bdot_overdamped}
\end{align}
The characteristic function is simplified into

\begin{align}
\hat{\chi}_{\mathrm{over}}^{w,q}(s,u) & =\frac{\lambda(0)}{\lambda(\tau)}\sqrt{\frac{\beta_{S}}{\hat{b}(\tau)}}e^{-\hat{\Lambda}(t)}.\label{eq:chitildeoverdamped}
\end{align}

It is worth mentioning that in the overdamped regime neglecting the
momentum degree of freedom does not affect the work statistics \citep{Pan2018},
but indeed affects to the heats statistics \citep{Murashita2016,Chen2021b,Paraguassu2022}.
To be consistent with Eq. (\ref{eq:chi_tauWQexact_numericaljointcharacteristic}),
we supplement the contribution from the momentum degree of freedom
(fast thermalization at the initial time) to the characteristic function
\begin{equation}
\chi_{\mathrm{over}}^{w,q}(s,u)=\frac{\hat{\chi}_{\mathrm{over}}^{w,q}(s,u)}{\sqrt{(1-\frac{iu}{\beta})(1+\frac{iu}{\beta_{S}})}}.\label{eq:chitauoverwithsupplement}
\end{equation}
We will compare Eq. (\ref{eq:chitauoverwithsupplement}) with the
exact result {[}Eq. (\ref{eq:chi_tauWQexact_numericaljointcharacteristic}){]}
in the overdamped regime in the next subsection.

Similar to the highly underdamped regime, we can even obtain analytical
results of the characteristic function under some specific protocols
of the control parameter. For example, we choose the protocol \citep{Kwon2013}
\begin{equation}
\lambda(t)=\frac{\lambda(0)}{\sqrt{1+\epsilon t}},
\end{equation}
where $\epsilon$ is a constant determining the tuning rate. The analytical
result of the characteristic function can be obtained as\begin{widetext}

\begin{equation}
\chi_{\mathrm{over}}^{w,q}(s,u)=\frac{(1+\epsilon\tau)^{\frac{\delta+\epsilon}{4\epsilon}}/\sqrt{(1-\frac{iu}{\beta})(1+\frac{iu}{\beta_{S}})}}{\left\{ \cosh[\frac{\theta}{\epsilon}\ln(1+\epsilon\tau)]+\sinh[\frac{\theta}{\epsilon}\ln(1+\epsilon\tau)]\left[\frac{\beta}{\beta_{S}}(\frac{\theta}{\delta}-\frac{(\delta+\epsilon)^{2}}{4\theta\delta})+(1+\frac{2iu}{\beta_{S}})\frac{\delta+\epsilon}{2\theta}-\frac{iu\delta}{\beta\theta}(1+\frac{iu}{\beta_{S}})\right]\right\} ^{\frac{1}{2}}},\label{eq:jointcharacteristicfunction_overdamped}
\end{equation}
\end{widetext}where $\delta=2\lambda^{2}(0)/\kappa$ and $\theta=\sqrt{i(s-u)\delta\epsilon/\beta+(\delta+\epsilon)^{2}/4}$.
The free energy difference is $\Delta F_{S}=-[\ln(1+\epsilon\tau)]/(2\beta_{S})$.
One can check that the analytical expression (\ref{eq:jointcharacteristicfunction_overdamped})
satisfies the differential fluctuation theorem (\ref{eq:crook_relation_by_characteristic_function}).

By setting $u=0$ in Eqs. (\ref{eq:jointcharacteristicfunction_underdamped})
and (\ref{eq:jointcharacteristicfunction_overdamped}), we recover
the known results of the work distribution in the highly underdamped
and the overdamped regimes \citep{Salazar2020,Kwon2013}. Actually,
we can calculate the joint distribution of work and heat in the generic
underdamped regime under an arbitrary protocol by numerically solving
Eqs. (\ref{eq:Lambdadot})-(\ref{eq:cdotequation}). In that sense,
our method substantially extends the range of applicability.

\subsection{Joint distribution of work and heat}

Previously, either the heat distribution or the work distribution
has been calculated for various systems \citep{Imparato2007,Chatterjee2010,Speck2011,Kwon2013,Gong2016,Denzler2018,Fogedby2009,Paraguassu2021a,Saha2014},
but the joint distribution of work and heat has not been calculated
so far. For arbitrary protocols of the control parameter $\lambda(t)$,
the results of the characteristic function $\chi^{w,q}(s,u)$ can
be numerically calculated via Eqs. (\ref{eq:Lambdadot})-(\ref{eq:cdotequation}).
As an example to show the effectiveness of our method, we calculate
the joint distribution $P(w,q)$ of work and heat. The joint distribution
$P(w,q)$ of work and heat is the inverse Fourier transform of the
characteristic function $\chi^{w,q}(s,u)$.

We consider a compression process under the exponential protocol $\lambda(t)=\lambda(0)\exp(\alpha t)$
($\alpha>0$) in the underdamped regime ($\kappa/\lambda(0)=1/10$)
and an expansion process under the specific protocol $\lambda(t)=\lambda(0)/\sqrt{1+\epsilon t}$
($\epsilon>0$) in the overdamped regime ($\kappa/\lambda(0)=10$).
In the numerical calculation, we set the initial frequency $\lambda(0)=1$
and the inverse temperatures $\beta_{S}=\beta=1$, i.e., the system
is initially in equilibrium with the heat reservoir. Both $\alpha$
and $\epsilon$ are set to be $0.05$ with the control time $\tau=20$.

\begin{figure}
\includegraphics[height=7.5cm]{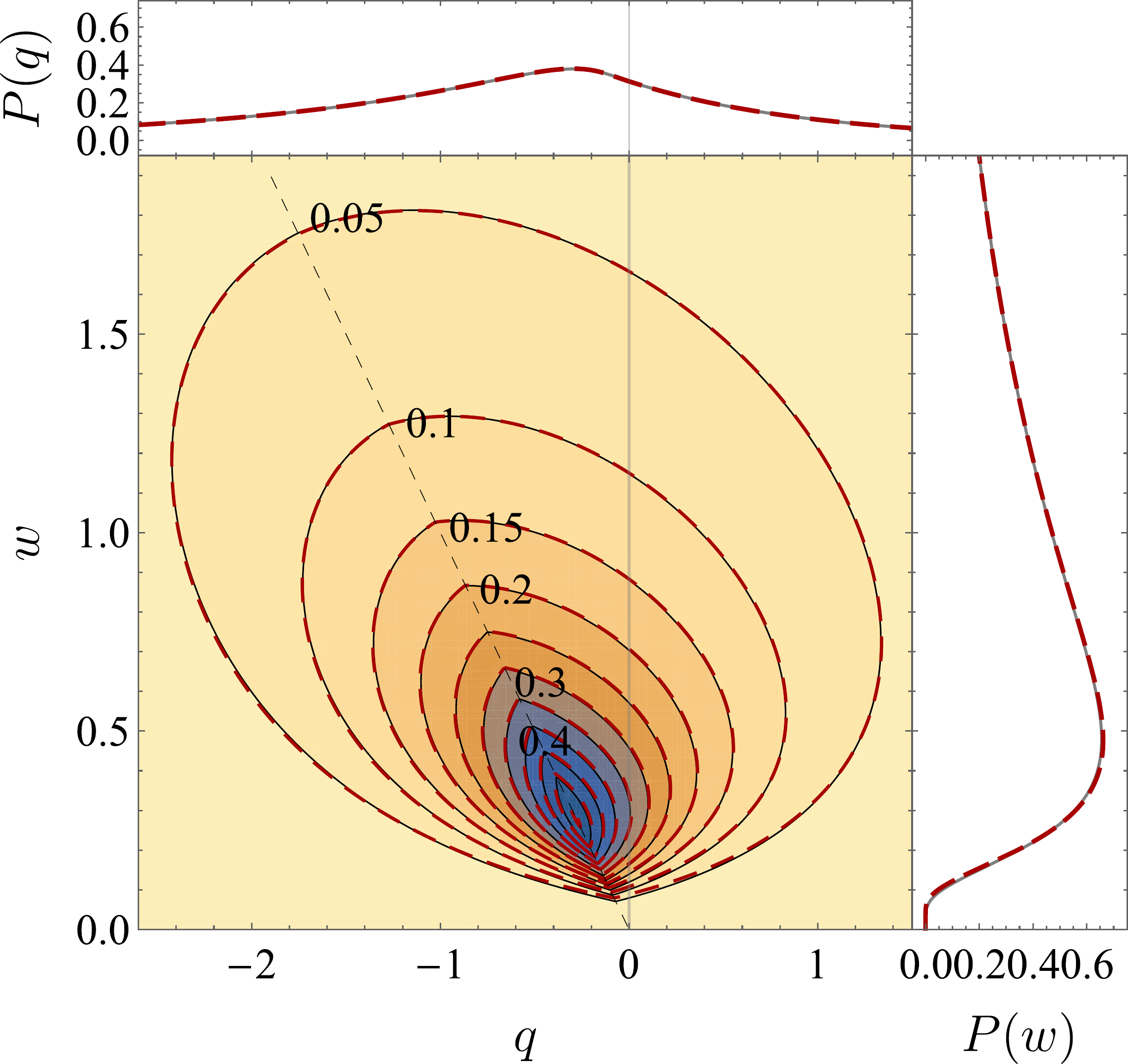}

\caption{The joint distribution $P(w,q)$ and the marginal distributions $P(w)$
and $P(q)$ at the end $\tau=20$ of the underdamped compression process
($\kappa=0.1$, $\lambda(0)=1$, and $\alpha=0.05$). The black dashed
line shows $w=q$. In both the contour map and the marginal distributions,
the gray solid (red dashed) contours illustrate the numerical (analytical)
results. \label{fig:underdampedresults}}
\end{figure}

Figure \ref{fig:underdampedresults} illustrates the joint distribution
$P(w,q)$ (contour map) of work and heat as well as the marginal distributions,
the work distribution $P(w)$ and the heat distribution $P(q)$, at
the end $\tau=20$ of the compression process in the underdamped regime
($\kappa=0.1$, $\lambda(0)=1$, and $\alpha=0.05$). The joint distribution
$P(w,q)$ is obtained via the two-dimensional discrete inverse Fourier
transform of the characteristic function $\chi^{w,q}(s,u)$, where
$s$ and $u$ range from $-400$ to $400$ with the interval $0.2$.
The red dashed contours are obtained from the analytical expression
(\ref{eq:jointcharacteristicfunction_underdamped}) in the highly
underdamped regime, and agree well with the gray solid contours obtained
from the exact numerical results. In the marginal distributions, the
approximate analytical results (red dashed curves) agree well with
the exact numerical results (gray solid curves).

\begin{figure}[h]
\includegraphics[height=7.5cm]{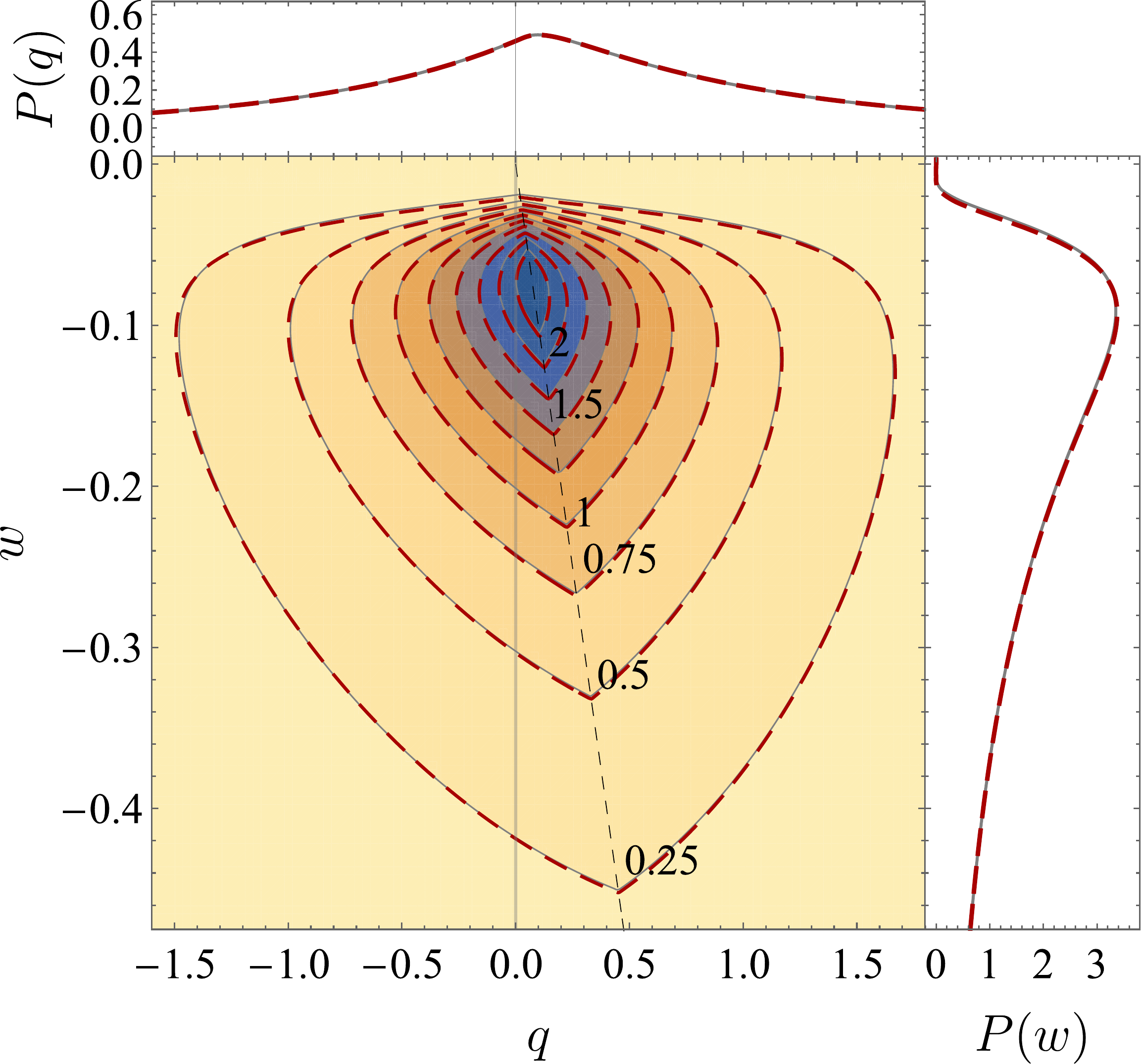}

\caption{The joint distribution $P(w,q)$ and the marginal distributions $P(w)$
and $P(q)$ at the end $\tau=20$ of the overdamped expansion process
($\kappa=10$, $\lambda(0)=1$, and $\epsilon=0.05$). \label{fig:overdampedresults}}
\end{figure}

Figure \ref{fig:overdampedresults} illustrates the joint distribution
$P(w,q)$ as well as the marginal distributions $P(w)$ and $P(q)$
of work and heat for the expansion process in the overdamped regime
($\kappa=10$, $\lambda(0)=1$, and $\epsilon=0.05$). The heat distribution
$P(q)$ is more disperse than the work distribution $P(w)$. In the
joint distribution, the agreement between the red dashed contours
and the gray solid contours shows the overdamped approximation is
perfect under the current parameters. Notice that we have included
the contribution from the momentum degree of freedom to the heat statistics
in Eq. (\ref{eq:chitauoverwithsupplement}). In the overdamped regime,
the thermalization of momentum only contributes to the heat statistics
but does not affect the work statistics.

\section{Conclusion\label{sec:Conclusion}}

In this article, we study the fluctuation theorems when a system is
in contact with multiple heat reservoirs and meanwhile is driven by
an external agent. In this circumstance, the marginal distributions
of work or heat do not satisfy any fluctuation theorem. But only the
joint distribution of work and heat satisfies a family of fluctuation
theorems. We discover a hierarchical structure of fluctuation theorems
for the joint distribution of work and heat in the situation of multiple
heat reservoirs (see Fig. \ref{fig:Fluctuation-theorems-(FTs)}).
This is an exhaustive list of fluctuation theorems concerning work
and heat for a driven system in contact with multiple heat reservoirs.
We demonstrate how these fluctuation theorems at different levels
can be derived from microreversibility \citep{Campisi2011} of the
dynamics by adopting a step-by-step coarse-graining procedure. Thus,
we put all fluctuation theorems into a unified framework. From Fig.
\ref{fig:Fluctuation-theorems-(FTs)}, one can also see how the previously
known fluctuation theorems and the new fluctuation theorems discovered
by us can be fitted into the hierarchical structure of fluctuation
theorems. The Jarzynski equality, the Crooks relation, the exchange
fluctuation theorem, and the Clausius inequality can be recovered
under specific conditions. The conventional statements of the second
law follow from the integral fluctuation theorems by utilizing Jensen's
inequality.

We propose a general method to calculate the joint statistics of work
and heat via the Feynman-Kac equation. The joint distribution of work
and heat encodes more detailed information of the nonequilibrium driving
processes compared to the marginal distributions of work or heat.
We exemplify our method with a classical Brownian particle moving
in a time-dependent potential, and obtain explicit results of the
joint distribution of work and heat for the breathing harmonic oscillator.
For the classical Brownian motion model, the system-bath interaction
can be strong, and a consistent thermodynamic structure is restored
by defining work and heat based on the renormalized system Hamiltonian
\citep{RevModPhys.92.041002}. In the highly underdamped and the overdamped
regimes, we obtain analytical expressions of the characteristic function
of work and heat under some specific protocols, and recover the known
results of the work distribution \citep{Salazar2020,Kwon2013}. In
addition, we can also calculate the joint statistics of work and heat
in the generic underdamped regime, which has not been reported previously.

The general method can be further employed to study many other problems
in stochastic thermodynamics, for example, to evaluate the work and
the heat statistics in shortcuts to isothermality \citep{Martinez2016,Li2017,Li2021,Chen2022}.
Also, it is intriguing to study the joint statistics of work and heat
for generic open quantum systems \citep{Liu2014,Funo2018a,Salazar2020}.
Extension of our method to calculate the joint distribution of work
and heat to driven open quantum systems is left for future exploration.
\begin{acknowledgments}
This work is supported by the National Natural Science Foundation
of China (NSFC) under Grants No. 12147157, No. 11775001 and No. 11825501.
\end{acknowledgments}

\appendix

\section{Fluctuation theorems: Quantum setup\label{sec:Joint-fluctuation-theorems_quantum-setup}}

In the quantum setup, the Hamiltonians $H_{S}(\lambda(t))$ and $H_{\nu}$
are Hermitian operators. The initial state of the total system is
represented by the density matrix $\rho_{\mathrm{tot}}^{\mathrm{i}}=\rho_{S}^{\mathrm{i}}\otimes\pi_{1}\otimes...\otimes\pi_{N}$
in the product form. We consider the initial distribution of the system
is also a canonical distribution $\rho_{S}^{\mathrm{i}}=\pi_{S}^{\mathrm{i}}=\sum_{m}p_{S,m}^{\mathrm{i}}\left|m\right\rangle \left\langle m\right|$.
The canonical distribution of the $\nu$-th heat reservoir $\pi_{\nu}=\sum_{n_{\nu}}p_{\nu,n_{\nu}}\left|n_{\nu}\right\rangle \left\langle n_{\nu}\right|$.
Here, $\left|m\right\rangle $ ($\left|n_{\nu}\right\rangle $) is
the eigenstate of the Hamiltonian of the system $H_{S}(\lambda(0))$
(the Hamiltonian of the $\nu$-th heat reservoir $H_{\nu}$), and
$p_{S,m}^{\mathrm{i}}$ ($p_{\nu,n_{\nu}}$) is the equilibrium population.
The evolution of the total system during the time interval $[0,\tau]$
is given by the unitary evolution $U_{\mathrm{tot}}=\mathrm{T}\exp(-\int_{0}^{\tau}iH_{\mathrm{tot}}(\lambda(t)))$
with the time-ordering operator $\mathrm{T}$ and the total Hamiltonian
$H_{\mathrm{tot}}(\lambda(t))=H_{S}(\lambda(t))+\sum_{\nu}H_{\nu}+h_{\mathrm{int}}$.
The interaction Hamiltonian $h_{\mathrm{int}}$ is weak and can be
neglected when implementing the two-point measurements for work and
heat.

We implement the joint measurements of energies over the system and
the heat reservoirs at the beginning (end) with the outcomes $E_{S,m}^{\mathrm{i}}$
and $E_{\nu,n_{\nu}}$ ($E_{S,m^{\prime}}^{\mathrm{f}}$ and $E_{\nu,n_{\nu}^{\prime}}$),
and obtain the trajectory of the transition $\Gamma=(m,\{n_{\nu}\}\rightarrow m^{\prime},\{n_{\nu}^{\prime}\})$.
Here, $E_{S,m}^{\mathrm{i}}$ and $E_{S,m^{\prime}}^{\mathrm{f}}$
are the eigenenergies of the system Hamiltonians $H_{S}(\lambda(0))$
and $H_{S}(\lambda(\tau))$ at the initial and the final time, and
$E_{\nu,n_{\nu}}$ is the eigenenergy of the Hamiltonian $H_{\nu}$
of the $\nu$-th heat reservoir. The transition probability is $\mathcal{P}(m^{\prime},\{n_{\nu}^{\prime}\}|m,\{n_{\nu}\})=\left|\left\langle m^{\prime},\{n_{\nu}^{\prime}\}\right|U_{\mathrm{tot}}\left|m,\{n_{\nu}\}\right\rangle \right|^{2}$,
where $\left|m,\{n_{\nu}\}\right\rangle $ is the direct product of
the eigenstates of the system and the heat reservoirs. The heat exchange
with the $\nu$-th heat reservoir is defined by \citep{Talkner2009}
\begin{equation}
q_{\nu}(\Gamma)\coloneqq E_{\nu,n_{\nu}}-E_{\nu,n_{\nu}^{\prime}}.
\end{equation}
The work performed by the external driving, according to the first
law, is \citep{Talkner2009}
\begin{equation}
w(\Gamma)\coloneqq E_{S,m^{\prime}}^{\mathrm{f}}-E_{S,m}^{\mathrm{i}}+\sum_{\nu}(E_{\nu,n_{\nu}^{\prime}}-E_{\nu,n_{\nu}}).
\end{equation}

In the quantum setup, microreversibility is guaranteed by the time-reversal
invariance of the Hamiltonian

\begin{equation}
H_{\mathrm{tot}}(\lambda(t))\Theta=\Theta H_{\mathrm{tot}}(\lambda(t)),\label{eq:time-reversal_invariant_qunatum}
\end{equation}
where $\Theta$ is the quantum mechanical time-reversal (anti-unitary)
operator \citep{Andrieux2008,Campisi2011}. For the reverse process,
the Hamiltonians are associated with the forward ones as $\tilde{H}_{S}(\tilde{\lambda}(t))=\Theta H_{S}(\lambda(\tau-t))\Theta^{\dagger}$
and $\tilde{H}_{\nu}=\Theta H_{\nu}\Theta^{\dagger}$, where the control
parameter is tuned as $\tilde{\lambda}(t)=\lambda(\tau-t)$. The initial
canonical states are $\tilde{\pi}_{S}^{\mathrm{i}}=\Theta\pi_{S}^{\mathrm{f}}\Theta^{\dagger}=\sum{}_{m^{\prime}}p_{S,m^{\prime}}^{\mathrm{f}}\Theta\left|m^{\prime}\right\rangle \left\langle m^{\prime}\right|\Theta^{\dagger}$
and $\tilde{\pi}_{\nu}=\Theta\pi_{\nu}\Theta^{\dagger}=\sum_{n_{\nu}^{\prime}}p_{\nu,n_{\nu}^{\prime}}\Theta\left|n_{\nu}^{\prime}\right\rangle \left\langle n_{\nu}^{\prime}\right|\Theta^{\dagger}$.
The corresponding evolution of the total system is $\tilde{U}_{\mathrm{tot}}=\mathrm{T}\exp(-\int_{0}^{\tau}i\tilde{H}_{\mathrm{tot}}(\tilde{\lambda}(t)))$.
The probability of the transition from $\Theta\left|m^{\prime},\{n_{j}^{\prime}\}\right\rangle $
to $\Theta\left|m,\{n_{j}\}\right\rangle $ is $\tilde{\mathcal{P}}(m,\{n_{\nu}\}|m^{\prime},\{n_{\nu}^{\prime}\})=\left|\left\langle m,\{n_{\nu}\}\left|\Theta^{\dagger}\tilde{U}_{\mathrm{tot}}\Theta\right|m^{\prime},\{n_{\nu}^{\prime}\}\right\rangle \right|^{2}$.
From Eq. (\ref{eq:time-reversal_invariant_qunatum}), one can verify
microreversibility of the evolution $\Theta^{\dagger}\tilde{U}_{\mathrm{tot}}\Theta=U_{\mathrm{tot}}^{\dagger}$.
Thus, the transition probabilities of the forward and the reverse
processes satisfy
\begin{equation}
\tilde{\mathcal{P}}(m,\{n_{\nu}\}|m^{\prime},\{n_{\nu}^{\prime}\})=\mathcal{P}(m^{\prime},\{n_{\nu}^{\prime}\}|m,\{n_{\nu}\}),\label{eq:micro-reversibility_quantum}
\end{equation}
which is the quantum counterpart of Eq. (\ref{eq:microreversibility}).

We prepare the heat reservoirs in their equilibrium states in both
the forward and the reverse processes. For the initial state $\left|m\right\rangle $
of the system, the conditional probability of observing the transition
$\Gamma$ in the forward process is $\mathcal{P}(m^{\prime},\{n_{\nu}^{\prime}\},\{n_{\nu}\}|m)\coloneqq\mathcal{P}(m^{\prime},\{n_{\nu}^{\prime}\}|m,\{n_{\nu}\})\prod_{\nu}p_{\nu,n_{\nu}}$.
We sum over the initial and the final states of the heat reservoirs
$\mathcal{P}_{S}(m^{\prime},\{q_{\nu}\}|m)=\sum_{\{n_{\nu}\},\{n_{\nu}^{\prime}\}}\mathcal{P}(m^{\prime},\{n_{\nu}^{\prime}\},\{n_{\nu}\}|m)\delta(q_{\nu}-E_{\nu,n_{\nu}}+E_{\nu,n_{\nu}^{\prime}})$,
and obtain

\begin{equation}
\frac{\tilde{\mathcal{P}}_{S}(m,\{-q_{\nu}\}|m^{\prime})}{\mathcal{P}_{S}(m^{\prime},\{q_{\nu}\}|m)}=e^{\sum_{\nu}\beta_{\nu}q_{\nu}}.\label{eq:differential_FT_quantum_conditional}
\end{equation}
Here, $\tilde{\mathcal{P}}_{S}(m,\{-q_{\nu}\}|m^{\prime})$ is similarly
defined in the reverse process with the initial state $\Theta\left|m^{\prime}\right\rangle $
of the system. Equation (\ref{eq:differential_FT_quantum_conditional})
is the quantum counterpart of Eq. (\ref{eq:trajectoryprobabilityconditional})
or Eq. (\ref{eq:differential_FT_classical_conditional}).

Including the initial canonical distribution of the system, we obtain
the probability $\mathcal{P}_{S}(m^{\prime},m,\{q_{\nu}\})=\mathcal{P}_{S}(m^{\prime},\{q_{\nu}\}|m)p_{S,m}^{\mathrm{i}}$
of observing the system jumping from $m$ to $m^{\prime}$ with the
heat exchange $q_{\nu}$. The ratio of probabilities becomes

\begin{equation}
\frac{\tilde{\mathcal{P}}_{S}(m,m^{\prime},\{-q_{\nu}\})}{\mathcal{P}_{S}(m^{\prime},m,\{q_{\nu}\})}=e^{-\beta_{S}[w(\Gamma)-\Delta F_{S}]+\sum_{\nu}(\beta_{\nu}-\beta_{S})q_{\nu}}.\label{eq:differential_FT_quantum_complete}
\end{equation}
This is the quantum counterpart of Eq. (\ref{eq:trajectoryprobabilityobserved})
or Eq. (\ref{eq:differential_FT_classical_complete}). By summing
over the initial and the final states of the system $P(w,\{q_{\nu}\})=\sum_{m,m^{\prime}}\mathcal{P}_{S}(m^{\prime},m,\{q_{\nu}\})\delta(w+\sum_{\nu}q_{\nu}-E_{S,m^{\prime}}^{\mathrm{f}}+E_{S,m}^{\mathrm{i}})$,
it can be verified that the ratio $\tilde{P}(-w,\{-q_{\nu}\})/P(w,\{q_{\nu}\})$
also satisfies the generalized Crooks relation (\ref{eq:crook_workandheat_classical}).
It is straightforward to derive the quantum counterparts of the differential
and the integral fluctuation theorems (\ref{eq:Hummer-Szabo_relation_classical}),
(\ref{eq:generalized_Jarzynski_equality_classical}) and (\ref{eq:integral_fluctuation_theorem_work=000026heat}).

\section{Fluctuation theorems for entropy production \label{sec:fluctuation_theorems_entropy}}

Based on Eq. (\ref{eq:fluctuationtheorem_complete}), we formulate
fluctuation theorems for entropy production in a hierarchy. The entropy
change, similar to work and heat, can also been defined along the
trajectory \citep{Seifert2005}. The entropy change of the $\nu$-th
heat reservoir is determined by the heat exchange $\Delta s_{\nu}=-\beta_{\nu}q_{\nu}$,
when the heat exchange is much smaller than the internal energy of
every heat reservoir. The entropy change of the system is related
to the initial and final phase-space points 
\begin{equation}
\Delta s_{S}=-\ln\tilde{\rho}_{S}^{\mathrm{i}}(\tilde{\gamma}_{S}(0))+\ln\rho_{S}^{\mathrm{i}}(\gamma_{S}(0)).\label{eq:system entropy}
\end{equation}
The initial distribution $\tilde{\rho}_{S}^{\mathrm{i}}$ in the reverse
process can be chosen as the time-reversal of the final distribution
in the forward process, i.e., $\tilde{\rho}_{S}^{\mathrm{i}}(\tilde{\gamma}_{S}(0))=\Theta[\rho_{S}^{\mathrm{f}}(\gamma_{S}(\tau))]$.
The total entropy change is $\Delta s_{\mathrm{tot}}=\Delta s_{S}+\sum_{\nu}\Delta s_{\nu}$.

Then, the detailed fluctuation theorem (\ref{eq:trajectoryprobabilityconditional})
can be written as 
\begin{equation}
\frac{\mathcal{\tilde{P}}_{S}(\tilde{\gamma}_{S};\{-\Delta s_{\nu}\}|\tilde{\gamma}_{S}(0))}{\mathcal{P}_{S}(\gamma_{S};\{\Delta s_{\nu}\}|\gamma_{S}(0))}=e^{-\sum_{\nu}\Delta s_{\nu}},\label{eq:detailed_FT_entropy_conditional}
\end{equation}
where the heat exchanges in the probability densities are replaced
by the entropy changes of the heat reservoirs. Together with the initial
distribution $\rho_{S}^{\mathrm{i}}(\gamma_{S}(0))$ of the system,
we obtain the complete trajectory probability density $\mathcal{P}_{S}(\gamma_{S};\Delta s_{S},\{\Delta s_{\nu}\})=\mathcal{P}_{S}(\gamma_{S};\{\Delta s_{\nu}\}|\gamma_{S}(0))\rho_{S}^{\mathrm{i}}(\gamma_{S}(0))$,
where the entropy change of the system is related to the initial and
the final phase-space points {[}Eq. (\ref{eq:system entropy}){]}.
Equation (\ref{eq:fluctuationtheorem_complete}) can also be written
as

\begin{equation}
\frac{\mathcal{\tilde{P}}_{S}(\tilde{\gamma}_{S};-\Delta s_{S},\{-\Delta s_{\nu}\})}{\mathcal{P}_{S}(\gamma_{S};\Delta s_{S},\{\Delta s_{\nu}\})}=e^{-\Delta s_{\mathrm{tot}}}.\label{eq:detailed_FT_entropy_complete}
\end{equation}
Equations (\ref{eq:detailed_FT_entropy_conditional}) and (\ref{eq:detailed_FT_entropy_complete})
are identical to Eqs. (\ref{eq:trajectoryprobabilityconditional})
and (\ref{eq:fluctuationtheorem_complete}), but are formulated in
terms of entropy changes.

\begin{figure*}
\includegraphics[width=9.24683cm]{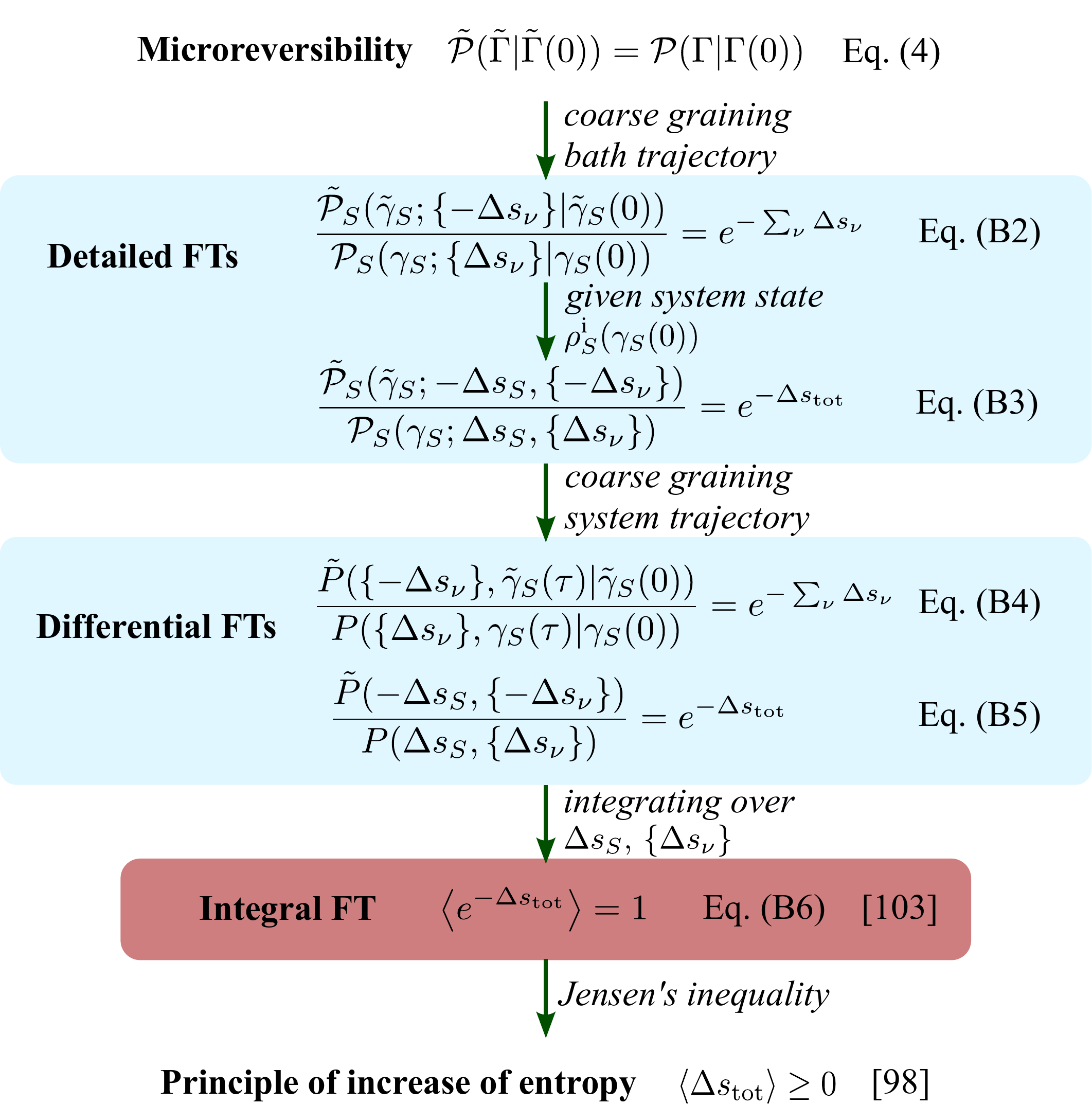}\caption{Hierarchical structure of fluctuation theorems for entropy production.
Fluctuation theorems at different levels can be derived from microreversibility
of the dynamics by adopting a step-by-step coarse-graining procedure.
The fluctuation theorem with red (dark) background has been obtained
previously, but these with blue (light) background have not been reported
so far. To derive these fluctuation theorems, we have assumed the
heat exchange is much smaller than the internal energy of every heat
reservoir. Please note that Eq. (\ref{eq:integral_fluctuation_entropy})
has been previously obtained in Ref. \citep{Jarzynski1999} (see Eq.
(26) therein), and a coarse-grained version of Eq. (\ref{eq:entropy_differential_fts})
has been previously obtained in Ref. \citep{Jarzynski2000} (see Eq.
(4) therein). \label{fig:Hierarchical-structure-of_entropy}}
\end{figure*}

We group the system trajectories according to the entropy changes
$\Delta s_{\nu}$ of the heat reservoirs, the initial and final values
$\gamma_{S}(0)$ and $\gamma_{S}(\tau)$ of the phase-space points,
and obtain the conditional joint distribution $P(\{\Delta s_{\nu}\},\gamma_{S}(\tau)|\gamma_{S}(0))\coloneqq\sum_{\gamma_{S}}\mathcal{P}_{S}(\gamma_{S};\{\Delta s_{\nu}\}|\gamma_{S}(0))$.
The differential fluctuation theorem for the conditional joint distribution
is obtained from Eq. (\ref{eq:detailed_FT_entropy_conditional}) as

\begin{equation}
\frac{\tilde{P}(\{-\Delta s_{\nu}\},\tilde{\gamma}_{S}(\tau)|\tilde{\gamma}_{S}(0))}{P(\{\Delta s_{\nu}\},\gamma_{S}(\tau)|\gamma_{S}(0))}=e^{-\sum_{\nu}\Delta s_{\nu}}.\label{eq:entropy_differential_fts}
\end{equation}
Please note that a coarse-grained version of Eq. (\ref{eq:entropy_differential_fts})
has been previously obtained in Ref. \citep{Jarzynski2000} (see Eq.
(4) therein). We can define the joint distribution of entropy changes
as $P(\Delta s_{S},\{\Delta s_{\nu}\})\coloneqq\sum_{\gamma_{S}(0),\gamma_{S}(\tau)}P(\{\Delta s_{\nu}\},\gamma_{S}(\tau)|\gamma_{S}(0))\rho_{S}^{\mathrm{i}}(\gamma_{S}(0))\delta[\Delta s_{S}+\ln\rho_{S}^{\mathrm{f}}(\gamma_{S}(\tau))-\ln\rho_{S}^{\mathrm{i}}(\gamma_{S}(0))]$,
and obtain the differential fluctuation theorem 
\begin{equation}
\frac{\tilde{P}(-\Delta s_{S},\{-\Delta s_{\nu}\})}{P(\Delta s_{S},\{\Delta s_{\nu}\})}=e^{-\Delta s_{\mathrm{tot}}}.
\end{equation}
By integrating over $\Delta s_{S}$ and $\Delta s_{\nu}$, it is straightforward
to obtain the integral fluctuation theorem

\begin{equation}
\left\langle e^{-\Delta s_{\mathrm{tot}}}\right\rangle =1,\label{eq:integral_fluctuation_entropy}
\end{equation}
which has been previously obtained in Ref. \citep{Jarzynski1999}
(see Eq. (26) therein). From Jensen's inequality, the integral fluctuation
theorem (\ref{eq:integral_fluctuation_entropy}) leads to the principle
of increase of entropy, $\left\langle \Delta s_{\mathrm{tot}}\right\rangle \geq0$
\citep{BlundellBook2009}. We illustrate the hierarchical structure
of fluctuation theorems for entropy production in Fig. \ref{fig:Hierarchical-structure-of_entropy}.

\begin{widetext}

\section{Proof of Eq. (\ref{eq:crook_relation_by_characteristic_function})
based on Kramers equation\label{sec:Proof-of-Eq.}}

We rewrite Eq. (\ref{eq:Feynman-Kac_equation_joint}) as

\begin{equation}
\frac{\partial\eta}{\partial t}=\mathscr{K}_{t}(s,\{u_{\nu}\})\left[\eta\right],\label{eq:Feynman-Kac_equation_joint-1}
\end{equation}
with the time-dependent operator

\begin{equation}
\mathscr{K}_{t}(s,\{u_{\nu}\})\left[\eta\right]=\mathscr{L}\left[\eta\right]+\sum_{\nu}e^{iu_{\nu}H_{S}}\mathscr{D}_{\nu}\left[e^{-iu_{\nu}H_{S}}\eta\right]+is\dot{\lambda}\frac{\partial\mathcal{U}}{\partial\lambda}\eta.
\end{equation}

For $\chi^{w,\{q_{\nu}\}}(s+i\beta_{S},\{u_{\nu}+i(\beta_{S}-\beta_{\nu})\})$,
the time-dependent operator becomes

\begin{equation}
\mathscr{K}_{t}(s+i\beta_{S},\{u_{\nu}+i(\beta_{S}-\beta_{\nu})\})\left[\eta\right]=\mathscr{L}\left[\eta\right]+\sum_{\nu}e^{\left(iu_{\nu}-\beta_{S}\right)H_{S}}\breve{\mathscr{D}}_{\nu}\left[e^{-\left(iu_{\nu}-\beta_{S}\right)H_{S}}\eta\right]+is\dot{\lambda}\frac{\partial\mathcal{U}}{\partial\lambda}\eta-\beta_{S}\dot{\lambda}\frac{\partial\mathcal{U}}{\partial\lambda}\eta,\label{eq:time-dependent_operator}
\end{equation}
where $\breve{\mathscr{D}}_{\nu}$ is defined as 
\begin{equation}
\breve{\mathscr{D}}_{\nu}\left[\cdot\right]\coloneqq e^{\beta_{\nu}H_{S}}\mathscr{D}_{\nu}\left[e^{-\beta_{\nu}H_{S}}\cdot\right]=-\kappa_{\nu}p\frac{\partial(\cdot)}{\partial p}+\frac{\kappa_{\nu}m}{\beta_{\nu}}\frac{\partial^{2}(\cdot)}{\partial p^{2}}.
\end{equation}
Let us define a new variable $\vartheta(x,p,t)\coloneqq\exp(\beta_{S}H_{S})\eta(x,p,t)$.
We rewrite the differential equation (\ref{eq:Feynman-Kac_equation_joint-1})
associated with the operator (\ref{eq:time-dependent_operator}) as

\begin{equation}
\frac{\partial\vartheta}{\partial t}=\mathscr{L}\left[\vartheta\right]+\sum_{\nu}e^{iu_{\nu}H_{S}}\breve{\mathscr{D}}_{\nu}\left[e^{-iu_{\nu}H_{S}}\vartheta\right]+is\dot{\lambda}\frac{\partial\mathcal{U}}{\partial\lambda}\vartheta.\label{eq:developing_equation_for_vartheta}
\end{equation}
The initial condition is $\vartheta(x,p,0)=1/Z_{S}^{\mathrm{i}}(\beta_{S})$.
At the final time $t=\tau$, the characteristic function can be rewritten
as

\begin{equation}
\chi^{w,\{q_{\nu}\}}(s+i\beta_{S},\{u_{\nu}+i(\beta_{S}-\beta_{\nu})\})=\iint_{-\infty}^{\infty}e^{-\beta_{S}H_{S}^{\mathrm{f}}}\vartheta(x,p,\tau)dxdp,\label{eq:characteristic_function_vartheta}
\end{equation}
where $\vartheta(x,p,t)$ is propagated according to Eq. (\ref{eq:developing_equation_for_vartheta}),
and the final Hamiltonian is $H_{S}^{\mathrm{f}}=H_{S}(x,p,\lambda(\tau))$.
One can instead consider the corresponding propagation over $\exp(-\beta_{S}H_{S}^{\mathrm{f}})$.
We rewrite $\vartheta(x,p,\tau)=\mathscr{U}_{\tau}[\vartheta(x,p,0)]$,
which is propagated by the evolution operator $\mathscr{U}_{\tau}$
generated by Eq. (\ref{eq:developing_equation_for_vartheta}). For
the integral $\iint_{-\infty}^{\infty}\varphi(x,p,\tau)\mathscr{U}_{\tau}[\vartheta(x,p,0)]dxdp$,
the conjugate evolution operator $\mathscr{U}_{\tau}^{\dagger}$ on
$\varphi(x,p,\tau)$ satisfies

\begin{equation}
\iint_{-\infty}^{\infty}\varphi(x,p,\tau)\mathscr{U}_{\tau}[\vartheta(x,p,0)]dxdp=\iint_{-\infty}^{\infty}\mathscr{U}_{\tau}^{\dagger}(\varphi(x,p,\tau))\vartheta(x,p,0)dxdp.\label{eq:54}
\end{equation}
In the following, we will show the right-hand side of Eq. (\ref{eq:54})
corresponds to the evolution in the reverse process, and thus prove
Eq. (\ref{eq:crook_relation_by_characteristic_function}).

The deterministic evolution satisfies

\begin{align}
\iint_{-\infty}^{\infty}\varphi\mathscr{L}\left[\vartheta\right]dxdp & =\iint_{-\infty}^{\infty}\left(\frac{p}{m}\vartheta\right)\frac{\partial\varphi}{\partial x}-\left(\frac{\partial\mathcal{U}}{\partial x}\vartheta\right)\frac{\partial\varphi}{\partial p}dxdp\\
 & =\iint_{-\infty}^{\infty}\mathscr{L}\left[\Theta(\varphi)\right]\Theta(\vartheta)dxdp.\label{eq:55}
\end{align}
The dissipation term satisfies

\begin{align}
\iint_{-\infty}^{\infty}\varphi\breve{\mathscr{D}}_{\nu}\left[\vartheta\right]dxdp & =\iint_{-\infty}^{\infty}\left[\kappa_{\nu}\frac{\partial(p\varphi)}{\partial p}+\frac{\kappa_{\nu}m}{\beta_{\nu}}\frac{\partial^{2}\varphi}{\partial p^{2}}\right]\vartheta dxdp\\
 & =\iint_{-\infty}^{\infty}\mathscr{D}_{\nu}\left[\varphi\right]\vartheta dxdp.\label{eq:57}
\end{align}
In the reverse process, the control parameter is tuned as $\tilde{\lambda}(t)=\lambda(\tau-t)$.
The performed work is rewritten as 

\begin{equation}
is\dot{\lambda}\frac{\partial\mathcal{U}}{\partial\lambda}=-is\dot{\tilde{\lambda}}\frac{\partial\mathcal{U}}{\partial\lambda}.\label{eq:58}
\end{equation}
Both the distributions $\exp(-\beta_{S}H_{S})$ and $\vartheta(x,p,t)$
in the phase space are unchanged under the time-reversal operation,
namely, $\Theta[\exp(-\beta_{S}H_{S})]=\exp(-\beta_{S}H_{S})$ and
$\Theta[\vartheta(x,p,t)]=\vartheta(x,p,t)$. Combing Eqs. (\ref{eq:55}),
(\ref{eq:57}) and (\ref{eq:58}), and the facts $\Theta[\vartheta]=\vartheta$
and $\Theta[\varphi]=\varphi$, we obtain the identity relation

\begin{equation}
\iint_{-\infty}^{\infty}\varphi\left(\mathscr{L}\left[\vartheta\right]+\sum_{\nu}e^{iu_{\nu}}\breve{\mathscr{D}}_{\nu}\left[e^{-iu_{\nu}}\vartheta\right]+is\dot{\lambda}\frac{\partial\mathcal{U}}{\partial\lambda}\vartheta\right)dxdp=\iint_{-\infty}^{\infty}\mathscr{\tilde{K}}_{\tau-t}(-s,\{-u_{\nu}\})\left[\varphi\right]\vartheta dxdp.
\end{equation}
Therefore, the propagation over $\vartheta$ in Eq. (\ref{eq:characteristic_function_vartheta})
can be replaced by the propagation over $\exp(-\beta_{S}H_{S}^{\mathrm{f}})$
generated by $\mathscr{\tilde{K}}_{\tau-t}(-s,\{-u_{\nu}\})$ in the
reverse process, i.e.

\begin{equation}
\iint_{-\infty}^{\infty}e^{-\beta_{S}H_{S}^{\mathrm{f}}}\vartheta(x,p,\tau)dxdp=\frac{Z_{S}^{\mathrm{f}}(\beta_{S})}{Z_{S}^{\mathrm{i}}(\beta_{S})}\iint_{-\infty}^{\infty}\tilde{\eta}(x,p,\tau)dxdp,
\end{equation}
where $\tilde{\eta}(x,p,\tau)$ is associated with the characteristic
function $\tilde{\chi}^{w,\{q_{\nu}\}}(-s,\{-u_{\nu}\})$ in the reverse
process, and the initial condition is $\tilde{\eta}(x,p,\tau)=\exp(-\beta_{S}H_{S}^{\mathrm{f}})/Z_{S}^{\mathrm{f}}(\beta_{S})$.
Thus, we prove Eq. (\ref{eq:crook_relation_by_characteristic_function})
based on the Kramers equation.

\end{widetext}

\bibliographystyle{apsrev4-1_addtitle}
\bibliography{heatref}

\begin{thebibliography}{119}%
\makeatletter
\providecommand \@ifxundefined [1]{%
 \@ifx{#1\undefined}
}%
\providecommand \@ifnum [1]{%
 \ifnum #1\expandafter \@firstoftwo
 \else \expandafter \@secondoftwo
 \fi
}%
\providecommand \@ifx [1]{%
 \ifx #1\expandafter \@firstoftwo
 \else \expandafter \@secondoftwo
 \fi
}%
\providecommand \natexlab [1]{#1}%
\providecommand \enquote  [1]{``#1''}%
\providecommand \bibnamefont  [1]{#1}%
\providecommand \bibfnamefont [1]{#1}%
\providecommand \citenamefont [1]{#1}%
\providecommand \href@noop [0]{\@secondoftwo}%
\providecommand \href [0]{\begingroup \@sanitize@url \@href}%
\providecommand \@href[1]{\@@startlink{#1}\@@href}%
\providecommand \@@href[1]{\endgroup#1\@@endlink}%
\providecommand \@sanitize@url [0]{\catcode `\\12\catcode `\$12\catcode
  `\&12\catcode `\#12\catcode `\^12\catcode `\_12\catcode `\%12\relax}%
\providecommand \@@startlink[1]{}%
\providecommand \@@endlink[0]{}%
\providecommand \url  [0]{\begingroup\@sanitize@url \@url }%
\providecommand \@url [1]{\endgroup\@href {#1}{\urlprefix }}%
\providecommand \urlprefix  [0]{URL }%
\providecommand \Eprint [0]{\href }%
\providecommand \doibase [0]{http://dx.doi.org/}%
\providecommand \selectlanguage [0]{\@gobble}%
\providecommand \bibinfo  [0]{\@secondoftwo}%
\providecommand \bibfield  [0]{\@secondoftwo}%
\providecommand \translation [1]{[#1]}%
\providecommand \BibitemOpen [0]{}%
\providecommand \bibitemStop [0]{}%
\providecommand \bibitemNoStop [0]{.\EOS\space}%
\providecommand \EOS [0]{\spacefactor3000\relax}%
\providecommand \BibitemShut  [1]{\csname bibitem#1\endcsname}%
\let\auto@bib@innerbib\@empty
\bibitem [{\citenamefont {Jarzynski}(1997{\natexlab{a}})}]{Jarzynski1997}%
  \BibitemOpen
  \bibfield  {author} {\bibinfo {author} {\bibfnamefont {C.}~\bibnamefont
  {Jarzynski}},\ }\bibinfo {title} {Nonequilibrium Equality for Free Energy
  Differences},\ \href {\doibase 10.1103/physrevlett.78.2690} {\bibfield
  {journal} {\bibinfo  {journal} {Phys. Rev. Lett.}\ }\textbf {\bibinfo
  {volume} {78}},\ \bibinfo {pages} {2690} (\bibinfo {year}
  {1997}{\natexlab{a}})}\BibitemShut {NoStop}%
\bibitem [{\citenamefont {Jarzynski}(1997{\natexlab{b}})}]{Jarzynski1997a}%
  \BibitemOpen
  \bibfield  {author} {\bibinfo {author} {\bibfnamefont {C.}~\bibnamefont
  {Jarzynski}},\ }\bibinfo {title} {Equilibrium free-energy differences from
  nonequilibrium measurements: A master-equation approach},\ \href {\doibase
  10.1103/physreve.56.5018} {\bibfield  {journal} {\bibinfo  {journal} {Phys.
  Rev. E}\ }\textbf {\bibinfo {volume} {56}},\ \bibinfo {pages} {5018}
  (\bibinfo {year} {1997}{\natexlab{b}})}\BibitemShut {NoStop}%
\bibitem [{\citenamefont {Sekimoto}(1998)}]{Sekimoto1998}%
  \BibitemOpen
  \bibfield  {author} {\bibinfo {author} {\bibfnamefont {K.}~\bibnamefont
  {Sekimoto}},\ }\bibinfo {title} {Langevin Equation and Thermodynamics},\
  \href {\doibase 10.1143/ptps.130.17} {\bibfield  {journal} {\bibinfo
  {journal} {Prog. Theor. Phys. Suppl.}\ }\textbf {\bibinfo {volume} {130}},\
  \bibinfo {pages} {17} (\bibinfo {year} {1998})}\BibitemShut {NoStop}%
\bibitem [{\citenamefont {Seifert}(2008)}]{Seifert2008}%
  \BibitemOpen
  \bibfield  {author} {\bibinfo {author} {\bibfnamefont {U.}~\bibnamefont
  {Seifert}},\ }\bibinfo {title} {Stochastic thermodynamics: Principles and
  perspectives},\ \href {\doibase 10.1140/epjb/e2008-00001-9} {\bibfield
  {journal} {\bibinfo  {journal} {Eur. Phys. J. B}\ }\textbf {\bibinfo {volume}
  {64}},\ \bibinfo {pages} {423} (\bibinfo {year} {2008})}\BibitemShut
  {NoStop}%
\bibitem [{\citenamefont {Esposito}\ \emph {et~al.}(2009)\citenamefont
  {Esposito}, \citenamefont {Harbola},\ and\ \citenamefont
  {Mukamel}}]{Esposito2009}%
  \BibitemOpen
  \bibfield  {author} {\bibinfo {author} {\bibfnamefont {M.}~\bibnamefont
  {Esposito}}, \bibinfo {author} {\bibfnamefont {U.}~\bibnamefont {Harbola}}, \
  and\ \bibinfo {author} {\bibfnamefont {S.}~\bibnamefont {Mukamel}},\
  }\bibinfo {title} {Nonequilibrium fluctuations, fluctuation theorems, and
  counting statistics in quantum systems},\ \href {\doibase
  10.1103/revmodphys.81.1665} {\bibfield  {journal} {\bibinfo  {journal} {Rev.
  Mod. Phys.}\ }\textbf {\bibinfo {volume} {81}},\ \bibinfo {pages} {1665}
  (\bibinfo {year} {2009})}\BibitemShut {NoStop}%
\bibitem [{\citenamefont {Campisi}\ \emph {et~al.}(2011)\citenamefont
  {Campisi}, \citenamefont {H\"{a}nggi},\ and\ \citenamefont
  {Talkner}}]{Campisi2011}%
  \BibitemOpen
  \bibfield  {author} {\bibinfo {author} {\bibfnamefont {M.}~\bibnamefont
  {Campisi}}, \bibinfo {author} {\bibfnamefont {P.}~\bibnamefont {H\"{a}nggi}},
  \ and\ \bibinfo {author} {\bibfnamefont {P.}~\bibnamefont {Talkner}},\
  }\bibinfo {title} {Colloquium: Quantum fluctuation relations: Foundations and
  applications},\ \href {\doibase 10.1103/revmodphys.83.771} {\bibfield
  {journal} {\bibinfo  {journal} {Rev. Mod. Phys.}\ }\textbf {\bibinfo {volume}
  {83}},\ \bibinfo {pages} {771} (\bibinfo {year} {2011})}\BibitemShut
  {NoStop}%
\bibitem [{\citenamefont {Sekimoto}(2010)}]{Sekimoto2011}%
  \BibitemOpen
  \bibfield  {author} {\bibinfo {author} {\bibfnamefont {K.}~\bibnamefont
  {Sekimoto}},\ }\href
  {https://www.ebook.de/de/product/9497560/ken_sekimoto_stochastic_energetics.html}
  {\emph {\bibinfo {title} {Stochastic Energetics}}}\ (\bibinfo  {publisher}
  {Springer, Berlin},\ \bibinfo {year} {2010})\BibitemShut {NoStop}%
\bibitem [{\citenamefont {Jarzynski}(2011)}]{Jarzynski2011}%
  \BibitemOpen
  \bibfield  {author} {\bibinfo {author} {\bibfnamefont {C.}~\bibnamefont
  {Jarzynski}},\ }\bibinfo {title} {Equalities and Inequalities:
  Irreversibility and the Second Law of Thermodynamics at the Nanoscale},\
  \href {\doibase 10.1146/annurev-conmatphys-062910-140506} {\bibfield
  {journal} {\bibinfo  {journal} {Annu. Rev. Condens. Matter Phys.}\ }\textbf
  {\bibinfo {volume} {2}},\ \bibinfo {pages} {329} (\bibinfo {year}
  {2011})}\BibitemShut {NoStop}%
\bibitem [{\citenamefont {Seifert}(2012)}]{Seifert2012}%
  \BibitemOpen
  \bibfield  {author} {\bibinfo {author} {\bibfnamefont {U.}~\bibnamefont
  {Seifert}},\ }\bibinfo {title} {Stochastic thermodynamics, fluctuation
  theorems and molecular machines},\ \href {\doibase
  10.1088/0034-4885/75/12/126001} {\bibfield  {journal} {\bibinfo  {journal}
  {Rep. Prog. Phys.}\ }\textbf {\bibinfo {volume} {75}},\ \bibinfo {pages}
  {126001} (\bibinfo {year} {2012})}\BibitemShut {NoStop}%
\bibitem [{\citenamefont {Peliti}\ and\ \citenamefont
  {Pigolotti}(2021)}]{LucaPeliti2021}%
  \BibitemOpen
  \bibfield  {author} {\bibinfo {author} {\bibfnamefont {L.}~\bibnamefont
  {Peliti}}\ and\ \bibinfo {author} {\bibfnamefont {S.}~\bibnamefont
  {Pigolotti}},\ }\href
  {https://www.ebook.de/de/product/39845980/luca_peliti_simone_pigolotti_stochastic_thermodynamics.html}
  {\emph {\bibinfo {title} {Stochastic Thermodynamics}}}\ (\bibinfo
  {publisher} {Princeton University Press, Princeton},\ \bibinfo {year}
  {2021})\BibitemShut {NoStop}%
\bibitem [{\citenamefont {Jarzynski}\ and\ \citenamefont
  {W{\'{o}}jcik}(2004)}]{Jarzynski_2004}%
  \BibitemOpen
  \bibfield  {author} {\bibinfo {author} {\bibfnamefont {C.}~\bibnamefont
  {Jarzynski}}\ and\ \bibinfo {author} {\bibfnamefont {D.~K.}\ \bibnamefont
  {W{\'{o}}jcik}},\ }\bibinfo {title} {Classical and Quantum Fluctuation
  Theorems for Heat Exchange},\ \href {\doibase 10.1103/physrevlett.92.230602}
  {\bibfield  {journal} {\bibinfo  {journal} {Phys. Rev. Lett.}\ }\textbf
  {\bibinfo {volume} {92}},\ \bibinfo {pages} {230602} (\bibinfo {year}
  {2004})}\BibitemShut {NoStop}%
\bibitem [{\citenamefont {Crooks}(1999)}]{Crooks1999}%
  \BibitemOpen
  \bibfield  {author} {\bibinfo {author} {\bibfnamefont {G.~E.}\ \bibnamefont
  {Crooks}},\ }\bibinfo {title} {Entropy production fluctuation theorem and the
  nonequilibrium work relation for free energy differences},\ \href {\doibase
  10.1103/physreve.60.2721} {\bibfield  {journal} {\bibinfo  {journal} {Phys.
  Rev. E}\ }\textbf {\bibinfo {volume} {60}},\ \bibinfo {pages} {2721}
  (\bibinfo {year} {1999})}\BibitemShut {NoStop}%
\bibitem [{\citenamefont {Seifert}(2005)}]{Seifert2005}%
  \BibitemOpen
  \bibfield  {author} {\bibinfo {author} {\bibfnamefont {U.}~\bibnamefont
  {Seifert}},\ }\bibinfo {title} {Entropy Production along a Stochastic
  Trajectory and an Integral Fluctuation Theorem},\ \href {\doibase
  10.1103/physrevlett.95.040602} {\bibfield  {journal} {\bibinfo  {journal}
  {Phys. Rev. Lett.}\ }\textbf {\bibinfo {volume} {95}},\ \bibinfo {pages}
  {040602} (\bibinfo {year} {2005})}\BibitemShut {NoStop}%
\bibitem [{\citenamefont {Mazonka}\ and\ \citenamefont
  {Jarzynski}(1999)}]{Mazonka1999}%
  \BibitemOpen
  \bibfield  {author} {\bibinfo {author} {\bibfnamefont {O.}~\bibnamefont
  {Mazonka}}\ and\ \bibinfo {author} {\bibfnamefont {C.}~\bibnamefont
  {Jarzynski}},\ }\bibinfo {title} {Exactly solvable model illustrating
  far-from-equilibrium predictions},\ \href@noop {} {\  (\bibinfo {year}
  {1999})},\ \Eprint {http://arxiv.org/abs/cond-mat/9912121}
  {arXiv:cond-mat/9912121 [cond-mat.stat-mech]} \BibitemShut {NoStop}%
\bibitem [{\citenamefont {Jarzynski}(2000)}]{Jarzynski2000}%
  \BibitemOpen
  \bibfield  {author} {\bibinfo {author} {\bibfnamefont {C.}~\bibnamefont
  {Jarzynski}},\ }\bibinfo {title} {Hamiltonian Derivation of a Detailed
  Fluctuation Theorem},\ \href {\doibase 10.1023/a:1018670721277} {\bibfield
  {journal} {\bibinfo  {journal} {J. Stat. Phys.}\ }\textbf {\bibinfo {volume}
  {98}},\ \bibinfo {pages} {77} (\bibinfo {year} {2000})}\BibitemShut {NoStop}%
\bibitem [{\citenamefont {Hummer}\ and\ \citenamefont
  {Szabo}(2001)}]{Hummer2001}%
  \BibitemOpen
  \bibfield  {author} {\bibinfo {author} {\bibfnamefont {G.}~\bibnamefont
  {Hummer}}\ and\ \bibinfo {author} {\bibfnamefont {A.}~\bibnamefont {Szabo}},\
  }\bibinfo {title} {Free energy reconstruction from nonequilibrium
  single-molecule pulling experiments},\ \href {\doibase
  10.1073/pnas.071034098} {\bibfield  {journal} {\bibinfo  {journal} {Proc.
  Natl. Acad. Sci. U. S. A.}\ }\textbf {\bibinfo {volume} {98}},\ \bibinfo
  {pages} {3658} (\bibinfo {year} {2001})}\BibitemShut {NoStop}%
\bibitem [{\citenamefont {van Zon}\ and\ \citenamefont
  {Cohen}(2003)}]{Zon2003}%
  \BibitemOpen
  \bibfield  {author} {\bibinfo {author} {\bibfnamefont {R.}~\bibnamefont {van
  Zon}}\ and\ \bibinfo {author} {\bibfnamefont {E.~G.~D.}\ \bibnamefont
  {Cohen}},\ }\bibinfo {title} {Stationary and transient work-fluctuation
  theorems for a dragged Brownian particle},\ \href {\doibase
  10.1103/physreve.67.046102} {\bibfield  {journal} {\bibinfo  {journal} {Phys.
  Rev. E}\ }\textbf {\bibinfo {volume} {67}},\ \bibinfo {pages} {046102}
  (\bibinfo {year} {2003})}\BibitemShut {NoStop}%
\bibitem [{\citenamefont {Narayan}\ and\ \citenamefont
  {Dhar}(2004)}]{Narayan2003}%
  \BibitemOpen
  \bibfield  {author} {\bibinfo {author} {\bibfnamefont {O.}~\bibnamefont
  {Narayan}}\ and\ \bibinfo {author} {\bibfnamefont {A.}~\bibnamefont {Dhar}},\
  }\bibinfo {title} {Reexamination of experimental tests of the fluctuation
  theorem},\ \href {\doibase 10.1088/0305-4470/37/1/004} {\bibfield  {journal}
  {\bibinfo  {journal} {J. Phys. A: Math. Gen.}\ }\textbf {\bibinfo {volume}
  {37}},\ \bibinfo {pages} {63} (\bibinfo {year} {2004})}\BibitemShut {NoStop}%
\bibitem [{\citenamefont {Speck}\ and\ \citenamefont
  {Seifert}(2004)}]{Speck2004}%
  \BibitemOpen
  \bibfield  {author} {\bibinfo {author} {\bibfnamefont {T.}~\bibnamefont
  {Speck}}\ and\ \bibinfo {author} {\bibfnamefont {U.}~\bibnamefont
  {Seifert}},\ }\bibinfo {title} {Distribution of work in isothermal
  nonequilibrium processes},\ \href {\doibase 10.1103/physreve.70.066112}
  {\bibfield  {journal} {\bibinfo  {journal} {Phys. Rev. E}\ }\textbf {\bibinfo
  {volume} {70}},\ \bibinfo {pages} {066112} (\bibinfo {year}
  {2004})}\BibitemShut {NoStop}%
\bibitem [{\citenamefont {Kawai}\ \emph {et~al.}(2007)\citenamefont {Kawai},
  \citenamefont {Parrondo},\ and\ \citenamefont {den Broeck}}]{Kawai2007}%
  \BibitemOpen
  \bibfield  {author} {\bibinfo {author} {\bibfnamefont {R.}~\bibnamefont
  {Kawai}}, \bibinfo {author} {\bibfnamefont {J.~M.~R.}\ \bibnamefont
  {Parrondo}}, \ and\ \bibinfo {author} {\bibfnamefont {C.~V.}\ \bibnamefont
  {den Broeck}},\ }\bibinfo {title} {Dissipation: The Phase-Space
  Perspective},\ \href {\doibase 10.1103/physrevlett.98.080602} {\bibfield
  {journal} {\bibinfo  {journal} {Phys. Rev. Lett.}\ }\textbf {\bibinfo
  {volume} {98}},\ \bibinfo {pages} {080602} (\bibinfo {year}
  {2007})}\BibitemShut {NoStop}%
\bibitem [{\citenamefont {Imparato}\ \emph {et~al.}(2007)\citenamefont
  {Imparato}, \citenamefont {Peliti}, \citenamefont {Pesce}, \citenamefont
  {Rusciano},\ and\ \citenamefont {Sasso}}]{Imparato2007}%
  \BibitemOpen
  \bibfield  {author} {\bibinfo {author} {\bibfnamefont {A.}~\bibnamefont
  {Imparato}}, \bibinfo {author} {\bibfnamefont {L.}~\bibnamefont {Peliti}},
  \bibinfo {author} {\bibfnamefont {G.}~\bibnamefont {Pesce}}, \bibinfo
  {author} {\bibfnamefont {G.}~\bibnamefont {Rusciano}}, \ and\ \bibinfo
  {author} {\bibfnamefont {A.}~\bibnamefont {Sasso}},\ }\bibinfo {title} {Work
  and heat probability distribution of an optically driven Brownian particle:
  Theory and experiments},\ \href {\doibase 10.1103/physreve.76.050101}
  {\bibfield  {journal} {\bibinfo  {journal} {Phys. Rev. E}\ }\textbf {\bibinfo
  {volume} {76}},\ \bibinfo {pages} {050101(R)} (\bibinfo {year}
  {2007})}\BibitemShut {NoStop}%
\bibitem [{\citenamefont {Maragakis}\ \emph {et~al.}(2008)\citenamefont
  {Maragakis}, \citenamefont {Spichty},\ and\ \citenamefont
  {Karplus}}]{Maragakis2008}%
  \BibitemOpen
  \bibfield  {author} {\bibinfo {author} {\bibfnamefont {P.}~\bibnamefont
  {Maragakis}}, \bibinfo {author} {\bibfnamefont {M.}~\bibnamefont {Spichty}},
  \ and\ \bibinfo {author} {\bibfnamefont {M.}~\bibnamefont {Karplus}},\
  }\bibinfo {title} {A Differential Fluctuation Theorem},\ \href {\doibase
  10.1021/jp077037r} {\bibfield  {journal} {\bibinfo  {journal} {J. Phys. Chem.
  B}\ }\textbf {\bibinfo {volume} {112}},\ \bibinfo {pages} {6168} (\bibinfo
  {year} {2008})}\BibitemShut {NoStop}%
\bibitem [{\citenamefont {Sagawa}\ and\ \citenamefont
  {Ueda}(2010)}]{Sagawa2010}%
  \BibitemOpen
  \bibfield  {author} {\bibinfo {author} {\bibfnamefont {T.}~\bibnamefont
  {Sagawa}}\ and\ \bibinfo {author} {\bibfnamefont {M.}~\bibnamefont {Ueda}},\
  }\bibinfo {title} {Generalized Jarzynski Equality under Nonequilibrium
  Feedback Control},\ \href {\doibase 10.1103/physrevlett.104.090602}
  {\bibfield  {journal} {\bibinfo  {journal} {Phys. Rev. Lett.}\ }\textbf
  {\bibinfo {volume} {104}},\ \bibinfo {pages} {090602} (\bibinfo {year}
  {2010})}\BibitemShut {NoStop}%
\bibitem [{\citenamefont {Speck}(2011)}]{Speck2011}%
  \BibitemOpen
  \bibfield  {author} {\bibinfo {author} {\bibfnamefont {T.}~\bibnamefont
  {Speck}},\ }\bibinfo {title} {Work distribution for the driven harmonic
  oscillator with time-dependent strength: exact solution and slow driving},\
  \href {\doibase 10.1088/1751-8113/44/30/305001} {\bibfield  {journal}
  {\bibinfo  {journal} {J. Phys. A: Math. Theor.}\ }\textbf {\bibinfo {volume}
  {44}},\ \bibinfo {pages} {305001} (\bibinfo {year} {2011})}\BibitemShut
  {NoStop}%
\bibitem [{\citenamefont {Kwon}\ \emph {et~al.}(2013)\citenamefont {Kwon},
  \citenamefont {Noh},\ and\ \citenamefont {Park}}]{Kwon2013}%
  \BibitemOpen
  \bibfield  {author} {\bibinfo {author} {\bibfnamefont {C.}~\bibnamefont
  {Kwon}}, \bibinfo {author} {\bibfnamefont {J.~D.}\ \bibnamefont {Noh}}, \
  and\ \bibinfo {author} {\bibfnamefont {H.}~\bibnamefont {Park}},\ }\bibinfo
  {title} {Work fluctuations in a time-dependent harmonic potential: Rigorous
  results beyond the overdamped limit},\ \href {\doibase
  10.1103/physreve.88.062102} {\bibfield  {journal} {\bibinfo  {journal} {Phys.
  Rev. E}\ }\textbf {\bibinfo {volume} {88}},\ \bibinfo {pages} {062102}
  (\bibinfo {year} {2013})}\BibitemShut {NoStop}%
\bibitem [{\citenamefont {Saha}\ and\ \citenamefont
  {Mukherji}(2014)}]{Saha2014}%
  \BibitemOpen
  \bibfield  {author} {\bibinfo {author} {\bibfnamefont {B.}~\bibnamefont
  {Saha}}\ and\ \bibinfo {author} {\bibfnamefont {S.}~\bibnamefont
  {Mukherji}},\ }\bibinfo {title} {Work and heat distributions for a Brownian
  particle subjected to an oscillatory drive},\ \href {\doibase
  10.1088/1742-5468/2014/08/p08014} {\bibfield  {journal} {\bibinfo  {journal}
  {J. Stat. Mech.: Theory Exp.}\ }\textbf {\bibinfo {volume} {2014}},\ \bibinfo
  {pages} {P08014} (\bibinfo {year} {2014})}\BibitemShut {NoStop}%
\bibitem [{\citenamefont {Holubec}\ \emph {et~al.}(2015)\citenamefont
  {Holubec}, \citenamefont {Dierl}, \citenamefont {Einax}, \citenamefont
  {Maass}, \citenamefont {Chvosta},\ and\ \citenamefont
  {Ryabov}}]{Holubec2015}%
  \BibitemOpen
  \bibfield  {author} {\bibinfo {author} {\bibfnamefont {V.}~\bibnamefont
  {Holubec}}, \bibinfo {author} {\bibfnamefont {M.}~\bibnamefont {Dierl}},
  \bibinfo {author} {\bibfnamefont {M.}~\bibnamefont {Einax}}, \bibinfo
  {author} {\bibfnamefont {P.}~\bibnamefont {Maass}}, \bibinfo {author}
  {\bibfnamefont {P.}~\bibnamefont {Chvosta}}, \ and\ \bibinfo {author}
  {\bibfnamefont {A.}~\bibnamefont {Ryabov}},\ }\bibinfo {title} {Asymptotics
  of work distribution for a Brownian particle in a time-dependent anharmonic
  potential},\ \href {\doibase 10.1088/0031-8949/2015/t165/014024} {\bibfield
  {journal} {\bibinfo  {journal} {Phys. Scr.}\ }\textbf {\bibinfo {volume}
  {T165}},\ \bibinfo {pages} {014024} (\bibinfo {year} {2015})}\BibitemShut
  {NoStop}%
\bibitem [{\citenamefont {Gong}\ and\ \citenamefont {Quan}(2015)}]{Gong2015}%
  \BibitemOpen
  \bibfield  {author} {\bibinfo {author} {\bibfnamefont {Z.}~\bibnamefont
  {Gong}}\ and\ \bibinfo {author} {\bibfnamefont {H.~T.}\ \bibnamefont
  {Quan}},\ }\bibinfo {title} {Jarzynski equality, Crooks fluctuation theorem,
  and the fluctuation theorems of heat for arbitrary initial states},\ \href
  {\doibase 10.1103/physreve.92.012131} {\bibfield  {journal} {\bibinfo
  {journal} {Phys. Rev. E}\ }\textbf {\bibinfo {volume} {92}},\ \bibinfo
  {pages} {012131} (\bibinfo {year} {2015})}\BibitemShut {NoStop}%
\bibitem [{\citenamefont {Gong}\ \emph {et~al.}(2016)\citenamefont {Gong},
  \citenamefont {Lan},\ and\ \citenamefont {Quan}}]{Gong2016}%
  \BibitemOpen
  \bibfield  {author} {\bibinfo {author} {\bibfnamefont {Z.}~\bibnamefont
  {Gong}}, \bibinfo {author} {\bibfnamefont {Y.}~\bibnamefont {Lan}}, \ and\
  \bibinfo {author} {\bibfnamefont {H.~T.}\ \bibnamefont {Quan}},\ }\bibinfo
  {title} {Stochastic Thermodynamics of a Particle in a Box},\ \href {\doibase
  10.1103/physrevlett.117.180603} {\bibfield  {journal} {\bibinfo  {journal}
  {Phys. Rev. Lett.}\ }\textbf {\bibinfo {volume} {117}},\ \bibinfo {pages}
  {180603} (\bibinfo {year} {2016})}\BibitemShut {NoStop}%
\bibitem [{\citenamefont {Hoang}\ \emph {et~al.}(2018)\citenamefont {Hoang},
  \citenamefont {Pan}, \citenamefont {Ahn}, \citenamefont {Bang}, \citenamefont
  {Quan},\ and\ \citenamefont {Li}}]{Hoang2018}%
  \BibitemOpen
  \bibfield  {author} {\bibinfo {author} {\bibfnamefont {T.~M.}\ \bibnamefont
  {Hoang}}, \bibinfo {author} {\bibfnamefont {R.}~\bibnamefont {Pan}}, \bibinfo
  {author} {\bibfnamefont {J.}~\bibnamefont {Ahn}}, \bibinfo {author}
  {\bibfnamefont {J.}~\bibnamefont {Bang}}, \bibinfo {author} {\bibfnamefont
  {H.~T.}\ \bibnamefont {Quan}}, \ and\ \bibinfo {author} {\bibfnamefont
  {T.}~\bibnamefont {Li}},\ }\bibinfo {title} {Experimental Test of the
  Differential Fluctuation Theorem and a Generalized Jarzynski Equality for
  Arbitrary Initial States},\ \href {\doibase 10.1103/physrevlett.120.080602}
  {\bibfield  {journal} {\bibinfo  {journal} {Phys. Rev. Lett.}\ }\textbf
  {\bibinfo {volume} {120}},\ \bibinfo {pages} {080602} (\bibinfo {year}
  {2018})}\BibitemShut {NoStop}%
\bibitem [{\citenamefont {Pagare}\ and\ \citenamefont
  {Cherayil}(2019)}]{Pagare2019}%
  \BibitemOpen
  \bibfield  {author} {\bibinfo {author} {\bibfnamefont {A.}~\bibnamefont
  {Pagare}}\ and\ \bibinfo {author} {\bibfnamefont {B.~J.}\ \bibnamefont
  {Cherayil}},\ }\bibinfo {title} {Stochastic thermodynamics of a harmonically
  trapped colloid in linear mixed flow},\ \href {\doibase
  10.1103/physreve.100.052124} {\bibfield  {journal} {\bibinfo  {journal}
  {Phys. Rev. E}\ }\textbf {\bibinfo {volume} {100}},\ \bibinfo {pages}
  {052124} (\bibinfo {year} {2019})}\BibitemShut {NoStop}%
\bibitem [{\citenamefont {Salazar}(2020)}]{Salazar2020}%
  \BibitemOpen
  \bibfield  {author} {\bibinfo {author} {\bibfnamefont {D.~S.~P.}\
  \bibnamefont {Salazar}},\ }\bibinfo {title} {Work distribution in thermal
  processes},\ \href {\doibase 10.1103/physreve.101.030101} {\bibfield
  {journal} {\bibinfo  {journal} {Phys. Rev. E}\ }\textbf {\bibinfo {volume}
  {101}},\ \bibinfo {pages} {030101(R)} (\bibinfo {year} {2020})}\BibitemShut
  {NoStop}%
\bibitem [{\citenamefont {Taniguchi}\ and\ \citenamefont
  {Cohen}(2007)}]{Taniguchi2007}%
  \BibitemOpen
  \bibfield  {author} {\bibinfo {author} {\bibfnamefont {T.}~\bibnamefont
  {Taniguchi}}\ and\ \bibinfo {author} {\bibfnamefont {E.~G.~D.}\ \bibnamefont
  {Cohen}},\ }\bibinfo {title} {Onsager-Machlup Theory for Nonequilibrium
  Steady States and Fluctuation Theorems},\ \href {\doibase
  10.1007/s10955-006-9252-2} {\bibfield  {journal} {\bibinfo  {journal} {J.
  Stat. Phys.}\ }\textbf {\bibinfo {volume} {126}},\ \bibinfo {pages} {1}
  (\bibinfo {year} {2007})}\BibitemShut {NoStop}%
\bibitem [{\citenamefont {Taniguchi}\ and\ \citenamefont
  {Cohen}(2008)}]{Taniguchi2008a}%
  \BibitemOpen
  \bibfield  {author} {\bibinfo {author} {\bibfnamefont {T.}~\bibnamefont
  {Taniguchi}}\ and\ \bibinfo {author} {\bibfnamefont {E.~G.~D.}\ \bibnamefont
  {Cohen}},\ }\bibinfo {title} {Inertial Effects in Nonequilibrium Work
  Fluctuations by a Path Integral Approach},\ \href {\doibase
  10.1007/s10955-007-9398-6} {\bibfield  {journal} {\bibinfo  {journal} {J.
  Stat. Phys.}\ }\textbf {\bibinfo {volume} {130}},\ \bibinfo {pages} {1}
  (\bibinfo {year} {2008})}\BibitemShut {NoStop}%
\bibitem [{\citenamefont {Imparato}\ and\ \citenamefont
  {Peliti}(2006)}]{PhysRevE.74.026106}%
  \BibitemOpen
  \bibfield  {author} {\bibinfo {author} {\bibfnamefont {A.}~\bibnamefont
  {Imparato}}\ and\ \bibinfo {author} {\bibfnamefont {L.}~\bibnamefont
  {Peliti}},\ }\bibinfo {title} {Fluctuation relations for a driven Brownian
  particle},\ \href {\doibase 10.1103/PhysRevE.74.026106} {\bibfield  {journal}
  {\bibinfo  {journal} {Phys. Rev. E}\ }\textbf {\bibinfo {volume} {74}},\
  \bibinfo {pages} {026106} (\bibinfo {year} {2006})}\BibitemShut {NoStop}%
\bibitem [{\citenamefont {Then}\ and\ \citenamefont {Engel}(2008)}]{Then2008}%
  \BibitemOpen
  \bibfield  {author} {\bibinfo {author} {\bibfnamefont {H.}~\bibnamefont
  {Then}}\ and\ \bibinfo {author} {\bibfnamefont {A.}~\bibnamefont {Engel}},\
  }\bibinfo {title} {Computing the optimal protocol for finite-time processes
  in stochastic thermodynamics},\ \href {\doibase 10.1103/physreve.77.041105}
  {\bibfield  {journal} {\bibinfo  {journal} {Phys. Rev. E}\ }\textbf {\bibinfo
  {volume} {77}},\ \bibinfo {pages} {041105} (\bibinfo {year}
  {2008})}\BibitemShut {NoStop}%
\bibitem [{\citenamefont {Engel}(2009)}]{Engel2009}%
  \BibitemOpen
  \bibfield  {author} {\bibinfo {author} {\bibfnamefont {A.}~\bibnamefont
  {Engel}},\ }\bibinfo {title} {Asymptotics of work distributions in
  nonequilibrium systems},\ \href {\doibase 10.1103/physreve.80.021120}
  {\bibfield  {journal} {\bibinfo  {journal} {Phys. Rev. E}\ }\textbf {\bibinfo
  {volume} {80}},\ \bibinfo {pages} {021120} (\bibinfo {year}
  {2009})}\BibitemShut {NoStop}%
\bibitem [{\citenamefont {Minh}\ and\ \citenamefont
  {Adib}(2009)}]{PhysRevE.79.021122}%
  \BibitemOpen
  \bibfield  {author} {\bibinfo {author} {\bibfnamefont {D.~D.~L.}\
  \bibnamefont {Minh}}\ and\ \bibinfo {author} {\bibfnamefont {A.~B.}\
  \bibnamefont {Adib}},\ }\bibinfo {title} {Path integral analysis of
  Jarzynski's equality: Analytical results},\ \href {\doibase
  10.1103/PhysRevE.79.021122} {\bibfield  {journal} {\bibinfo  {journal} {Phys.
  Rev. E}\ }\textbf {\bibinfo {volume} {79}},\ \bibinfo {pages} {021122}
  (\bibinfo {year} {2009})}\BibitemShut {NoStop}%
\bibitem [{\citenamefont {Baiesi}\ \emph {et~al.}(2006)\citenamefont {Baiesi},
  \citenamefont {Jacobs}, \citenamefont {Maes},\ and\ \citenamefont
  {Skantzos}}]{Baiesi2006}%
  \BibitemOpen
  \bibfield  {author} {\bibinfo {author} {\bibfnamefont {M.}~\bibnamefont
  {Baiesi}}, \bibinfo {author} {\bibfnamefont {T.}~\bibnamefont {Jacobs}},
  \bibinfo {author} {\bibfnamefont {C.}~\bibnamefont {Maes}}, \ and\ \bibinfo
  {author} {\bibfnamefont {N.~S.}\ \bibnamefont {Skantzos}},\ }\bibinfo {title}
  {Fluctuation symmetries for work and heat},\ \href {\doibase
  10.1103/physreve.74.021111} {\bibfield  {journal} {\bibinfo  {journal} {Phys.
  Rev. E}\ }\textbf {\bibinfo {volume} {74}},\ \bibinfo {pages} {021111}
  (\bibinfo {year} {2006})}\BibitemShut {NoStop}%
\bibitem [{\citenamefont {Saha}\ \emph {et~al.}(2011)\citenamefont {Saha},
  \citenamefont {Bhattacharjee},\ and\ \citenamefont
  {Chakraborty}}]{PhysRevE.83.011104}%
  \BibitemOpen
  \bibfield  {author} {\bibinfo {author} {\bibfnamefont {A.}~\bibnamefont
  {Saha}}, \bibinfo {author} {\bibfnamefont {J.~K.}\ \bibnamefont
  {Bhattacharjee}}, \ and\ \bibinfo {author} {\bibfnamefont {S.}~\bibnamefont
  {Chakraborty}},\ }\bibinfo {title} {Work probability distribution and tossing
  a biased coin},\ \href {\doibase 10.1103/PhysRevE.83.011104} {\bibfield
  {journal} {\bibinfo  {journal} {Phys. Rev. E}\ }\textbf {\bibinfo {volume}
  {83}},\ \bibinfo {pages} {011104} (\bibinfo {year} {2011})}\BibitemShut
  {NoStop}%
\bibitem [{\citenamefont {Rana}\ \emph {et~al.}(2014)\citenamefont {Rana},
  \citenamefont {Pal}, \citenamefont {Saha},\ and\ \citenamefont
  {Jayannavar}}]{PhysRevE.90.042146}%
  \BibitemOpen
  \bibfield  {author} {\bibinfo {author} {\bibfnamefont {S.}~\bibnamefont
  {Rana}}, \bibinfo {author} {\bibfnamefont {P.~S.}\ \bibnamefont {Pal}},
  \bibinfo {author} {\bibfnamefont {A.}~\bibnamefont {Saha}}, \ and\ \bibinfo
  {author} {\bibfnamefont {A.~M.}\ \bibnamefont {Jayannavar}},\ }\bibinfo
  {title} {Single-particle stochastic heat engine},\ \href {\doibase
  10.1103/PhysRevE.90.042146} {\bibfield  {journal} {\bibinfo  {journal} {Phys.
  Rev. E}\ }\textbf {\bibinfo {volume} {90}},\ \bibinfo {pages} {042146}
  (\bibinfo {year} {2014})}\BibitemShut {NoStop}%
\bibitem [{\citenamefont {Saha}\ and\ \citenamefont
  {Jayannavar}(2008)}]{PhysRevE.77.022105}%
  \BibitemOpen
  \bibfield  {author} {\bibinfo {author} {\bibfnamefont {A.}~\bibnamefont
  {Saha}}\ and\ \bibinfo {author} {\bibfnamefont {A.~M.}\ \bibnamefont
  {Jayannavar}},\ }\bibinfo {title} {Nonequilibrium work distributions for a
  trapped Brownian particle in a time-dependent magnetic field},\ \href
  {\doibase 10.1103/PhysRevE.77.022105} {\bibfield  {journal} {\bibinfo
  {journal} {Phys. Rev. E}\ }\textbf {\bibinfo {volume} {77}},\ \bibinfo
  {pages} {022105} (\bibinfo {year} {2008})}\BibitemShut {NoStop}%
\bibitem [{\citenamefont {Li}\ and\ \citenamefont
  {Tu}(2019)}]{PhysRevE.100.012127}%
  \BibitemOpen
  \bibfield  {author} {\bibinfo {author} {\bibfnamefont {G.}~\bibnamefont
  {Li}}\ and\ \bibinfo {author} {\bibfnamefont {Z.~C.}\ \bibnamefont {Tu}},\
  }\bibinfo {title} {Stochastic thermodynamics with odd controlling
  parameters},\ \href {\doibase 10.1103/PhysRevE.100.012127} {\bibfield
  {journal} {\bibinfo  {journal} {Phys. Rev. E}\ }\textbf {\bibinfo {volume}
  {100}},\ \bibinfo {pages} {012127} (\bibinfo {year} {2019})}\BibitemShut
  {NoStop}%
\bibitem [{\citenamefont {Talkner}\ \emph {et~al.}(2009)\citenamefont
  {Talkner}, \citenamefont {Campisi},\ and\ \citenamefont
  {H\"{a}nggi}}]{Talkner2009}%
  \BibitemOpen
  \bibfield  {author} {\bibinfo {author} {\bibfnamefont {P.}~\bibnamefont
  {Talkner}}, \bibinfo {author} {\bibfnamefont {M.}~\bibnamefont {Campisi}}, \
  and\ \bibinfo {author} {\bibfnamefont {P.}~\bibnamefont {H\"{a}nggi}},\
  }\bibinfo {title} {Fluctuation theorems in driven open quantum systems},\
  \href {\doibase 10.1088/1742-5468/2009/02/p02025} {\bibfield  {journal}
  {\bibinfo  {journal} {J. Stat. Mech. Theory Exp.}\ }\textbf {\bibinfo
  {volume} {2009}},\ \bibinfo {pages} {P02025} (\bibinfo {year}
  {2009})}\BibitemShut {NoStop}%
\bibitem [{\citenamefont {Nicolis}\ and\ \citenamefont
  {Decker}(2017)}]{Nicolis2017}%
  \BibitemOpen
  \bibfield  {author} {\bibinfo {author} {\bibfnamefont {G.}~\bibnamefont
  {Nicolis}}\ and\ \bibinfo {author} {\bibfnamefont {Y.~D.}\ \bibnamefont
  {Decker}},\ }\bibinfo {title} {Stochastic Thermodynamics of Brownian
  Motion},\ \href {\doibase 10.3390/e19090434} {\bibfield  {journal} {\bibinfo
  {journal} {Entropy}\ }\textbf {\bibinfo {volume} {19}},\ \bibinfo {pages}
  {434} (\bibinfo {year} {2017})}\BibitemShut {NoStop}%
\bibitem [{\citenamefont {Kurchan}(2000)}]{Kurchan2000}%
  \BibitemOpen
  \bibfield  {author} {\bibinfo {author} {\bibfnamefont {J.}~\bibnamefont
  {Kurchan}},\ }\bibinfo {title} {A quantum fluctuation theorem},\ \href@noop
  {} {\  (\bibinfo {year} {2000})},\ \Eprint
  {http://arxiv.org/abs/cond-mat/0007360} {arXiv:cond-mat/0007360
  [cond-mat.stat-mech]} \BibitemShut {NoStop}%
\bibitem [{\citenamefont {Tasaki}(2000)}]{Tasaki2000}%
  \BibitemOpen
  \bibfield  {author} {\bibinfo {author} {\bibfnamefont {H.}~\bibnamefont
  {Tasaki}},\ }\bibinfo {title} {Jarzynski relations for quantum systems and
  some applications},\ \href@noop {} {\  (\bibinfo {year} {2000})},\ \Eprint
  {http://arxiv.org/abs/cond-mat/0009244} {arXiv:cond-mat/0009244
  [cond-mat.stat-mech]} \BibitemShut {NoStop}%
\bibitem [{\citenamefont {Talkner}\ \emph {et~al.}(2007)\citenamefont
  {Talkner}, \citenamefont {Lutz},\ and\ \citenamefont
  {H\"{a}nggi}}]{Talkner2007}%
  \BibitemOpen
  \bibfield  {author} {\bibinfo {author} {\bibfnamefont {P.}~\bibnamefont
  {Talkner}}, \bibinfo {author} {\bibfnamefont {E.}~\bibnamefont {Lutz}}, \
  and\ \bibinfo {author} {\bibfnamefont {P.}~\bibnamefont {H\"{a}nggi}},\
  }\bibinfo {title} {Fluctuation theorems: Work is not an observable},\ \href
  {\doibase 10.1103/physreve.75.050102} {\bibfield  {journal} {\bibinfo
  {journal} {Phys. Rev. E}\ }\textbf {\bibinfo {volume} {75}},\ \bibinfo
  {pages} {050102(R)} (\bibinfo {year} {2007})}\BibitemShut {NoStop}%
\bibitem [{\citenamefont {Deffner}\ and\ \citenamefont
  {Lutz}(2008)}]{Deffner2008}%
  \BibitemOpen
  \bibfield  {author} {\bibinfo {author} {\bibfnamefont {S.}~\bibnamefont
  {Deffner}}\ and\ \bibinfo {author} {\bibfnamefont {E.}~\bibnamefont {Lutz}},\
  }\bibinfo {title} {Nonequilibrium work distribution of a quantum harmonic
  oscillator},\ \href {\doibase 10.1103/physreve.77.021128} {\bibfield
  {journal} {\bibinfo  {journal} {Phys. Rev. E}\ }\textbf {\bibinfo {volume}
  {77}},\ \bibinfo {pages} {021128} (\bibinfo {year} {2008})}\BibitemShut
  {NoStop}%
\bibitem [{\citenamefont {Andrieux}\ and\ \citenamefont
  {Gaspard}(2008)}]{Andrieux2008}%
  \BibitemOpen
  \bibfield  {author} {\bibinfo {author} {\bibfnamefont {D.}~\bibnamefont
  {Andrieux}}\ and\ \bibinfo {author} {\bibfnamefont {P.}~\bibnamefont
  {Gaspard}},\ }\bibinfo {title} {Quantum Work Relations and Response Theory},\
  \href {\doibase 10.1103/physrevlett.100.230404} {\bibfield  {journal}
  {\bibinfo  {journal} {Phys. Rev. Lett.}\ }\textbf {\bibinfo {volume} {100}},\
  \bibinfo {pages} {230404} (\bibinfo {year} {2008})}\BibitemShut {NoStop}%
\bibitem [{\citenamefont {Campisi}\ \emph {et~al.}(2009)\citenamefont
  {Campisi}, \citenamefont {Talkner},\ and\ \citenamefont
  {H\"{a}nggi}}]{Campisi2009}%
  \BibitemOpen
  \bibfield  {author} {\bibinfo {author} {\bibfnamefont {M.}~\bibnamefont
  {Campisi}}, \bibinfo {author} {\bibfnamefont {P.}~\bibnamefont {Talkner}}, \
  and\ \bibinfo {author} {\bibfnamefont {P.}~\bibnamefont {H\"{a}nggi}},\
  }\bibinfo {title} {Fluctuation Theorem for Arbitrary Open Quantum Systems},\
  \href {\doibase 10.1103/physrevlett.102.210401} {\bibfield  {journal}
  {\bibinfo  {journal} {Phys. Rev. Lett.}\ }\textbf {\bibinfo {volume} {102}},\
  \bibinfo {pages} {210401} (\bibinfo {year} {2009})}\BibitemShut {NoStop}%
\bibitem [{\citenamefont {Hekking}\ and\ \citenamefont
  {Pekola}(2013)}]{Hekking2013}%
  \BibitemOpen
  \bibfield  {author} {\bibinfo {author} {\bibfnamefont {F.~W.~J.}\
  \bibnamefont {Hekking}}\ and\ \bibinfo {author} {\bibfnamefont {J.~P.}\
  \bibnamefont {Pekola}},\ }\bibinfo {title} {Quantum Jump Approach for Work
  and Dissipation in a Two-Level System},\ \href {\doibase
  10.1103/physrevlett.111.093602} {\bibfield  {journal} {\bibinfo  {journal}
  {Phys. Rev. Lett.}\ }\textbf {\bibinfo {volume} {111}},\ \bibinfo {pages}
  {093602} (\bibinfo {year} {2013})}\BibitemShut {NoStop}%
\bibitem [{\citenamefont {Liu}(2014)}]{Liu2014}%
  \BibitemOpen
  \bibfield  {author} {\bibinfo {author} {\bibfnamefont {F.}~\bibnamefont
  {Liu}},\ }\bibinfo {title} {Calculating work in adiabatic two-level quantum
  Markovian master equations: A characteristic function method},\ \href
  {\doibase 10.1103/physreve.90.032121} {\bibfield  {journal} {\bibinfo
  {journal} {Phys. Rev. E}\ }\textbf {\bibinfo {volume} {90}},\ \bibinfo
  {pages} {032121} (\bibinfo {year} {2014})}\BibitemShut {NoStop}%
\bibitem [{\citenamefont {Funo}\ and\ \citenamefont
  {Quan}(2018{\natexlab{a}})}]{Funo2018a}%
  \BibitemOpen
  \bibfield  {author} {\bibinfo {author} {\bibfnamefont {K.}~\bibnamefont
  {Funo}}\ and\ \bibinfo {author} {\bibfnamefont {H.~T.}\ \bibnamefont
  {Quan}},\ }\bibinfo {title} {Path Integral Approach to Quantum
  Thermodynamics},\ \href {\doibase 10.1103/physrevlett.121.040602} {\bibfield
  {journal} {\bibinfo  {journal} {Phys. Rev. Lett.}\ }\textbf {\bibinfo
  {volume} {121}},\ \bibinfo {pages} {040602} (\bibinfo {year}
  {2018}{\natexlab{a}})}\BibitemShut {NoStop}%
\bibitem [{\citenamefont {Jarzynski}\ \emph {et~al.}(2015)\citenamefont
  {Jarzynski}, \citenamefont {Quan},\ and\ \citenamefont
  {Rahav}}]{Jarzynski2015}%
  \BibitemOpen
  \bibfield  {author} {\bibinfo {author} {\bibfnamefont {C.}~\bibnamefont
  {Jarzynski}}, \bibinfo {author} {\bibfnamefont {H.~T.}\ \bibnamefont {Quan}},
  \ and\ \bibinfo {author} {\bibfnamefont {S.}~\bibnamefont {Rahav}},\
  }\bibinfo {title} {Quantum-Classical Correspondence Principle for Work
  Distributions},\ \href {\doibase 10.1103/physrevx.5.031038} {\bibfield
  {journal} {\bibinfo  {journal} {Phys. Rev. X}\ }\textbf {\bibinfo {volume}
  {5}},\ \bibinfo {pages} {031038} (\bibinfo {year} {2015})}\BibitemShut
  {NoStop}%
\bibitem [{\citenamefont {Garc{\'{\i}}a-Mata}\ \emph
  {et~al.}(2017)\citenamefont {Garc{\'{\i}}a-Mata}, \citenamefont {Roncaglia},\
  and\ \citenamefont {Wisniacki}}]{GarciaMata2017}%
  \BibitemOpen
  \bibfield  {author} {\bibinfo {author} {\bibfnamefont {I.}~\bibnamefont
  {Garc{\'{\i}}a-Mata}}, \bibinfo {author} {\bibfnamefont {A.~J.}\ \bibnamefont
  {Roncaglia}}, \ and\ \bibinfo {author} {\bibfnamefont {D.~A.}\ \bibnamefont
  {Wisniacki}},\ }\bibinfo {title} {Quantum-to-classical transition in the work
  distribution for chaotic systems},\ \href {\doibase
  10.1103/physreve.95.050102} {\bibfield  {journal} {\bibinfo  {journal} {Phys.
  Rev. E}\ }\textbf {\bibinfo {volume} {95}},\ \bibinfo {pages} {050102(R)}
  (\bibinfo {year} {2017})}\BibitemShut {NoStop}%
\bibitem [{\citenamefont {Brodier}\ \emph {et~al.}(2020)\citenamefont
  {Brodier}, \citenamefont {Mallick},\ and\ \citenamefont
  {de~Almeida}}]{Brodier2020}%
  \BibitemOpen
  \bibfield  {author} {\bibinfo {author} {\bibfnamefont {O.}~\bibnamefont
  {Brodier}}, \bibinfo {author} {\bibfnamefont {K.}~\bibnamefont {Mallick}}, \
  and\ \bibinfo {author} {\bibfnamefont {A.~M.~O.}\ \bibnamefont
  {de~Almeida}},\ }\bibinfo {title} {Semiclassical work and quantum work
  identities in Weyl representation},\ \href {\doibase
  10.1088/1751-8121/ab8110} {\bibfield  {journal} {\bibinfo  {journal} {J.
  Phys. A: Math. Theor.}\ }\textbf {\bibinfo {volume} {53}},\ \bibinfo {pages}
  {325001} (\bibinfo {year} {2020})}\BibitemShut {NoStop}%
\bibitem [{\citenamefont {Zhu}\ \emph {et~al.}(2016)\citenamefont {Zhu},
  \citenamefont {Gong}, \citenamefont {Wu},\ and\ \citenamefont
  {Quan}}]{PhysRevE.93.062108}%
  \BibitemOpen
  \bibfield  {author} {\bibinfo {author} {\bibfnamefont {L.}~\bibnamefont
  {Zhu}}, \bibinfo {author} {\bibfnamefont {Z.}~\bibnamefont {Gong}}, \bibinfo
  {author} {\bibfnamefont {B.}~\bibnamefont {Wu}}, \ and\ \bibinfo {author}
  {\bibfnamefont {H.~T.}\ \bibnamefont {Quan}},\ }\bibinfo {title}
  {Quantum-classical correspondence principle for work distributions in a
  chaotic system},\ \href {\doibase 10.1103/PhysRevE.93.062108} {\bibfield
  {journal} {\bibinfo  {journal} {Phys. Rev. E}\ }\textbf {\bibinfo {volume}
  {93}},\ \bibinfo {pages} {062108} (\bibinfo {year} {2016})}\BibitemShut
  {NoStop}%
\bibitem [{\citenamefont {Fei}\ \emph {et~al.}(2018)\citenamefont {Fei},
  \citenamefont {Quan},\ and\ \citenamefont {Liu}}]{PhysRevE.98.012132}%
  \BibitemOpen
  \bibfield  {author} {\bibinfo {author} {\bibfnamefont {Z.}~\bibnamefont
  {Fei}}, \bibinfo {author} {\bibfnamefont {H.~T.}\ \bibnamefont {Quan}}, \
  and\ \bibinfo {author} {\bibfnamefont {F.}~\bibnamefont {Liu}},\ }\bibinfo
  {title} {Quantum corrections of work statistics in closed quantum systems},\
  \href {\doibase 10.1103/PhysRevE.98.012132} {\bibfield  {journal} {\bibinfo
  {journal} {Phys. Rev. E}\ }\textbf {\bibinfo {volume} {98}},\ \bibinfo
  {pages} {012132} (\bibinfo {year} {2018})}\BibitemShut {NoStop}%
\bibitem [{\citenamefont {Saito}\ and\ \citenamefont {Dhar}(2007)}]{Saito2007}%
  \BibitemOpen
  \bibfield  {author} {\bibinfo {author} {\bibfnamefont {K.}~\bibnamefont
  {Saito}}\ and\ \bibinfo {author} {\bibfnamefont {A.}~\bibnamefont {Dhar}},\
  }\bibinfo {title} {Fluctuation Theorem in Quantum Heat Conduction},\ \href
  {\doibase 10.1103/physrevlett.99.180601} {\bibfield  {journal} {\bibinfo
  {journal} {Phys. Rev. Lett.}\ }\textbf {\bibinfo {volume} {99}},\ \bibinfo
  {pages} {180601} (\bibinfo {year} {2007})}\BibitemShut {NoStop}%
\bibitem [{\citenamefont {Dubi}\ and\ \citenamefont
  {Di~Ventra}(2011)}]{Dubi2011}%
  \BibitemOpen
  \bibfield  {author} {\bibinfo {author} {\bibfnamefont {Y.}~\bibnamefont
  {Dubi}}\ and\ \bibinfo {author} {\bibfnamefont {M.}~\bibnamefont
  {Di~Ventra}},\ }\bibinfo {title} {Colloquium: Heat flow and thermoelectricity
  in atomic and molecular junctions},\ \href {\doibase
  10.1103/revmodphys.83.131} {\bibfield  {journal} {\bibinfo  {journal} {Rev.
  Mod. Phys.}\ }\textbf {\bibinfo {volume} {83}},\ \bibinfo {pages} {131}
  (\bibinfo {year} {2011})}\BibitemShut {NoStop}%
\bibitem [{\citenamefont {Ren}\ \emph {et~al.}(2010)\citenamefont {Ren},
  \citenamefont {H\"{a}nggi},\ and\ \citenamefont {Li}}]{Ren2010}%
  \BibitemOpen
  \bibfield  {author} {\bibinfo {author} {\bibfnamefont {J.}~\bibnamefont
  {Ren}}, \bibinfo {author} {\bibfnamefont {P.}~\bibnamefont {H\"{a}nggi}}, \
  and\ \bibinfo {author} {\bibfnamefont {B.}~\bibnamefont {Li}},\ }\bibinfo
  {title} {Berry-Phase-Induced Heat Pumping and Its Impact on the Fluctuation
  Theorem},\ \href {\doibase 10.1103/physrevlett.104.170601} {\bibfield
  {journal} {\bibinfo  {journal} {Phys. Rev. Lett.}\ }\textbf {\bibinfo
  {volume} {104}},\ \bibinfo {pages} {170601} (\bibinfo {year}
  {2010})}\BibitemShut {NoStop}%
\bibitem [{\citenamefont {Ren}\ \emph {et~al.}(2012)\citenamefont {Ren},
  \citenamefont {Liu},\ and\ \citenamefont {Li}}]{Ren2012}%
  \BibitemOpen
  \bibfield  {author} {\bibinfo {author} {\bibfnamefont {J.}~\bibnamefont
  {Ren}}, \bibinfo {author} {\bibfnamefont {S.}~\bibnamefont {Liu}}, \ and\
  \bibinfo {author} {\bibfnamefont {B.}~\bibnamefont {Li}},\ }\bibinfo {title}
  {Geometric Heat Flux for Classical Thermal Transport in Interacting Open
  Systems},\ \href {\doibase 10.1103/physrevlett.108.210603} {\bibfield
  {journal} {\bibinfo  {journal} {Phys. Rev. Lett.}\ }\textbf {\bibinfo
  {volume} {108}},\ \bibinfo {pages} {210603} (\bibinfo {year}
  {2012})}\BibitemShut {NoStop}%
\bibitem [{\citenamefont {Thingna}\ \emph {et~al.}(2012)\citenamefont
  {Thingna}, \citenamefont {Garc{\'{\i}}a-Palacios},\ and\ \citenamefont
  {Wang}}]{Thingna2012}%
  \BibitemOpen
  \bibfield  {author} {\bibinfo {author} {\bibfnamefont {J.}~\bibnamefont
  {Thingna}}, \bibinfo {author} {\bibfnamefont {J.~L.}\ \bibnamefont
  {Garc{\'{\i}}a-Palacios}}, \ and\ \bibinfo {author} {\bibfnamefont {J.-S.}\
  \bibnamefont {Wang}},\ }\bibinfo {title} {Steady-state thermal transport in
  anharmonic systems: Application to molecular junctions},\ \href {\doibase
  10.1103/physrevb.85.195452} {\bibfield  {journal} {\bibinfo  {journal} {Phys.
  Rev. B}\ }\textbf {\bibinfo {volume} {85}},\ \bibinfo {pages} {195452}
  (\bibinfo {year} {2012})}\BibitemShut {NoStop}%
\bibitem [{\citenamefont {Wang}\ \emph {et~al.}(2014)\citenamefont {Wang},
  \citenamefont {Agarwalla}, \citenamefont {Li},\ and\ \citenamefont
  {Thingna}}]{Wang2013}%
  \BibitemOpen
  \bibfield  {author} {\bibinfo {author} {\bibfnamefont {J.-S.}\ \bibnamefont
  {Wang}}, \bibinfo {author} {\bibfnamefont {B.~K.}\ \bibnamefont {Agarwalla}},
  \bibinfo {author} {\bibfnamefont {H.}~\bibnamefont {Li}}, \ and\ \bibinfo
  {author} {\bibfnamefont {J.}~\bibnamefont {Thingna}},\ }\bibinfo {title}
  {Nonequilibrium Green's function method for quantum thermal transport},\
  \href {\doibase 10.1007/s11467-013-0340-x} {\bibfield  {journal} {\bibinfo
  {journal} {Front. Phys.}\ }\textbf {\bibinfo {volume} {9}},\ \bibinfo {pages}
  {673} (\bibinfo {year} {2014})}\BibitemShut {NoStop}%
\bibitem [{\citenamefont {Fogedby}\ and\ \citenamefont
  {Imparato}(2014)}]{Fogedby2014}%
  \BibitemOpen
  \bibfield  {author} {\bibinfo {author} {\bibfnamefont {H.~C.}\ \bibnamefont
  {Fogedby}}\ and\ \bibinfo {author} {\bibfnamefont {A.}~\bibnamefont
  {Imparato}},\ }\bibinfo {title} {Heat fluctuations and fluctuation theorems
  in the case of multiple reservoirs},\ \href {\doibase
  10.1088/1742-5468/2014/11/p11011} {\bibfield  {journal} {\bibinfo  {journal}
  {J. Stat. Mech.: Theory Exp.}\ }\textbf {\bibinfo {volume} {2014}},\ \bibinfo
  {pages} {P11011} (\bibinfo {year} {2014})}\BibitemShut {NoStop}%
\bibitem [{\citenamefont {Li}\ \emph {et~al.}(2015)\citenamefont {Li},
  \citenamefont {Cai},\ and\ \citenamefont {Sun}}]{Li2015}%
  \BibitemOpen
  \bibfield  {author} {\bibinfo {author} {\bibfnamefont {S.-W.}\ \bibnamefont
  {Li}}, \bibinfo {author} {\bibfnamefont {C.}~\bibnamefont {Cai}}, \ and\
  \bibinfo {author} {\bibfnamefont {C.}~\bibnamefont {Sun}},\ }\bibinfo {title}
  {Steady quantum coherence in non-equilibrium environment},\ \href {\doibase
  10.1016/j.aop.2015.05.004} {\bibfield  {journal} {\bibinfo  {journal} {Ann.
  Phys. (NY)}\ }\textbf {\bibinfo {volume} {360}},\ \bibinfo {pages} {19}
  (\bibinfo {year} {2015})}\BibitemShut {NoStop}%
\bibitem [{\citenamefont {Kilgour}\ \emph {et~al.}(2019)\citenamefont
  {Kilgour}, \citenamefont {Agarwalla},\ and\ \citenamefont
  {Segal}}]{Kilgour2019}%
  \BibitemOpen
  \bibfield  {author} {\bibinfo {author} {\bibfnamefont {M.}~\bibnamefont
  {Kilgour}}, \bibinfo {author} {\bibfnamefont {B.~K.}\ \bibnamefont
  {Agarwalla}}, \ and\ \bibinfo {author} {\bibfnamefont {D.}~\bibnamefont
  {Segal}},\ }\bibinfo {title} {Path-integral methodology and simulations of
  quantum thermal transport: Full counting statistics approach},\ \href
  {\doibase 10.1063/1.5084949} {\bibfield  {journal} {\bibinfo  {journal} {J.
  Chem. Phys.}\ }\textbf {\bibinfo {volume} {150}},\ \bibinfo {pages} {084111}
  (\bibinfo {year} {2019})}\BibitemShut {NoStop}%
\bibitem [{\citenamefont {Aurell}\ \emph {et~al.}(2020)\citenamefont {Aurell},
  \citenamefont {Donvil},\ and\ \citenamefont {Mallick}}]{Aurell2020}%
  \BibitemOpen
  \bibfield  {author} {\bibinfo {author} {\bibfnamefont {E.}~\bibnamefont
  {Aurell}}, \bibinfo {author} {\bibfnamefont {B.}~\bibnamefont {Donvil}}, \
  and\ \bibinfo {author} {\bibfnamefont {K.}~\bibnamefont {Mallick}},\
  }\bibinfo {title} {Large deviations and fluctuation theorem for the quantum
  heat current in the spin-boson model},\ \href {\doibase
  10.1103/physreve.101.052116} {\bibfield  {journal} {\bibinfo  {journal}
  {Phys. Rev. E}\ }\textbf {\bibinfo {volume} {101}},\ \bibinfo {pages}
  {052116} (\bibinfo {year} {2020})}\BibitemShut {NoStop}%
\bibitem [{\citenamefont {Santos}\ \emph {et~al.}(2020)\citenamefont {Santos},
  \citenamefont {Timpanaro},\ and\ \citenamefont {Landi}}]{Santos2020}%
  \BibitemOpen
  \bibfield  {author} {\bibinfo {author} {\bibfnamefont {J.}~\bibnamefont
  {Santos}}, \bibinfo {author} {\bibfnamefont {A.}~\bibnamefont {Timpanaro}}, \
  and\ \bibinfo {author} {\bibfnamefont {G.}~\bibnamefont {Landi}},\ }\bibinfo
  {title} {Joint Fluctuation Theorems for Sequential Heat Exchange},\ \href
  {\doibase 10.3390/e22070763} {\bibfield  {journal} {\bibinfo  {journal}
  {Entropy}\ }\textbf {\bibinfo {volume} {22}},\ \bibinfo {pages} {763}
  (\bibinfo {year} {2020})}\BibitemShut {NoStop}%
\bibitem [{\citenamefont {Levy}\ and\ \citenamefont
  {Lostaglio}(2020)}]{Levy2020}%
  \BibitemOpen
  \bibfield  {author} {\bibinfo {author} {\bibfnamefont {A.}~\bibnamefont
  {Levy}}\ and\ \bibinfo {author} {\bibfnamefont {M.}~\bibnamefont
  {Lostaglio}},\ }\bibinfo {title} {Quasiprobability Distribution for Heat
  Fluctuations in the Quantum Regime},\ \href {\doibase
  10.1103/prxquantum.1.010309} {\bibfield  {journal} {\bibinfo  {journal}
  {{PRX} Quantum}\ }\textbf {\bibinfo {volume} {1}},\ \bibinfo {pages} {010309}
  (\bibinfo {year} {2020})}\BibitemShut {NoStop}%
\bibitem [{\citenamefont {Gupta}\ and\ \citenamefont
  {Sivak}(2021)}]{Gupta2021}%
  \BibitemOpen
  \bibfield  {author} {\bibinfo {author} {\bibfnamefont {D.}~\bibnamefont
  {Gupta}}\ and\ \bibinfo {author} {\bibfnamefont {D.~A.}\ \bibnamefont
  {Sivak}},\ }\bibinfo {title} {Heat fluctuations in a harmonic chain of active
  particles},\ \href {\doibase 10.1103/physreve.104.024605} {\bibfield
  {journal} {\bibinfo  {journal} {Phys. Rev. E}\ }\textbf {\bibinfo {volume}
  {104}},\ \bibinfo {pages} {024605} (\bibinfo {year} {2021})}\BibitemShut
  {NoStop}%
\bibitem [{\citenamefont {Gallavotti}\ and\ \citenamefont
  {Cohen}(1995)}]{Gallavotti1995}%
  \BibitemOpen
  \bibfield  {author} {\bibinfo {author} {\bibfnamefont {G.}~\bibnamefont
  {Gallavotti}}\ and\ \bibinfo {author} {\bibfnamefont {E.~G.~D.}\ \bibnamefont
  {Cohen}},\ }\bibinfo {title} {Dynamical ensembles in stationary states},\
  \href {\doibase 10.1007/bf02179860} {\bibfield  {journal} {\bibinfo
  {journal} {J. Stat. Phys.}\ }\textbf {\bibinfo {volume} {80}},\ \bibinfo
  {pages} {931} (\bibinfo {year} {1995})}\BibitemShut {NoStop}%
\bibitem [{\citenamefont {van Zon}\ and\ \citenamefont
  {Cohen}(2004)}]{Zon2004}%
  \BibitemOpen
  \bibfield  {author} {\bibinfo {author} {\bibfnamefont {R.}~\bibnamefont {van
  Zon}}\ and\ \bibinfo {author} {\bibfnamefont {E.~G.~D.}\ \bibnamefont
  {Cohen}},\ }\bibinfo {title} {Extended heat-fluctuation theorems for a system
  with deterministic and stochastic forces},\ \href {\doibase
  10.1103/physreve.69.056121} {\bibfield  {journal} {\bibinfo  {journal} {Phys.
  Rev. E}\ }\textbf {\bibinfo {volume} {69}},\ \bibinfo {pages} {056121}
  (\bibinfo {year} {2004})}\BibitemShut {NoStop}%
\bibitem [{\citenamefont {Fogedby}\ and\ \citenamefont
  {Imparato}(2009)}]{Fogedby2009}%
  \BibitemOpen
  \bibfield  {author} {\bibinfo {author} {\bibfnamefont {H.~C.}\ \bibnamefont
  {Fogedby}}\ and\ \bibinfo {author} {\bibfnamefont {A.}~\bibnamefont
  {Imparato}},\ }\bibinfo {title} {Heat distribution function for motion in a
  general potential at low temperature},\ \href {\doibase
  10.1088/1751-8113/42/47/475004} {\bibfield  {journal} {\bibinfo  {journal}
  {J. Phys. A: Math. Theor.}\ }\textbf {\bibinfo {volume} {42}},\ \bibinfo
  {pages} {475004} (\bibinfo {year} {2009})}\BibitemShut {NoStop}%
\bibitem [{\citenamefont {Chatterjee}\ and\ \citenamefont
  {Cherayil}(2010)}]{Chatterjee2010}%
  \BibitemOpen
  \bibfield  {author} {\bibinfo {author} {\bibfnamefont {D.}~\bibnamefont
  {Chatterjee}}\ and\ \bibinfo {author} {\bibfnamefont {B.~J.}\ \bibnamefont
  {Cherayil}},\ }\bibinfo {title} {Exact path-integral evaluation of the heat
  distribution function of a trapped Brownian oscillator},\ \href {\doibase
  10.1103/physreve.82.051104} {\bibfield  {journal} {\bibinfo  {journal} {Phys.
  Rev. E}\ }\textbf {\bibinfo {volume} {82}},\ \bibinfo {pages} {051104}
  (\bibinfo {year} {2010})}\BibitemShut {NoStop}%
\bibitem [{\citenamefont {Gomez-Solano}\ \emph {et~al.}(2011)\citenamefont
  {Gomez-Solano}, \citenamefont {Petrosyan},\ and\ \citenamefont
  {Ciliberto}}]{GomezSolano2011}%
  \BibitemOpen
  \bibfield  {author} {\bibinfo {author} {\bibfnamefont {J.~R.}\ \bibnamefont
  {Gomez-Solano}}, \bibinfo {author} {\bibfnamefont {A.}~\bibnamefont
  {Petrosyan}}, \ and\ \bibinfo {author} {\bibfnamefont {S.}~\bibnamefont
  {Ciliberto}},\ }\bibinfo {title} {Heat Fluctuations in a Nonequilibrium
  Bath},\ \href {\doibase 10.1103/physrevlett.106.200602} {\bibfield  {journal}
  {\bibinfo  {journal} {Phys. Rev. Lett.}\ }\textbf {\bibinfo {volume} {106}},\
  \bibinfo {pages} {200602} (\bibinfo {year} {2011})}\BibitemShut {NoStop}%
\bibitem [{\citenamefont {Salazar}\ and\ \citenamefont
  {Lira}(2016)}]{Salazar2016}%
  \BibitemOpen
  \bibfield  {author} {\bibinfo {author} {\bibfnamefont {D.~S.~P.}\
  \bibnamefont {Salazar}}\ and\ \bibinfo {author} {\bibfnamefont {S.~A.}\
  \bibnamefont {Lira}},\ }\bibinfo {title} {Exactly solvable nonequilibrium
  Langevin relaxation of a trapped nanoparticle},\ \href {\doibase
  10.1088/1751-8113/49/46/465001} {\bibfield  {journal} {\bibinfo  {journal}
  {J. Phys. A: Math. Theor.}\ }\textbf {\bibinfo {volume} {49}},\ \bibinfo
  {pages} {465001} (\bibinfo {year} {2016})}\BibitemShut {NoStop}%
\bibitem [{\citenamefont {Funo}\ and\ \citenamefont
  {Quan}(2018{\natexlab{b}})}]{Funo2018}%
  \BibitemOpen
  \bibfield  {author} {\bibinfo {author} {\bibfnamefont {K.}~\bibnamefont
  {Funo}}\ and\ \bibinfo {author} {\bibfnamefont {H.~T.}\ \bibnamefont
  {Quan}},\ }\bibinfo {title} {Path integral approach to heat in quantum
  thermodynamics},\ \href {\doibase 10.1103/physreve.98.012113} {\bibfield
  {journal} {\bibinfo  {journal} {Phys. Rev. E}\ }\textbf {\bibinfo {volume}
  {98}},\ \bibinfo {pages} {012113} (\bibinfo {year}
  {2018}{\natexlab{b}})}\BibitemShut {NoStop}%
\bibitem [{\citenamefont {Denzler}\ and\ \citenamefont
  {Lutz}(2018)}]{Denzler2018}%
  \BibitemOpen
  \bibfield  {author} {\bibinfo {author} {\bibfnamefont {T.}~\bibnamefont
  {Denzler}}\ and\ \bibinfo {author} {\bibfnamefont {E.}~\bibnamefont {Lutz}},\
  }\bibinfo {title} {Heat distribution of a quantum harmonic oscillator},\
  \href {\doibase 10.1103/physreve.98.052106} {\bibfield  {journal} {\bibinfo
  {journal} {Phys. Rev. E}\ }\textbf {\bibinfo {volume} {98}},\ \bibinfo
  {pages} {052106} (\bibinfo {year} {2018})}\BibitemShut {NoStop}%
\bibitem [{\citenamefont {Salazar}\ \emph {et~al.}(2019)\citenamefont
  {Salazar}, \citenamefont {Mac\^edo},\ and\ \citenamefont
  {Vasconcelos}}]{Salazar2019}%
  \BibitemOpen
  \bibfield  {author} {\bibinfo {author} {\bibfnamefont {D.~S.~P.}\
  \bibnamefont {Salazar}}, \bibinfo {author} {\bibfnamefont {A.~M.~S.}\
  \bibnamefont {Mac\^edo}}, \ and\ \bibinfo {author} {\bibfnamefont {G.~L.}\
  \bibnamefont {Vasconcelos}},\ }\bibinfo {title} {Quantum heat distribution in
  thermal relaxation processes},\ \href {\doibase 10.1103/physreve.99.022133}
  {\bibfield  {journal} {\bibinfo  {journal} {Phys. Rev. E}\ }\textbf {\bibinfo
  {volume} {99}},\ \bibinfo {pages} {022133} (\bibinfo {year}
  {2019})}\BibitemShut {NoStop}%
\bibitem [{\citenamefont {Fogedby}(2020)}]{Fogedby2020}%
  \BibitemOpen
  \bibfield  {author} {\bibinfo {author} {\bibfnamefont {H.~C.}\ \bibnamefont
  {Fogedby}},\ }\bibinfo {title} {Heat fluctuations in equilibrium},\ \href
  {\doibase 10.1088/1742-5468/aba7b2} {\bibfield  {journal} {\bibinfo
  {journal} {J. Stat. Mech.: Theory Exp.}\ }\textbf {\bibinfo {volume}
  {2020}},\ \bibinfo {pages} {083208} (\bibinfo {year} {2020})}\BibitemShut
  {NoStop}%
\bibitem [{\citenamefont {Popovic}\ \emph {et~al.}(2021)\citenamefont
  {Popovic}, \citenamefont {Mitchison}, \citenamefont {Strathearn},
  \citenamefont {Lovett}, \citenamefont {Goold},\ and\ \citenamefont
  {Eastham}}]{Popovic2021}%
  \BibitemOpen
  \bibfield  {author} {\bibinfo {author} {\bibfnamefont {M.}~\bibnamefont
  {Popovic}}, \bibinfo {author} {\bibfnamefont {M.~T.}\ \bibnamefont
  {Mitchison}}, \bibinfo {author} {\bibfnamefont {A.}~\bibnamefont
  {Strathearn}}, \bibinfo {author} {\bibfnamefont {B.~W.}\ \bibnamefont
  {Lovett}}, \bibinfo {author} {\bibfnamefont {J.}~\bibnamefont {Goold}}, \
  and\ \bibinfo {author} {\bibfnamefont {P.~R.}\ \bibnamefont {Eastham}},\
  }\bibinfo {title} {Quantum Heat Statistics with Time-Evolving Matrix Product
  Operators},\ \href {\doibase 10.1103/prxquantum.2.020338} {\bibfield
  {journal} {\bibinfo  {journal} {{PRX} Quantum}\ }\textbf {\bibinfo {volume}
  {2}},\ \bibinfo {pages} {020338} (\bibinfo {year} {2021})}\BibitemShut
  {NoStop}%
\bibitem [{\citenamefont {Paraguass\'{u}}\ \emph {et~al.}(2021)\citenamefont
  {Paraguass\'{u}}, \citenamefont {Aquino},\ and\ \citenamefont
  {Morgado}}]{Paraguassu2021a}%
  \BibitemOpen
  \bibfield  {author} {\bibinfo {author} {\bibfnamefont {P.~V.}\ \bibnamefont
  {Paraguass\'{u}}}, \bibinfo {author} {\bibfnamefont {R.}~\bibnamefont
  {Aquino}}, \ and\ \bibinfo {author} {\bibfnamefont {W.~A.~M.}\ \bibnamefont
  {Morgado}},\ }\bibinfo {title} {The heat distribution of the underdamped
  Langevin equation},\ \href@noop {} {\  (\bibinfo {year} {2021})},\ \Eprint
  {http://arxiv.org/abs/2102.09115} {arXiv:2102.09115 [cond-mat.stat-mech]}
  \BibitemShut {NoStop}%
\bibitem [{\citenamefont {Chen}\ \emph {et~al.}(2021)\citenamefont {Chen},
  \citenamefont {Qiu},\ and\ \citenamefont {Quan}}]{Chen2021b}%
  \BibitemOpen
  \bibfield  {author} {\bibinfo {author} {\bibfnamefont {J.-F.}\ \bibnamefont
  {Chen}}, \bibinfo {author} {\bibfnamefont {T.}~\bibnamefont {Qiu}}, \ and\
  \bibinfo {author} {\bibfnamefont {H.-T.}\ \bibnamefont {Quan}},\ }\bibinfo
  {title} {Quantum{\textendash}Classical Correspondence Principle for Heat
  Distribution in Quantum Brownian Motion},\ \href {\doibase 10.3390/e23121602}
  {\bibfield  {journal} {\bibinfo  {journal} {Entropy}\ }\textbf {\bibinfo
  {volume} {23}},\ \bibinfo {pages} {1602} (\bibinfo {year}
  {2021})}\BibitemShut {NoStop}%
\bibitem [{\citenamefont {Crisanti}\ \emph {et~al.}(2017)\citenamefont
  {Crisanti}, \citenamefont {Sarracino},\ and\ \citenamefont
  {Zannetti}}]{PhysRevE.95.052138}%
  \BibitemOpen
  \bibfield  {author} {\bibinfo {author} {\bibfnamefont {A.}~\bibnamefont
  {Crisanti}}, \bibinfo {author} {\bibfnamefont {A.}~\bibnamefont {Sarracino}},
  \ and\ \bibinfo {author} {\bibfnamefont {M.}~\bibnamefont {Zannetti}},\
  }\bibinfo {title} {Heat fluctuations of Brownian oscillators in nonstationary
  processes: Fluctuation theorem and condensation transition},\ \href {\doibase
  10.1103/PhysRevE.95.052138} {\bibfield  {journal} {\bibinfo  {journal} {Phys.
  Rev. E}\ }\textbf {\bibinfo {volume} {95}},\ \bibinfo {pages} {052138}
  (\bibinfo {year} {2017})}\BibitemShut {NoStop}%
\bibitem [{\citenamefont {Murashita}\ and\ \citenamefont
  {Esposito}(2016)}]{Murashita2016}%
  \BibitemOpen
  \bibfield  {author} {\bibinfo {author} {\bibfnamefont {Y.}~\bibnamefont
  {Murashita}}\ and\ \bibinfo {author} {\bibfnamefont {M.}~\bibnamefont
  {Esposito}},\ }\bibinfo {title} {Overdamped stochastic thermodynamics with
  multiple reservoirs},\ \href {\doibase 10.1103/physreve.94.062148} {\bibfield
   {journal} {\bibinfo  {journal} {Phys. Rev. E}\ }\textbf {\bibinfo {volume}
  {94}},\ \bibinfo {pages} {062148} (\bibinfo {year} {2016})}\BibitemShut
  {NoStop}%
\bibitem [{\citenamefont {Pal}\ \emph {et~al.}(2017)\citenamefont {Pal},
  \citenamefont {Lahiri},\ and\ \citenamefont {Jayannavar}}]{Pal2017}%
  \BibitemOpen
  \bibfield  {author} {\bibinfo {author} {\bibfnamefont {P.~S.}\ \bibnamefont
  {Pal}}, \bibinfo {author} {\bibfnamefont {S.}~\bibnamefont {Lahiri}}, \ and\
  \bibinfo {author} {\bibfnamefont {A.~M.}\ \bibnamefont {Jayannavar}},\
  }\bibinfo {title} {Transient exchange fluctuation theorems for heat using a
  Hamiltonian framework: Classical and quantum regimes},\ \href {\doibase
  10.1103/physreve.95.042124} {\bibfield  {journal} {\bibinfo  {journal} {Phys.
  Rev. E}\ }\textbf {\bibinfo {volume} {95}},\ \bibinfo {pages} {042124}
  (\bibinfo {year} {2017})}\BibitemShut {NoStop}%
\bibitem [{\citenamefont {Lee}\ and\ \citenamefont {Park}(2018)}]{Lee2018}%
  \BibitemOpen
  \bibfield  {author} {\bibinfo {author} {\bibfnamefont {J.~S.}\ \bibnamefont
  {Lee}}\ and\ \bibinfo {author} {\bibfnamefont {H.}~\bibnamefont {Park}},\
  }\bibinfo {title} {Stochastic thermodynamics and hierarchy of fluctuation
  theorems with multiple reservoirs},\ \href {\doibase
  10.1088/1367-2630/aad6d3} {\bibfield  {journal} {\bibinfo  {journal} {New J.
  Phys.}\ }\textbf {\bibinfo {volume} {20}},\ \bibinfo {pages} {083010}
  (\bibinfo {year} {2018})}\BibitemShut {NoStop}%
\bibitem [{\citenamefont {Kurchan}(1998)}]{Kurchan1998}%
  \BibitemOpen
  \bibfield  {author} {\bibinfo {author} {\bibfnamefont {J.}~\bibnamefont
  {Kurchan}},\ }\bibinfo {title} {Fluctuation theorem for stochastic
  dynamics},\ \href {\doibase 10.1088/0305-4470/31/16/003} {\bibfield
  {journal} {\bibinfo  {journal} {J. Phys. A: Math. Gen.}\ }\textbf {\bibinfo
  {volume} {31}},\ \bibinfo {pages} {3719} (\bibinfo {year}
  {1998})}\BibitemShut {NoStop}%
\bibitem [{\citenamefont {Crooks}(2000)}]{Crooks2000}%
  \BibitemOpen
  \bibfield  {author} {\bibinfo {author} {\bibfnamefont {G.~E.}\ \bibnamefont
  {Crooks}},\ }\bibinfo {title} {Path-ensemble averages in systems driven far
  from equilibrium},\ \href {\doibase 10.1103/physreve.61.2361} {\bibfield
  {journal} {\bibinfo  {journal} {Phys. Rev. E}\ }\textbf {\bibinfo {volume}
  {61}},\ \bibinfo {pages} {2361} (\bibinfo {year} {2000})}\BibitemShut
  {NoStop}%
\bibitem [{\citenamefont {Maes}\ and\ \citenamefont
  {Neto{\v{c}}n{\'{y}}}(2003)}]{Maes2003}%
  \BibitemOpen
  \bibfield  {author} {\bibinfo {author} {\bibfnamefont {C.}~\bibnamefont
  {Maes}}\ and\ \bibinfo {author} {\bibfnamefont {K.}~\bibnamefont
  {Neto{\v{c}}n{\'{y}}}},\ }\bibinfo {title} {Time-reversal and entropy},\
  \href {\doibase 10.1023/a:1021026930129} {\bibfield  {journal} {\bibinfo
  {journal} {J. Stat. Phys.}\ }\textbf {\bibinfo {volume} {110}},\ \bibinfo
  {pages} {269} (\bibinfo {year} {2003})}\BibitemShut {NoStop}%
\bibitem [{\citenamefont {Maes}(2004)}]{Maes2004}%
  \BibitemOpen
  \bibfield  {author} {\bibinfo {author} {\bibfnamefont {C.}~\bibnamefont
  {Maes}},\ }in\ \href {\doibase 10.1007/978-3-0348-7932-3_8} {\emph {\bibinfo
  {booktitle} {Poincar{\'{e}} Seminar 2003}}},\ Vol.~\bibinfo {volume} {2}\
  (\bibinfo  {publisher} {Birkh\"{a}user Basel},\ \bibinfo {year} {2004})\
  p.~\bibinfo {pages} {29}\BibitemShut {NoStop}%
\bibitem [{\citenamefont {Andrieux}\ \emph {et~al.}(2007)\citenamefont
  {Andrieux}, \citenamefont {Gaspard}, \citenamefont {Ciliberto}, \citenamefont
  {Garnier}, \citenamefont {Joubaud},\ and\ \citenamefont
  {Petrosyan}}]{Andrieux2007}%
  \BibitemOpen
  \bibfield  {author} {\bibinfo {author} {\bibfnamefont {D.}~\bibnamefont
  {Andrieux}}, \bibinfo {author} {\bibfnamefont {P.}~\bibnamefont {Gaspard}},
  \bibinfo {author} {\bibfnamefont {S.}~\bibnamefont {Ciliberto}}, \bibinfo
  {author} {\bibfnamefont {N.}~\bibnamefont {Garnier}}, \bibinfo {author}
  {\bibfnamefont {S.}~\bibnamefont {Joubaud}}, \ and\ \bibinfo {author}
  {\bibfnamefont {A.}~\bibnamefont {Petrosyan}},\ }\bibinfo {title} {Entropy
  Production and Time Asymmetry in Nonequilibrium Fluctuations},\ \href
  {\doibase 10.1103/physrevlett.98.150601} {\bibfield  {journal} {\bibinfo
  {journal} {Phys. Rev. Lett.}\ }\textbf {\bibinfo {volume} {98}},\ \bibinfo
  {pages} {150601} (\bibinfo {year} {2007})}\BibitemShut {NoStop}%
\bibitem [{\citenamefont {Gomez-Marin}\ \emph {et~al.}(2008)\citenamefont
  {Gomez-Marin}, \citenamefont {Parrondo},\ and\ \citenamefont {den
  Broeck}}]{GomezMarin2008a}%
  \BibitemOpen
  \bibfield  {author} {\bibinfo {author} {\bibfnamefont {A.}~\bibnamefont
  {Gomez-Marin}}, \bibinfo {author} {\bibfnamefont {J.~M.~R.}\ \bibnamefont
  {Parrondo}}, \ and\ \bibinfo {author} {\bibfnamefont {C.~V.}\ \bibnamefont
  {den Broeck}},\ }\bibinfo {title} {The
  {\textquotedblleft}footprints{\textquotedblright} of irreversibility},\ \href
  {\doibase 10.1209/0295-5075/82/50002} {\bibfield  {journal} {\bibinfo
  {journal} {Europhys. Lett.}\ }\textbf {\bibinfo {volume} {82}},\ \bibinfo
  {pages} {50002} (\bibinfo {year} {2008})}\BibitemShut {NoStop}%
\bibitem [{\citenamefont {Parrondo}\ \emph {et~al.}(2009)\citenamefont
  {Parrondo}, \citenamefont {den Broeck},\ and\ \citenamefont
  {Kawai}}]{Parrondo2009}%
  \BibitemOpen
  \bibfield  {author} {\bibinfo {author} {\bibfnamefont {J.~M.~R.}\
  \bibnamefont {Parrondo}}, \bibinfo {author} {\bibfnamefont {C.~V.}\
  \bibnamefont {den Broeck}}, \ and\ \bibinfo {author} {\bibfnamefont
  {R.}~\bibnamefont {Kawai}},\ }\bibinfo {title} {Entropy production and the
  arrow of time},\ \href {\doibase 10.1088/1367-2630/11/7/073008} {\bibfield
  {journal} {\bibinfo  {journal} {New J. Phys.}\ }\textbf {\bibinfo {volume}
  {11}},\ \bibinfo {pages} {073008} (\bibinfo {year} {2009})}\BibitemShut
  {NoStop}%
\bibitem [{\citenamefont {Crooks}(2011)}]{Crooks2011}%
  \BibitemOpen
  \bibfield  {author} {\bibinfo {author} {\bibfnamefont {G.~E.}\ \bibnamefont
  {Crooks}},\ }\bibinfo {title} {On thermodynamic and microscopic
  reversibility},\ \href {\doibase 10.1088/1742-5468/2011/07/p07008} {\bibfield
   {journal} {\bibinfo  {journal} {J. Stat. Mech.: Theory Exp.}\ }\textbf
  {\bibinfo {volume} {2011}},\ \bibinfo {pages} {P07008} (\bibinfo {year}
  {2011})}\BibitemShut {NoStop}%
\bibitem [{\citenamefont {Blundell}\ and\ \citenamefont
  {Blundell}(2009)}]{BlundellBook2009}%
  \BibitemOpen
  \bibfield  {author} {\bibinfo {author} {\bibfnamefont {S.~J.}\ \bibnamefont
  {Blundell}}\ and\ \bibinfo {author} {\bibfnamefont {K.~M.}\ \bibnamefont
  {Blundell}},\ }\href
  {https://www.ebook.de/de/product/8623920/katherine_m_university_of_oxford_uk_blundell_concepts_in_thermal_physics.html}
  {\emph {\bibinfo {title} {Concepts in Thermal Physics}}}\ (\bibinfo
  {publisher} {Oxford University Press},\ \bibinfo {year} {2009})\BibitemShut
  {NoStop}%
\bibitem [{\citenamefont {G{\'{o}}mez}\ \emph {et~al.}(2021)\citenamefont
  {G{\'{o}}mez}, \citenamefont {Staudenmaier}, \citenamefont {Campisi},\ and\
  \citenamefont {Fabbri}}]{Gomez2021}%
  \BibitemOpen
  \bibfield  {author} {\bibinfo {author} {\bibfnamefont {S.~H.}\ \bibnamefont
  {G{\'{o}}mez}}, \bibinfo {author} {\bibfnamefont {N.}~\bibnamefont
  {Staudenmaier}}, \bibinfo {author} {\bibfnamefont {M.}~\bibnamefont
  {Campisi}}, \ and\ \bibinfo {author} {\bibfnamefont {N.}~\bibnamefont
  {Fabbri}},\ }\bibinfo {title} {Experimental test of fluctuation relations for
  driven open quantum systems with an {NV} center},\ \href {\doibase
  10.1088/1367-2630/abfc6a} {\bibfield  {journal} {\bibinfo  {journal} {New J.
  Phys.}\ }\textbf {\bibinfo {volume} {23}},\ \bibinfo {pages} {065004}
  (\bibinfo {year} {2021})}\BibitemShut {NoStop}%
\bibitem [{\citenamefont {Sinitsyn}(2011)}]{Sinitsyn2011}%
  \BibitemOpen
  \bibfield  {author} {\bibinfo {author} {\bibfnamefont {N.~A.}\ \bibnamefont
  {Sinitsyn}},\ }\bibinfo {title} {Fluctuation relation for heat engines},\
  \href {\doibase 10.1088/1751-8113/44/40/405001} {\bibfield  {journal}
  {\bibinfo  {journal} {J. Phys. A: Math. Theor.}\ }\textbf {\bibinfo {volume}
  {44}},\ \bibinfo {pages} {405001} (\bibinfo {year} {2011})}\BibitemShut
  {NoStop}%
\bibitem [{\citenamefont {Campisi}(2014)}]{Campisi2014}%
  \BibitemOpen
  \bibfield  {author} {\bibinfo {author} {\bibfnamefont {M.}~\bibnamefont
  {Campisi}},\ }\bibinfo {title} {Fluctuation relation for quantum heat engines
  and refrigerators},\ \href {\doibase 10.1088/1751-8113/47/24/245001}
  {\bibfield  {journal} {\bibinfo  {journal} {J. Phys. A.}\ }\textbf {\bibinfo
  {volume} {47}},\ \bibinfo {pages} {245001} (\bibinfo {year}
  {2014})}\BibitemShut {NoStop}%
\bibitem [{\citenamefont {Chen}\ \emph {et~al.}(2022)\citenamefont {Chen},
  \citenamefont {Chen}, \citenamefont {Fei},\ and\ \citenamefont
  {Quan}}]{Chen2021a}%
  \BibitemOpen
  \bibfield  {author} {\bibinfo {author} {\bibfnamefont {Y.~H.}\ \bibnamefont
  {Chen}}, \bibinfo {author} {\bibfnamefont {J.-F.}\ \bibnamefont {Chen}},
  \bibinfo {author} {\bibfnamefont {Z.}~\bibnamefont {Fei}}, \ and\ \bibinfo
  {author} {\bibfnamefont {H.~T.}\ \bibnamefont {Quan}},\ }\bibinfo {title}
  {Microscopic theory of the Curzon-Ahlborn heat engine based on a Brownian
  particle},\ \href {\doibase 10.1103/physreve.106.024105} {\bibfield
  {journal} {\bibinfo  {journal} {Phys. Rev. E}\ }\textbf {\bibinfo {volume}
  {106}},\ \bibinfo {pages} {024105} (\bibinfo {year} {2022})}\BibitemShut
  {NoStop}%
\bibitem [{\citenamefont {Jarzynski}(1999)}]{Jarzynski1999}%
  \BibitemOpen
  \bibfield  {author} {\bibinfo {author} {\bibfnamefont {C.}~\bibnamefont
  {Jarzynski}},\ }\bibinfo {title} {Microscopic analysis of Clausius-Duhem
  processes},\ \href {\doibase 10.1023/a:1004541004050} {\bibfield  {journal}
  {\bibinfo  {journal} {J. Stat. Phys.}\ }\textbf {\bibinfo {volume} {96}},\
  \bibinfo {pages} {415} (\bibinfo {year} {1999})}\BibitemShut {NoStop}%
\bibitem [{\citenamefont {Yang}\ and\ \citenamefont {Qian}(2020)}]{Yang2020}%
  \BibitemOpen
  \bibfield  {author} {\bibinfo {author} {\bibfnamefont {Y.-J.}\ \bibnamefont
  {Yang}}\ and\ \bibinfo {author} {\bibfnamefont {H.}~\bibnamefont {Qian}},\
  }\bibinfo {title} {Unified formalism for entropy production and fluctuation
  relations},\ \href {\doibase 10.1103/physreve.101.022129} {\bibfield
  {journal} {\bibinfo  {journal} {Phys. Rev. E}\ }\textbf {\bibinfo {volume}
  {101}},\ \bibinfo {pages} {022129} (\bibinfo {year} {2020})}\BibitemShut
  {NoStop}%
\bibitem [{\citenamefont {Manzano}\ \emph {et~al.}(2018)\citenamefont
  {Manzano}, \citenamefont {Horowitz},\ and\ \citenamefont
  {Parrondo}}]{Manzano2018}%
  \BibitemOpen
  \bibfield  {author} {\bibinfo {author} {\bibfnamefont {G.}~\bibnamefont
  {Manzano}}, \bibinfo {author} {\bibfnamefont {J.~M.}\ \bibnamefont
  {Horowitz}}, \ and\ \bibinfo {author} {\bibfnamefont {J.~M.~R.}\ \bibnamefont
  {Parrondo}},\ }\bibinfo {title} {Quantum Fluctuation Theorems for Arbitrary
  Environments: Adiabatic and Nonadiabatic Entropy Production},\ \href
  {\doibase 10.1103/physrevx.8.031037} {\bibfield  {journal} {\bibinfo
  {journal} {Phys. Rev. X}\ }\textbf {\bibinfo {volume} {8}},\ \bibinfo {pages}
  {031037} (\bibinfo {year} {2018})}\BibitemShut {NoStop}%
\bibitem [{\citenamefont {Rao}\ and\ \citenamefont {Esposito}(2018)}]{Rao2018}%
  \BibitemOpen
  \bibfield  {author} {\bibinfo {author} {\bibfnamefont {R.}~\bibnamefont
  {Rao}}\ and\ \bibinfo {author} {\bibfnamefont {M.}~\bibnamefont {Esposito}},\
  }\bibinfo {title} {Detailed Fluctuation Theorems: A Unifying Perspective},\
  \href {\doibase 10.3390/e20090635} {\bibfield  {journal} {\bibinfo  {journal}
  {Entropy}\ }\textbf {\bibinfo {volume} {20}},\ \bibinfo {pages} {635}
  (\bibinfo {year} {2018})}\BibitemShut {NoStop}%
\bibitem [{\citenamefont {Garc{\'{\i}}a-Garc{\'{\i}}a}\ \emph
  {et~al.}(2012)\citenamefont {Garc{\'{\i}}a-Garc{\'{\i}}a}, \citenamefont
  {Lecomte}, \citenamefont {Kolton},\ and\ \citenamefont
  {Dom{\'{\i}}nguez}}]{GarciaGarcia2012}%
  \BibitemOpen
  \bibfield  {author} {\bibinfo {author} {\bibfnamefont {R.}~\bibnamefont
  {Garc{\'{\i}}a-Garc{\'{\i}}a}}, \bibinfo {author} {\bibfnamefont
  {V.}~\bibnamefont {Lecomte}}, \bibinfo {author} {\bibfnamefont {A.~B.}\
  \bibnamefont {Kolton}}, \ and\ \bibinfo {author} {\bibfnamefont
  {D.}~\bibnamefont {Dom{\'{\i}}nguez}},\ }\bibinfo {title} {Joint probability
  distributions and fluctuation theorems},\ \href {\doibase
  10.1088/1742-5468/2012/02/p02009} {\bibfield  {journal} {\bibinfo  {journal}
  {J. Stat. Mech.: Theory Exp.}\ }\textbf {\bibinfo {volume} {2012}},\ \bibinfo
  {pages} {P02009} (\bibinfo {year} {2012})}\BibitemShut {NoStop}%
\bibitem [{\citenamefont {Miller}\ \emph {et~al.}(2021)\citenamefont {Miller},
  \citenamefont {Mohammady}, \citenamefont {Perarnau-Llobet},\ and\
  \citenamefont {Guarnieri}}]{Miller2021}%
  \BibitemOpen
  \bibfield  {author} {\bibinfo {author} {\bibfnamefont {H.~J.~D.}\
  \bibnamefont {Miller}}, \bibinfo {author} {\bibfnamefont {M.~H.}\
  \bibnamefont {Mohammady}}, \bibinfo {author} {\bibfnamefont {M.}~\bibnamefont
  {Perarnau-Llobet}}, \ and\ \bibinfo {author} {\bibfnamefont {G.}~\bibnamefont
  {Guarnieri}},\ }\bibinfo {title} {Joint statistics of work and entropy
  production along quantum trajectories},\ \href {\doibase
  10.1103/physreve.103.052138} {\bibfield  {journal} {\bibinfo  {journal}
  {Phys. Rev. E}\ }\textbf {\bibinfo {volume} {103}},\ \bibinfo {pages}
  {052138} (\bibinfo {year} {2021})}\BibitemShut {NoStop}%
\bibitem [{\citenamefont {Denzler}\ \emph {et~al.}(2021)\citenamefont
  {Denzler}, \citenamefont {Santos}, \citenamefont {Lutz},\ and\ \citenamefont
  {Serra}}]{Denzler2021}%
  \BibitemOpen
  \bibfield  {author} {\bibinfo {author} {\bibfnamefont {T.}~\bibnamefont
  {Denzler}}, \bibinfo {author} {\bibfnamefont {J.~F.~G.}\ \bibnamefont
  {Santos}}, \bibinfo {author} {\bibfnamefont {E.}~\bibnamefont {Lutz}}, \ and\
  \bibinfo {author} {\bibfnamefont {R.}~\bibnamefont {Serra}},\ }\bibinfo
  {title} {Nonequilibrium fluctuations of a quantum heat engine},\ \href@noop
  {} {\  (\bibinfo {year} {2021})},\ \Eprint {http://arxiv.org/abs/2104.13427}
  {arXiv:2104.13427 [quant-ph]} \BibitemShut {NoStop}%
\bibitem [{\citenamefont {Kramers}(1940)}]{Kramers1940}%
  \BibitemOpen
  \bibfield  {author} {\bibinfo {author} {\bibfnamefont {H.}~\bibnamefont
  {Kramers}},\ }\bibinfo {title} {Brownian motion in a field of force and the
  diffusion model of chemical reactions},\ \href {\doibase
  10.1016/s0031-8914(40)90098-2} {\bibfield  {journal} {\bibinfo  {journal}
  {Physica}\ }\textbf {\bibinfo {volume} {7}},\ \bibinfo {pages} {284}
  (\bibinfo {year} {1940})}\BibitemShut {NoStop}%
\bibitem [{\citenamefont {Talkner}\ and\ \citenamefont
  {H\"anggi}(2020)}]{RevModPhys.92.041002}%
  \BibitemOpen
  \bibfield  {author} {\bibinfo {author} {\bibfnamefont {P.}~\bibnamefont
  {Talkner}}\ and\ \bibinfo {author} {\bibfnamefont {P.}~\bibnamefont
  {H\"anggi}},\ }\bibinfo {title} {Colloquium: Statistical mechanics and
  thermodynamics at strong coupling: Quantum and classical},\ \href {\doibase
  10.1103/RevModPhys.92.041002} {\bibfield  {journal} {\bibinfo  {journal}
  {Rev. Mod. Phys.}\ }\textbf {\bibinfo {volume} {92}},\ \bibinfo {pages}
  {041002} (\bibinfo {year} {2020})}\BibitemShut {NoStop}%
\bibitem [{\citenamefont {Smoluchowski}(1916)}]{Smoluchowski1916}%
  \BibitemOpen
  \bibfield  {author} {\bibinfo {author} {\bibfnamefont {M.~V.}\ \bibnamefont
  {Smoluchowski}},\ }\bibinfo {title} {\"{U}ber Brownsche Molekularbewegung
  unter Einwirkung \"{a}u{\ss}erer Kr\"{a}fte und deren Zusammenhang mit der
  verallgemeinerten Diffusionsgleichung},\ \href {\doibase
  10.1002/andp.19163532408} {\bibfield  {journal} {\bibinfo  {journal} {Ann.
  Phys. (Berlin)}\ }\textbf {\bibinfo {volume} {353}},\ \bibinfo {pages} {1103}
  (\bibinfo {year} {1916})}\BibitemShut {NoStop}%
\bibitem [{\citenamefont {Bocquet}(1997)}]{Bocquet1997}%
  \BibitemOpen
  \bibfield  {author} {\bibinfo {author} {\bibfnamefont {L.}~\bibnamefont
  {Bocquet}},\ }\bibinfo {title} {High friction limit of the Kramers equation:
  The multiple time-scale approach},\ \href {\doibase 10.1119/1.18507}
  {\bibfield  {journal} {\bibinfo  {journal} {Am. J. Phys.}\ }\textbf {\bibinfo
  {volume} {65}},\ \bibinfo {pages} {140} (\bibinfo {year} {1997})}\BibitemShut
  {NoStop}%
\bibitem [{\citenamefont {Pan}\ \emph {et~al.}(2018)\citenamefont {Pan},
  \citenamefont {Hoang}, \citenamefont {Fei}, \citenamefont {Qiu},
  \citenamefont {Ahn}, \citenamefont {Li},\ and\ \citenamefont
  {Quan}}]{Pan2018}%
  \BibitemOpen
  \bibfield  {author} {\bibinfo {author} {\bibfnamefont {R.}~\bibnamefont
  {Pan}}, \bibinfo {author} {\bibfnamefont {T.~M.}\ \bibnamefont {Hoang}},
  \bibinfo {author} {\bibfnamefont {Z.}~\bibnamefont {Fei}}, \bibinfo {author}
  {\bibfnamefont {T.}~\bibnamefont {Qiu}}, \bibinfo {author} {\bibfnamefont
  {J.}~\bibnamefont {Ahn}}, \bibinfo {author} {\bibfnamefont {T.}~\bibnamefont
  {Li}}, \ and\ \bibinfo {author} {\bibfnamefont {H.~T.}\ \bibnamefont
  {Quan}},\ }\bibinfo {title} {Quantifying the validity and breakdown of the
  overdamped approximation in stochastic thermodynamics: Theory and
  experiment},\ \href {\doibase 10.1103/physreve.98.052105} {\bibfield
  {journal} {\bibinfo  {journal} {Phys. Rev. E}\ }\textbf {\bibinfo {volume}
  {98}},\ \bibinfo {pages} {052105} (\bibinfo {year} {2018})}\BibitemShut
  {NoStop}%
\bibitem [{\citenamefont {Paraguass{\'{u}}}\ \emph {et~al.}(2022)\citenamefont
  {Paraguass{\'{u}}}, \citenamefont {Aquino}, \citenamefont {Defaveri},\ and\
  \citenamefont {Morgado}}]{Paraguassu2022}%
  \BibitemOpen
  \bibfield  {author} {\bibinfo {author} {\bibfnamefont {P.~V.}\ \bibnamefont
  {Paraguass{\'{u}}}}, \bibinfo {author} {\bibfnamefont {R.}~\bibnamefont
  {Aquino}}, \bibinfo {author} {\bibfnamefont {L.}~\bibnamefont {Defaveri}}, \
  and\ \bibinfo {author} {\bibfnamefont {W.~A.~M.}\ \bibnamefont {Morgado}},\
  }\bibinfo {title} {Effects of the kinetic energy in heat for overdamped
  systems},\ \href {\doibase 10.1103/physreve.106.044106} {\bibfield  {journal}
  {\bibinfo  {journal} {Phys. Rev. E}\ }\textbf {\bibinfo {volume} {106}},\
  \bibinfo {pages} {044106} (\bibinfo {year} {2022})}\BibitemShut {NoStop}%
\bibitem [{\citenamefont {Mart{\'{\i}}nez}\ \emph {et~al.}(2016)\citenamefont
  {Mart{\'{\i}}nez}, \citenamefont {Petrosyan}, \citenamefont
  {Gu{\'{e}}ry-Odelin}, \citenamefont {Trizac},\ and\ \citenamefont
  {Ciliberto}}]{Martinez2016}%
  \BibitemOpen
  \bibfield  {author} {\bibinfo {author} {\bibfnamefont {I.~A.}\ \bibnamefont
  {Mart{\'{\i}}nez}}, \bibinfo {author} {\bibfnamefont {A.}~\bibnamefont
  {Petrosyan}}, \bibinfo {author} {\bibfnamefont {D.}~\bibnamefont
  {Gu{\'{e}}ry-Odelin}}, \bibinfo {author} {\bibfnamefont {E.}~\bibnamefont
  {Trizac}}, \ and\ \bibinfo {author} {\bibfnamefont {S.}~\bibnamefont
  {Ciliberto}},\ }\bibinfo {title} {Engineered swift equilibration of a
  Brownian~particle},\ \href {\doibase 10.1038/nphys3758} {\bibfield  {journal}
  {\bibinfo  {journal} {Nat. Phys.}\ }\textbf {\bibinfo {volume} {12}},\
  \bibinfo {pages} {843} (\bibinfo {year} {2016})}\BibitemShut {NoStop}%
\bibitem [{\citenamefont {Li}\ \emph {et~al.}(2017)\citenamefont {Li},
  \citenamefont {Quan},\ and\ \citenamefont {Tu}}]{Li2017}%
  \BibitemOpen
  \bibfield  {author} {\bibinfo {author} {\bibfnamefont {G.}~\bibnamefont
  {Li}}, \bibinfo {author} {\bibfnamefont {H.~T.}\ \bibnamefont {Quan}}, \ and\
  \bibinfo {author} {\bibfnamefont {Z.~C.}\ \bibnamefont {Tu}},\ }\bibinfo
  {title} {Shortcuts to isothermality and nonequilibrium work relations},\
  \href {\doibase 10.1103/physreve.96.012144} {\bibfield  {journal} {\bibinfo
  {journal} {Phys. Rev. E}\ }\textbf {\bibinfo {volume} {96}},\ \bibinfo
  {pages} {012144} (\bibinfo {year} {2017})}\BibitemShut {NoStop}%
\bibitem [{\citenamefont {Li}\ \emph {et~al.}(2022)\citenamefont {Li},
  \citenamefont {Chen}, \citenamefont {Sun},\ and\ \citenamefont
  {Dong}}]{Li2021}%
  \BibitemOpen
  \bibfield  {author} {\bibinfo {author} {\bibfnamefont {G.}~\bibnamefont
  {Li}}, \bibinfo {author} {\bibfnamefont {J.-F.}\ \bibnamefont {Chen}},
  \bibinfo {author} {\bibfnamefont {C.~P.}\ \bibnamefont {Sun}}, \ and\
  \bibinfo {author} {\bibfnamefont {H.}~\bibnamefont {Dong}},\ }\bibinfo
  {title} {Geodesic Path for the Minimal Energy Cost in Shortcuts to
  Isothermality},\ \href {\doibase 10.1103/PhysRevLett.128.230603} {\bibfield
  {journal} {\bibinfo  {journal} {Phys. Rev. Lett.}\ }\textbf {\bibinfo
  {volume} {128}},\ \bibinfo {pages} {230603} (\bibinfo {year}
  {2022})}\BibitemShut {NoStop}%
\bibitem [{\citenamefont {Chen}(2022)}]{Chen2022}%
  \BibitemOpen
  \bibfield  {author} {\bibinfo {author} {\bibfnamefont {J.-F.}\ \bibnamefont
  {Chen}},\ }\bibinfo {title} {Optimizing Brownian heat engine with shortcut
  strategy},\ \href {\doibase 10.1103/physreve.106.054108} {\bibfield
  {journal} {\bibinfo  {journal} {Phys. Rev. E}\ }\textbf {\bibinfo {volume}
  {106}},\ \bibinfo {pages} {054108} (\bibinfo {year} {2022})}\BibitemShut
  {NoStop}%
\end{thebibliography}%

\end{document}